\documentclass[iop,apj]{emulateapj}
\pdfoutput=1
\usepackage{amsmath,amssymb,amstext}
\usepackage{graphicx}
\usepackage{epstopdf}
\usepackage[breaklinks,colorlinks,citecolor=blue,linkcolor=magenta]{hyperref}

\usepackage[all]{hypcap}
\usepackage{natbib}
\newcommand\kms{\ifmmode{\rm km\thinspace s^{-1}}\else km\thinspace s$^{-1}$\fi}
\newcommand\ms{\ifmmode{\rm m\thinspace s^{-1}}\else m\thinspace s$^{-1}$\fi}
\newcommand\msun{\ifmmode{M_{\odot}}\else $M_{\odot}$\fi}
\newcommand\rsun{\ifmmode{R_{\odot}}\else $R_{\odot}$\fi}
\newcommand\lsun{\ifmmode{L_{\odot}}\else $L_{\odot}$\fi}
\newcommand\mjup{\ifmmode{M_{\rm Jup}}\else $M_{\rm Jup}$\fi}
\newcommand\rjup{\ifmmode{R_{\rm Jup}}\else $R_{\rm Jup}$\fi}
\newcommand\mearth{\ifmmode{M_\oplus}\else $M_\oplus$\fi}
\newcommand\rearth{\ifmmode{R_\oplus}\else $R_\oplus$\fi}
\newcommand\mpl{\ifmmode{M_{\rm P}}\else $M_{\rm P}$\fi}
\newcommand\rp{\ifmmode{R_{\rm P}}\else $R_{\rm P}$\fi}
\newcommand\mpb{\ifmmode{M_{\rm b}}\else $M_{\rm b}$\fi}
\newcommand\rpb{\ifmmode{R_{\rm b}}\else $R_{\rm b}$\fi}
\newcommand\mpc{\ifmmode{M_{\rm c}}\else $M_{\rm c}$\fi}
\newcommand\rpc{\ifmmode{R_{\rm c}}\else $R_{\rm c}$\fi}
\def\numax{$\nu_{\rm max}$}
\def\Dnu{$\Delta\nu$}

\def\uhz{$\mu\rm{Hz}$}
\def\teff{$T_{\rm eff}$}
\def\logg{$\log(g)$}
\newcommand{\feh}{\mbox{$\rm{[Fe/H]}$}}
\newcommand\this{{\em Kepler}-432}

\newcommand\mysim{\mathord{\sim}}
\newcommand\dg{\ensuremath{^\circ}}
\def\lsim{\mathrel{\rlap{\lower4pt\hbox{\hskip1pt$\sim$}}
    \raise1pt\hbox{$<$}}}
\def\gsim{\mathrel{\rlap{\lower4pt\hbox{\hskip1pt$\sim$}}
    \raise1pt\hbox{$>$}}}
\newcommand{\reffigl}[1]{Figure~\ref{fig:#1}}
\newcommand{\refsecl}[1]{\mbox{Section \ref{sec:#1}}}
\newcommand{\refsecs}[2]{\mbox{Sections \ref{sec:#1} and \ref{sec:#2}}}

\newcommand{\reftabl}[1]{Table~\ref{tab:#1}}
\submitted{Accepted to ApJ Jan 24, 2015}

\shorttitle{\this: a red giant multi-planet host}
\shortauthors{Quinn et al.}

\begin{document}

\title{\this: a Red Giant Interacting with One of its Two Long Period
Giant Planets}

\author{
  Samuel N. Quinn\altaffilmark{1,18},
  Timothy. R. White\altaffilmark{2},
  David W. Latham\altaffilmark{3},
  William J. Chaplin\altaffilmark{4,5},
  Rasmus Handberg\altaffilmark{4,5},
  Daniel Huber\altaffilmark{6,7,5},
  David M. Kipping\altaffilmark{3,17},
  Matthew J. Payne\altaffilmark{3},
  Chen Jiang\altaffilmark{5},
  Victor Silva Aguirre\altaffilmark{5},
  Dennis Stello\altaffilmark{6,5},
  David H. Sliski\altaffilmark{3,8},
  David R. Ciardi\altaffilmark{9},
  Lars A. Buchhave\altaffilmark{3,10},
  Timothy R. Bedding\altaffilmark{6,5},
  Guy R. Davies\altaffilmark{4,5},
  Saskia Hekker\altaffilmark{11,5},
  Hans Kjeldsen\altaffilmark{5},
  James S. Kuszlewicz\altaffilmark{4,5},
  Mark E. Everett\altaffilmark{12}
  Steve B. Howell\altaffilmark{13},
  Sarbani Basu\altaffilmark{14},
  Tiago L. Campante\altaffilmark{4,5},
  J{\o}rgen Christensen-Dalsgaard\altaffilmark{5},
  Yvonne P. Elsworth\altaffilmark{4,5},
  Christoffer Karoff\altaffilmark{5,15},
  Steven D. Kawaler\altaffilmark{16},
  Mikkel N. Lund\altaffilmark{5,4},
  Mia Lundkvist\altaffilmark{5},
  Gilbert A. Esquerdo\altaffilmark{3},
  Michael L. Calkins\altaffilmark{3}, and
  Perry Berlind\altaffilmark{3}
}

\altaffiltext{1}{Department of Physics and Astronomy, Georgia State
  University, 25 Park Place Suite 605, Atlanta, GA 30303, USA}
\altaffiltext{2}{Institut f\"ur Astrophysik,
  Georg-August-Universit\"at G\"ottingen, Friedrich-Hund-Platz 1,
  D-37077 G\"ottingen, Germany}
\altaffiltext{3}{Harvard-Smithsonian Center for Astrophysics, 60
  Garden St, Cambridge, MA 02138, USA}
\altaffiltext{4}{School of Physics and Astronomy, University of
  Birmingham, Birmingham, B15 2TT, UK}
\altaffiltext{5}{Stellar Astrophysics Centre, Department of Physics
  and Astronomy, Aarhus University, Ny Munkegade 120, DK-8000 Aarhus
  C, Denmark}
\altaffiltext{6}{Sydney Institute for Astronomy (SIfA), School of
  Physics, University of Sydney, NSW 2006, Australia}
\altaffiltext{7}{SETI Institute, 189 Bernardo Avenue, Mountain View,
  CA 94043, USA}
\altaffiltext{8}{Department of Physics and Astronomy, University of
  Pennsylvania, 219 S. 33rd St., Philadelphia, PA 19104, USA}
\altaffiltext{9}{NASA Exoplanet Science Institute/Caltech, Pasdena, CA
  91125, USA}
\altaffiltext{10}{Centre for Star and Planet Formation, Natural History
  Museum of Denmark, University of Copenhagen, DK-1350 Copenhagen,
  Denmark}
\altaffiltext{11}{Max-Planck-Institut f\"ur Sonnensystemforschung,
  Justus-von-Liebig-Weg 3, D-37077 G\"ottingen, Germany}
\altaffiltext{12}{National Optical Astronomy Observatories, 950 North
  Cherry Avenue, Tucson, AZ 85719, USA}
\altaffiltext{13}{NASA Ames Research Center, Moffet Field, Mountain
  View, CA 94035, USA}
\altaffiltext{14}{Department of Astronomy, Yale University, PO Box
  208101, New Haven, CT 06520, USA}
\altaffiltext{15}{Department of Geoscience, Aarhus University,
  H{\oe}egh-Guldbergs Gade 2, 8000, Aarhus C, Denmark}
\altaffiltext{16}{Department of Physics and Astronomy, Iowa State
  University, Ames, IA 50011, USA}
\altaffiltext{17}{NASA Carl Sagan Fellow}
\altaffiltext{18}{NSF Graduate Research Fellow}

\begin{abstract}

We report the discovery of \this{b}, a giant planet ($\mpb =
5.41^{+0.32}_{-0.18}~\mjup$, $\rpb = 1.145^{+0.036}_{-0.039}~\rjup$)
transiting an evolved star ($M_\star = 1.32^{+0.10}_{-0.07}~\msun,
R_\star = 4.06^{+0.12}_{-0.08}~\rsun$) with an orbital period of
$P_{\rm b} = 52.501129^{+0.000067}_{-0.000053}$~days. Radial
velocities (RVs) reveal that \this{b} orbits its parent star with an
eccentricity of $e=0.5134^{+0.0098}_{-0.0089}$, which we also measure
independently with asterodensity profiling (AP;
$e=0.507^{+0.039}_{-0.114}$), thereby confirming the validity of AP on
this particular evolved star. The well determined planetary properties
and unusually large mass also make this planet an important benchmark
for theoretical models of super-Jupiter formation. Long-term RV
monitoring detected the presence of a non-transiting outer planet
(\this{c}; $\mpc\sin{i_{\rm c}} = 2.43^{+0.22}_{-0.24}~\mjup$, $P_{\rm
  c} = 406.2^{+3.9}_{-2.5}$~days), and adaptive optics imaging
revealed a nearby ($0\farcs87$), faint companion (\this{B}) that is a
physically bound M dwarf. The host star exhibits high S/N
asteroseismic oscillations, which enable precise measurements of the
stellar mass, radius and age. Analysis of the rotational splitting of
the oscillation modes additionally reveals the stellar spin axis to be
nearly edge-on, which suggests that the stellar spin is likely
well-aligned with the orbit of the transiting planet. Despite its long
period, the obliquity of the $52.5$-day orbit may have been shaped by
star--planet interaction in a manner similar to hot Jupiter systems,
and we present observational and theoretical evidence to support this
scenario. Finally, as a short-period outlier among giant planets
orbiting giant stars, study of \this{b} may help explain the
distribution of massive planets orbiting giant stars interior to
$1$\,AU.

\end{abstract}

\keywords{ asteroseismology --- planets and satellites: dynamical
evolution and stability --- planets and satellites: formation ---
planets and satellites: gaseous planets --- planet-star interactions
--- stars: individual (\this, KIC 10864656, KOI-1299)}

\section{Introduction}

The NASA {\em Kepler} mission \citep{borucki:2010}, at its heart a
statistical endeavor, has provided a rich dataset that enables
ensemble studies of planetary populations, from gas giants to
Earth-sized planets. Such investigations can yield valuable
statistical constraints for theories of planetary formation and
subsequent dynamical evolution
\citep[e.g.,][]{buchhave:2014,steffen:2012}. Individual discoveries,
however, provide important case studies to explore these processes in
detail, especially in parameter space for which populations remain
small. Because of its unprecedented photometric sensitivity, duty
cycle, and time coverage, companions that are intrinsically rare or
otherwise difficult to detect are expected to be found by {\em
  Kepler}, and detailed study of such discoveries can lead to
characterization of poorly understood classes of objects and physical
processes.

Planets orbiting red giants are of interest because they trace the
planetary population around their progenitors, many of which are
massive and can be hard to survey while they reside on the main
sequence. Stars more massive than about $1.3~\msun$ \citep[the
so-called Kraft break;][]{kraft:1967} have negligible convective
envelopes, which prevents the generation of the magnetic winds that
drive angular momentum loss in smaller stars. Their rapid rotation and
high temperatures---resulting in broad spectral features that are
sparse in the optical---make precise radial velocities (RVs) extremely
difficult with current techniques. However, as they evolve to become
giant stars, they cool and spin down, making them ideal targets for
precise radial velocity work \citep[see, e.g.,][and references
therein]{johnson:2011}. There is already good evidence that planetary
populations around intermediate mass stars are substantially different
from those around their low mass counterparts. Higher mass stars seem
to harbor more Jupiters than do Sun-like stars \citep{johnson:2010},
and the typical planetary mass correlates with the stellar mass
\citep{lovis:2007, dollinger:2009, bowler:2010}, but there are not
many planets within $1$\,AU of more massive stars
\citep{johnson:2007,sato:2008}, and their orbits tend to be less
eccentric than Jupiters orbiting low mass stars
\citep{jones:2014}. Due to the observational difficulties associated
with massive and intermediate-mass main-sequence stars, many of the
more massive stars known to host planets have already reached an
advanced evolutionary state, and it is not yet clear whether most of
the orbital differences can be attributed to mass-dependent formation
and migration, or if planetary engulfment and/or tidal evolution as
the star swells on the giant branch plays a more important role.

While the number of planets known to orbit evolved stars has become
substantial, because of their typically long periods, not many
transit, and thus very few are amenable to detailed study. In fact,
{\em Kepler}-91b \citep{lillo:2014a,lillo:2014b,barclay:2014} and {\em Kepler}-56c
\citep{huber:2013b} are the only two massive planets ($M_p>0.5~\mjup$)
orbiting giant stars ($\log{g}<3.9$) that are known to
transit\footnote{Among the other $95$ such planets listed in The
Exoplanet Orbit Database (exoplanets.org), none transit.}. A transit
leads to a radius measurement, enabling investigation of interior
structure and composition via bulk density constraints and theoretical
models \citep[e.g.,][]{fortney:2007}, and also opens up the
possibility of atmospheric studies
\citep[e.g.,][]{charbonneau:2002,knutson:2008,berta:2012,
poppenhaeger:2013}, which can yield more specific details about
planetary structure, weather, or atomic and molecular abundances
within the atmosphere of the planet. Such information can provide
additional clues about the process of planet formation around hot
stars.

In studies of orbital migration of giant planets, the stellar
obliquity---the angle between the stellar spin axis and the orbital
angular momentum vector---has proven to be a valuable measurement, as
it holds clues about the dynamical history of the planetary system
\citep[see, e.g.,][]{albrecht:2012}. Assuming that the protoplanetary
disk is coplanar with the stellar equator, and thus the rotational and
orbital angular momenta start out well-aligned, some migration
mechanisms--e.g., Type II migration
\citep{goldreich:1980,lin:1986}---are expected to preserve low
obliquities, while others---for example, Kozai cycles
\citep[e.g.,][]{wu:2003,fabrycky:2007}, secular chaos \citep{wu:2011},
or planet-planet scattering
\citep[e.g.,][]{rasio:1996,juric:2008}---may excite large orbital
inclinations. Measurements of stellar obliquity can thus potentially
distinguish between classes of planetary migration.

The Rossiter--McLaughlin effect \citep{mclaughlin:1924,rossiter:1924}
has been the main source of (projected) obliquity measurements,
although at various times starspot crossings
\citep[e.g.,][]{desert:2011,nutzman:2011,sanchis:2011}, Doppler
tomography of planetary transits
\citep[e.g.,][]{collier:2010,johnson:2014}, gravity-darkened models of
the stellar disk during transit \citep[e.g.,][]{barnes:2011}, and the
rotational splitting of asteroseismic oscillation modes
\citep[e.g.,][]{huber:2013b} have all been used to determine absolute
or projected stellar obliquities. Most obliquity measurements to date
have been for hot Jupiters orbiting Sun-like stars, but to get a full
picture of planetary migration, we must study the dynamical histories
of planets across a range of separations and in a variety of
environments. Luckily, the diversity of techniques with which we can
gather this information allows us to begin investigating spin--orbit
misalignment for long period planets and around stars of various
masses and evolutionary states. For long-period planets orbiting
slowly rotating giant stars, the Rossiter--McLaughlin amplitudes are
small because they scale with the stellar rotational velocity and the
planet-to-star area ratio, and few transits occur, which limits the
opportunities to obtain follow-up measurements or to identify the
transit geometry from spot crossings. Fortunately, the detection of
asteroseismic modes does not require rapid rotation, and is
independent of the planetary properties, so it becomes a valuable tool
for long period planets orbiting evolved stars. The high precision,
high duty-cycle, long timespan, photometric observations of {\em
Kepler} are ideal for both identifying long period transiting planets
and examining the asteroseismic properties of their host stars.

In this paper, we highlight the discovery of \this{b} and c, a pair of
giant planets in long period orbits ($>50$~days) around an
oscillating, intermediate mass red giant. We present the photometric
observations and transit light curve analysis of \this~in
\refsecl{phot} and the follow-up imaging and spectroscopy in
\refsecs{im}{spec}, followed by the asteroseismic and radial velocity
analyses in \refsecs{seis}{rv}. False positive scenarios and orbital
stability are investigated in \refsecs{fp}{nbody}, and we
discuss the system in the context of planet formation and migration in
\refsecl{disc}, paying particular attention to star--planet
interactions (SPIs) and orbital evolution during the red giant phase. We
provide a summary in \refsecl{summary}.

\section{Photometry}
\label{sec:phot}
\subsection{Kepler Observations}
\label{sec:kepler}

\begin{figure}[!tb]
\includegraphics[width=0.48\textwidth]{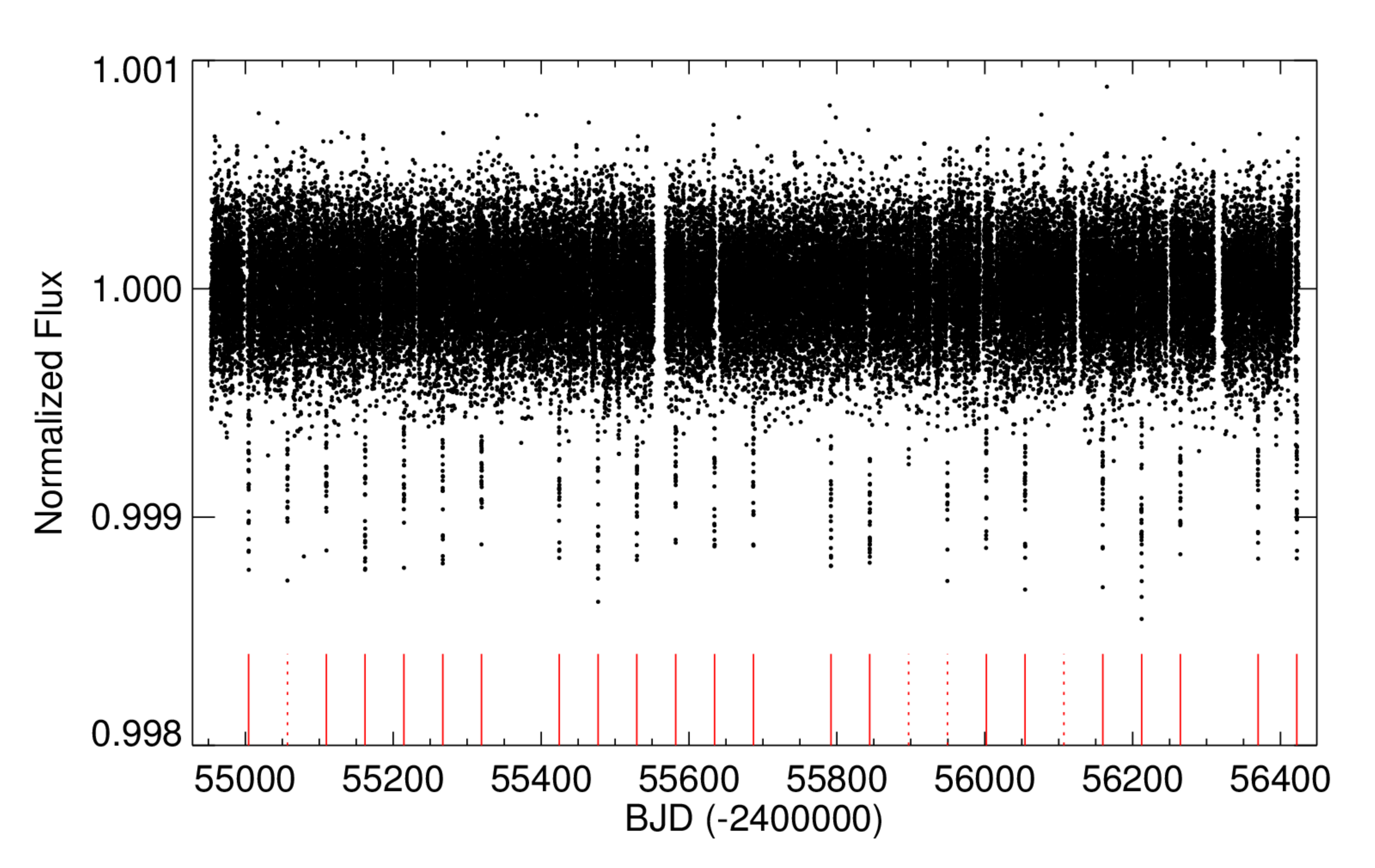}\hfill
\caption{ Detrended and normalized long cadence {\em Kepler} time
  series for \this, spanning $1470.5$~days. The transits, which are
  clearly visible even in the unfolded data, are indicated by red
  lines. Solid lines denote full transits, while dotted lines denote
  that only a partial transit was observed. Three of the $28$~expected
  transits occurred entirely during data gaps.
\label{fig:kepts}}
\end{figure}

The {\em Kepler} mission and its photometric performance are described
in \citet{borucki:2010}, and the characteristics of the detector on
board the spacecraft are described in \citet{koch:2010} and
\citet{vancleve:2008}. The photometric observations of \this~span {\em
Kepler} observation Quarters 0 through 17 (JD 2454953.5 to 2456424.0),
a total of $1470.5$~days. \this{b} was published by the {\em Kepler}
team as a {\em Kepler} Object of Interest (KOI) and planetary
candidate \citep[designated KOI-1299; see][]{batalha:2013}, and after
also being identified as a promising asteroseismic target, it was
observed in short cadence (SC) mode for $8$~quarters. We note that a
pair of recent papers have now confirmed the planetary nature of this
transiting companion via radial velocity measurements
\citep{ciceri:2015,ortiz:2015}.

The full photometric timeseries, normalized in each quarter, is shown
in \reffigl{kepts}. A transit signature with a period of $\mysim
52.5$~days is apparent in the data, and our investigation of the
transits is described in the following section.

\subsection{Light Curve Analysis}
\label{sec:lc}

A transit light curve analysis of \this{b} was performed previously by
\citet{sliski:2014}. In that work, the authors first detrended the
Simple Aperture Photometry (SAP) {\em Kepler}
data\footnote{Observations labeled as SAP\_FLUX in FITS files
retrieved from the Barbara A. Mikulski Archive for Space Telescopes
(MAST).}  for quarters $1$--$17$ using the CoFiAM (Cosine Filtering
with Autocorrelation Minimization) algorithm and then regressed the
cleaned data with the multimodal nested sampling algorithm MultiNest
\citep{feroz:2009} coupled to a \citet{mandel:2002} planetary transit
model. Details on the priors employed and treatment of limb darkening
are described in \citet{sliski:2014}. The authors compared the light
curve derived stellar density, $\rho_{\star,\rm{obs}}$, to that from
asteroseismology, $\rho_{\star,\rm{astero}}$, in a procedure dubbed
``Asterodensity Profiling'' \citep[AP;][]{kipping:2012,kipping:2014a}
to constrain the planet's minimum orbital eccentricity as being
$e_{\rm min} = 0.488_{-0.051}^{+0.025}$. The minimum eccentricity is
most easily retrieved with AP but the proper eccentricity (and
argument of periastron, $\omega$) can be estimated by including $e$
and $\omega$ as free parameters in the light curve fit and
marginalizing over $\omega$. In order to estimate the proper
eccentricity, we were motivated to re-visit the {\em Kepler} data, as
described below.

\begin{figure}[!tb]
\includegraphics[width=0.48\textwidth]{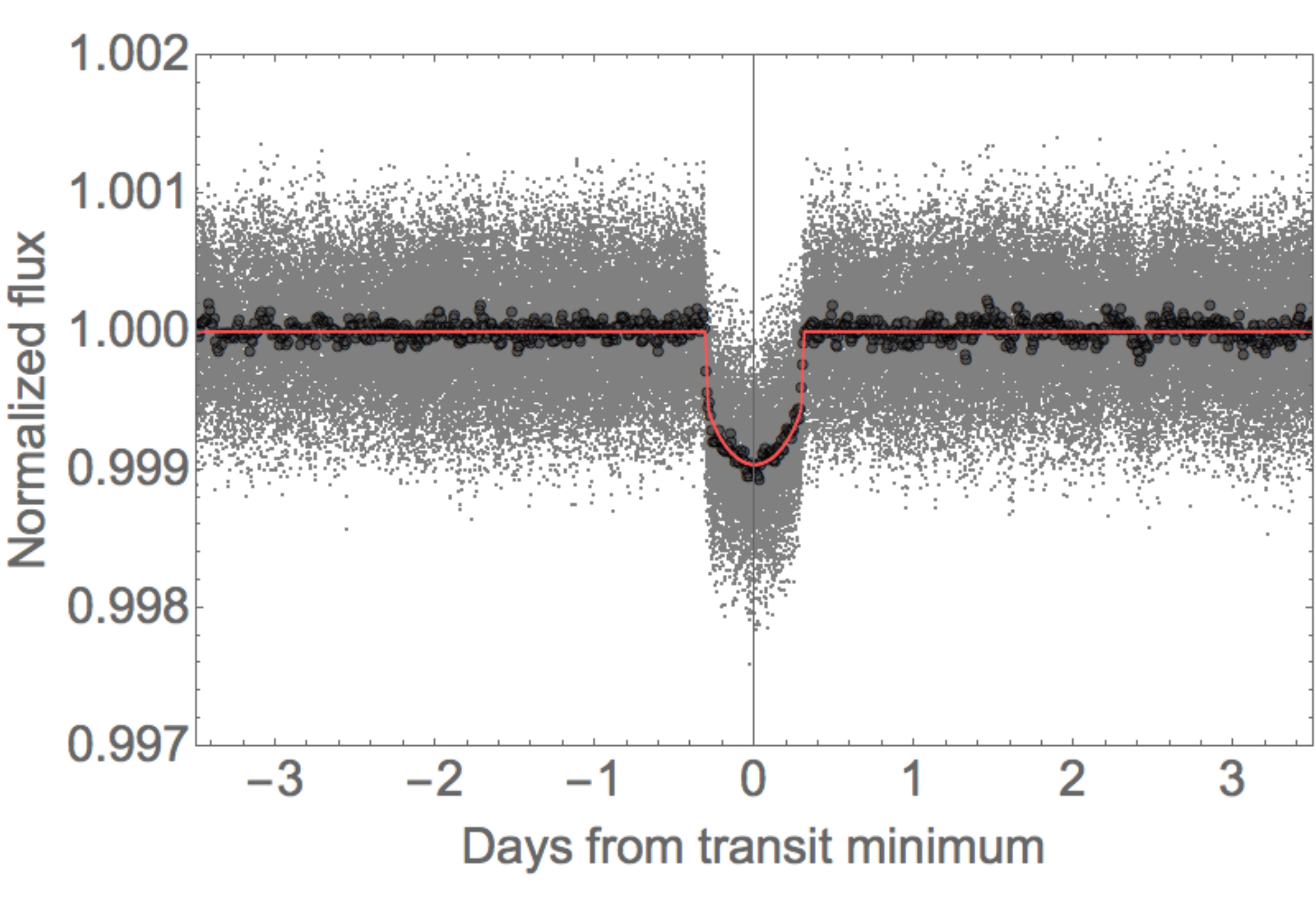}\hfill
\caption{ Folded short cadence {\em Kepler} light curve, shown as gray
points. For clarity, the binned data (every $100$~points) are
overplotted as large dark circles, and the best fit, with parameters
reported in \reftabl{transit}, is indicated by the solid red line.
\label{fig:folded_transit}}
\end{figure}

We first detrended the {\em Kepler} SAP data as was done in
\citet{sliski:2014}, by using the CoFiAM algorithm, which is described
in detail in \citet{kipping:2013}. CoFiAM acts like a harmonic filter,
removing any long term periodicities in the data but protecting those
variations occurring on the timescale of the transit or shorter, so as
to retain the true light curve shape. The algorithm requires an
estimate of the times of transit minimum, orbital period, and full
transit duration. Since \citet{sliski:2014} provided refined values
for these quantities, we used these updated values to conduct a
revised CoFiAM detrending of the {\em Kepler} data. As with the
previous analysis, the final light curves are optimized for a window
within three transit durations of the transit minima.

Due to the effects of stellar granulation on the photometry, we find
that the light curve scatter clearly exceeds the typical photometric
uncertainties. In order to obtain more realistic parameter
uncertainties, we added a ``jitter'' term in quadrature to the
photometric uncertainties to yield a reduced chi-squared of unity for
the out-of-transit data. This was done independently for the long- and
short-cadence data, although the photometric jitter terms were (as
expected) nearly identical at $177.4$~and $175.9\,$ppm for the
short- and long-cadence data, respectively.

The $13$~long-cadence transits for which no SC data was available and
the $11$~SC transits were stitched together and regressed to a transit
model using MultiNest. Our light curve model employs the quadratic
limb darkening \citet{mandel:2002} routine with the
\citet{kipping:2010} ``resampling'' prescription for accounting for
the smearing of the long-cadence data. The seven basic parameters in our
light curve fit were ratio-of-radii, $R_{\rm b}/R_\star$, stellar
density, $\rho_{\star}$, impact parameter, $b$, time of transit
minimum, $T_0$, orbital period, $P$, and the limb darkening
coefficients $q_1$~and $q_2$ described in \citet{kipping:2013a}. In
addition, we included an 8th parameter for the log of the contaminated
light fraction from a blend source,
$\log_{10}\beta=\log_{10}(F_{\mathrm{blend}}/F_{\star})$. This was
constrained from adaptive optics imaging (AO; see \refsecl{ao}) to be
$\log \beta = -2.647\pm0.042$ with a Gaussian prior, assuming Gaussian
uncertainties on the magnitudes measured from AO.

Ordinarily, a transit light curve contains very little information on
the orbital eccentricity and thus it is not possible to reach a
converged eccentricity solution with photometry alone
\citep{kipping:2008}. However, in cases where the parent star's mean
density is independently constrained, a transit light curve can be
used to constrain the orbital eccentricity and argument of periastron
\citep{dawson:2012,kipping:2014a}. This technique, an example of AP,
enables us to include $e$~and $\omega$~as our 9th and 10th transit
model parameters.

To enable the use of AP, we impose an informative Gaussian prior on
the mean stellar density given by the asteroseismology constraint
($\rho_{\star,\mathrm{astero}} = 27.94^{+0.55}_{-0.58}$\,kg\,m$^{-3}$;
\refsecl{seis}). We use the ECCSAMPLES code \citep{kipping:2014b} to
draw samples from an appropriate joint $e$--$\omega$ prior. This code
describes the eccentricity distribution as following a Beta
distribution (a weakly informative prior) and then accounts for the
bias in both $e$ and $\omega$ caused by the fact that the planet is
known to transit. For the Beta distribution shape parameters, we use
the ``short'' period calibration ($P<380$d) of \citet{kipping:2013b}:
$a_{\beta}=0.697$ and $b_{\beta}=3.27$.

The maximum a posteriori folded transit light curve is presented in
\reffigl{folded_transit}.  The transit parameters and associated
68.3\% uncertainties, derived solely from this photometric fit, are
reported in \reftabl{transit}.

\capstartfalse
\begin{deluxetable}{lr}
  \tablewidth{0pt}
  \tablecaption{\this~Transit Parameters
    \label{tab:transit}
  }
  \tablehead{
    \colhead{Parameter} &
    \colhead{Value}
  }
  \startdata
  $R_{\rm b}/R_\star$              & $0.02914^{+0.00038}_{-0.00093}$     \\
  $\rho_\star~({\rm kg\,m}^{-3})$  & $27.94^{+0.54}_{-0.58}$             \\
  $b$                              & $0.503^{+0.090}_{-0.168}$           \\
  $T_{\rm 0,b}$~(BJD)              & $2455949.5374^{+0.0018}_{-0.0016}$  \\
  $P_{\rm b}$~(days)               & $52.501134^{+0.000070}_{-0.000107}$ \\
  $q_1$                            & $0.309^{+0.047}_{-0.042}$           \\
  $q_2$                            & $0.674^{+0.151}_{-0.098}$           \\
  $\log\beta$                      & $-2.647 \pm 0.042$                  \\
  $e_{\rm b}$                      & $0.507^{+0.039}_{-0.114}$           \\
  $\omega_{\rm b}$~(degrees)       & $76^{+59}_{-24}$                    \\
  [-2.4ex]
  \enddata
\end{deluxetable}
\capstarttrue

\section{High Spatial Resolution Imaging}
\label{sec:im}
\subsection{Speckle Imaging}
\label{sec:speck}

Speckle imaging observations of \this~were performed on UT 2011 June
16 at the $3.5$~m WIYN telescope on Kitt Peak, AZ, using the
Differential Speckle Survey Instrument
\citep[DSSI;][]{horch:2010}. DSSI provides simultaneous images in two
filters using a dichroic beam splitter and two identical EMCCDs. These
images were obtained in the $R$~($6920$~\AA) and $I$~($8800$~\AA)
bands. Data reduction and analysis of these images is described in
\citet{torres:2011}, \citet{horch:2010}, and \citet{howell:2011}. The
reconstructed $R$- and $I$-band images reveal no stellar companions
brighter than $\Delta R \mysim 4.5$~magnitudes and $\Delta I \mysim
3.5$~magnitudes, within the annulus from $0\farcs05$~to
$2\arcsec$. The contrasts achieved as a function of distance are
plotted in \reffigl{sens}, and represent $5$-$\sigma$ detection
thresholds.

\subsection{Adaptive Optics Imaging}
\label{sec:ao}

\begin{figure}[!tb]
\epsscale{1.2}
\plotone{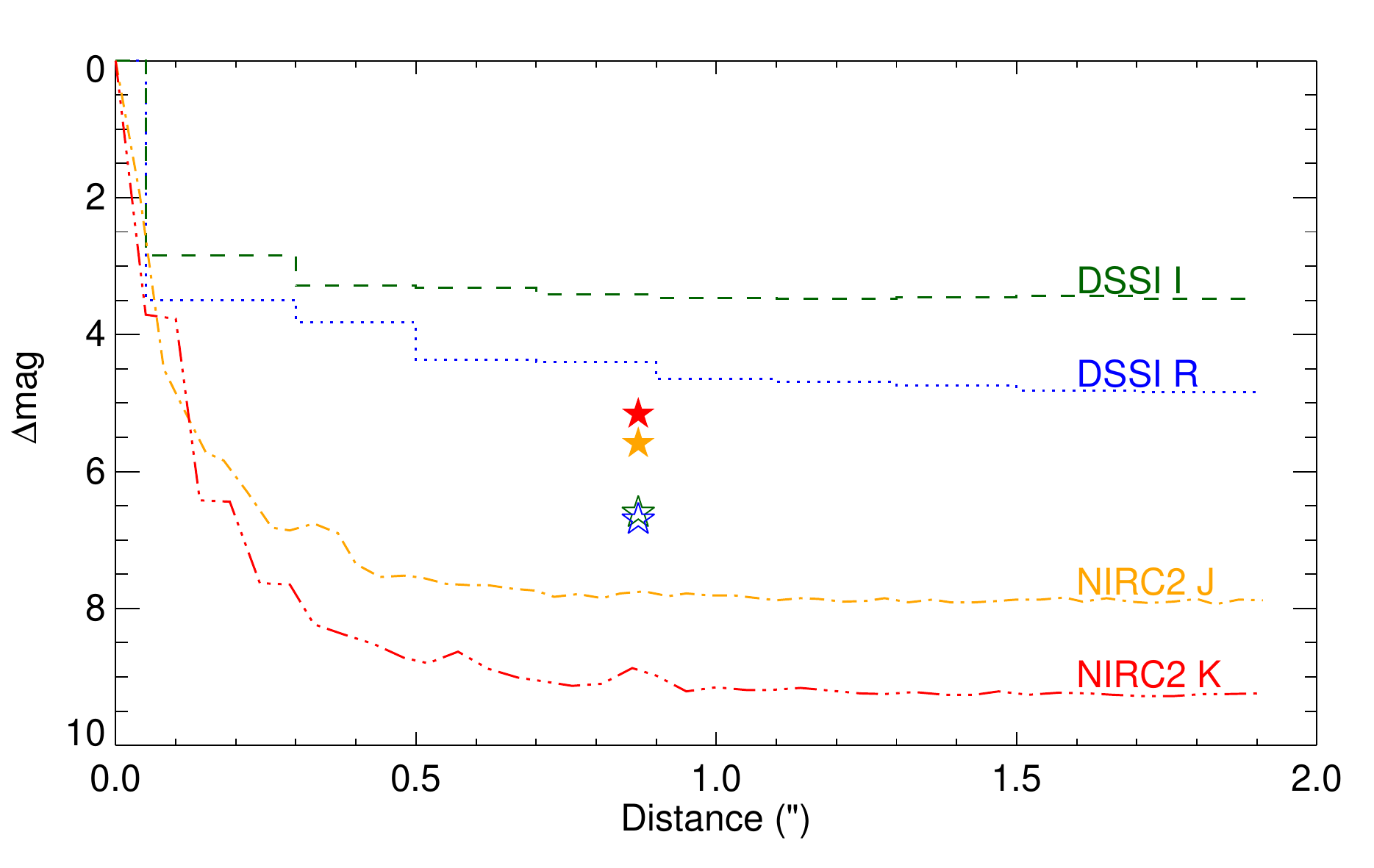}
\caption{ Contrasts as a function of distance for all high resolution
  follow up, and the magnitude difference of the visual companion in
  each filter. Based on its infrared magnitudes from the NIRC2 $J$-
  and $K$-band detections (filled stars), we estimate the companion
  contrast in the $R$- and $I$-bands (open stars) to be $6.7$~and
  $6.6$~magnitudes, respectively. These magnitudes are consistent with
  our non-detections in the DSSI images.
\label{fig:sens}}
\end{figure}

\begin{figure}[!tb]
\centering
\includegraphics[width=0.23\textwidth]{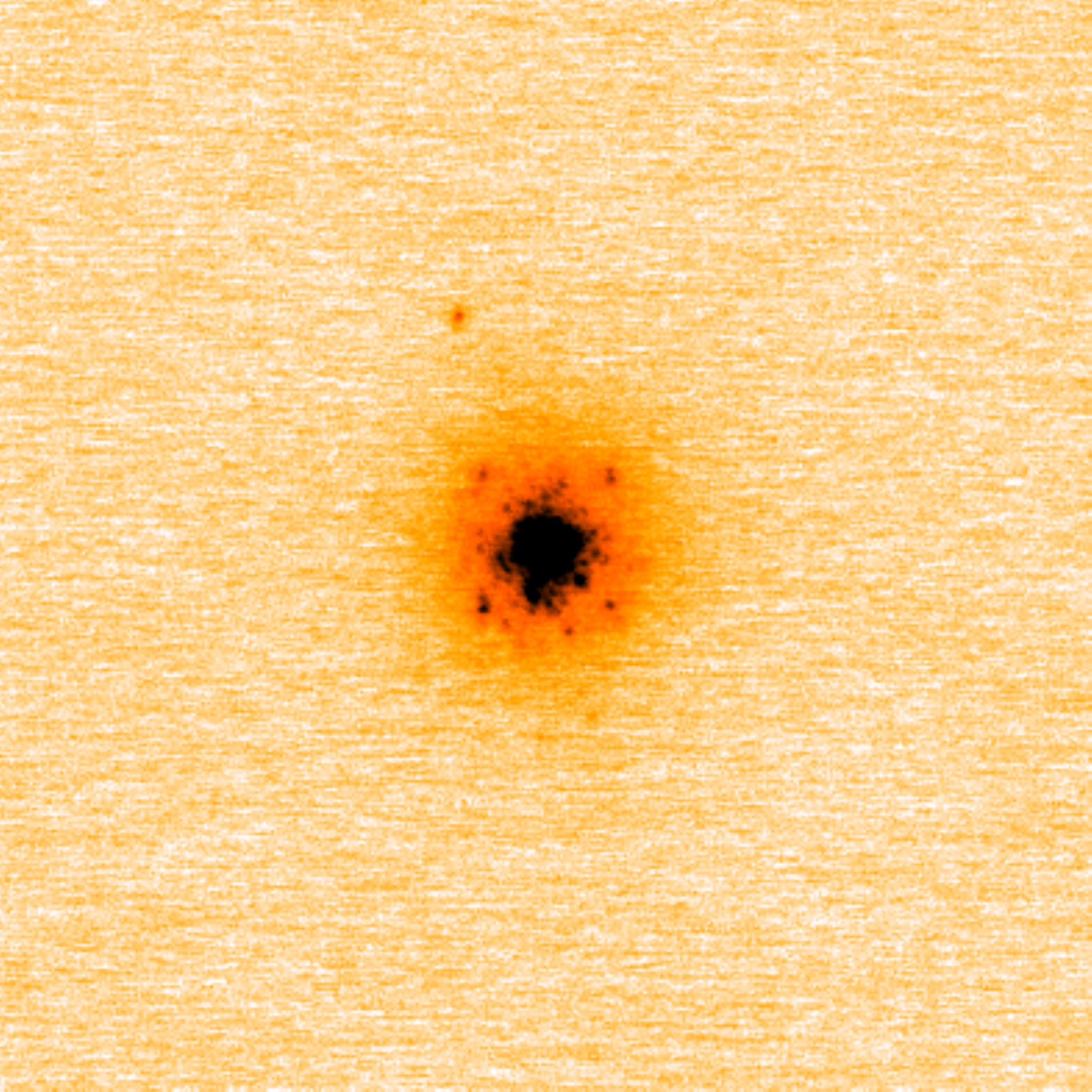}\hfill
\includegraphics[width=0.23\textwidth]{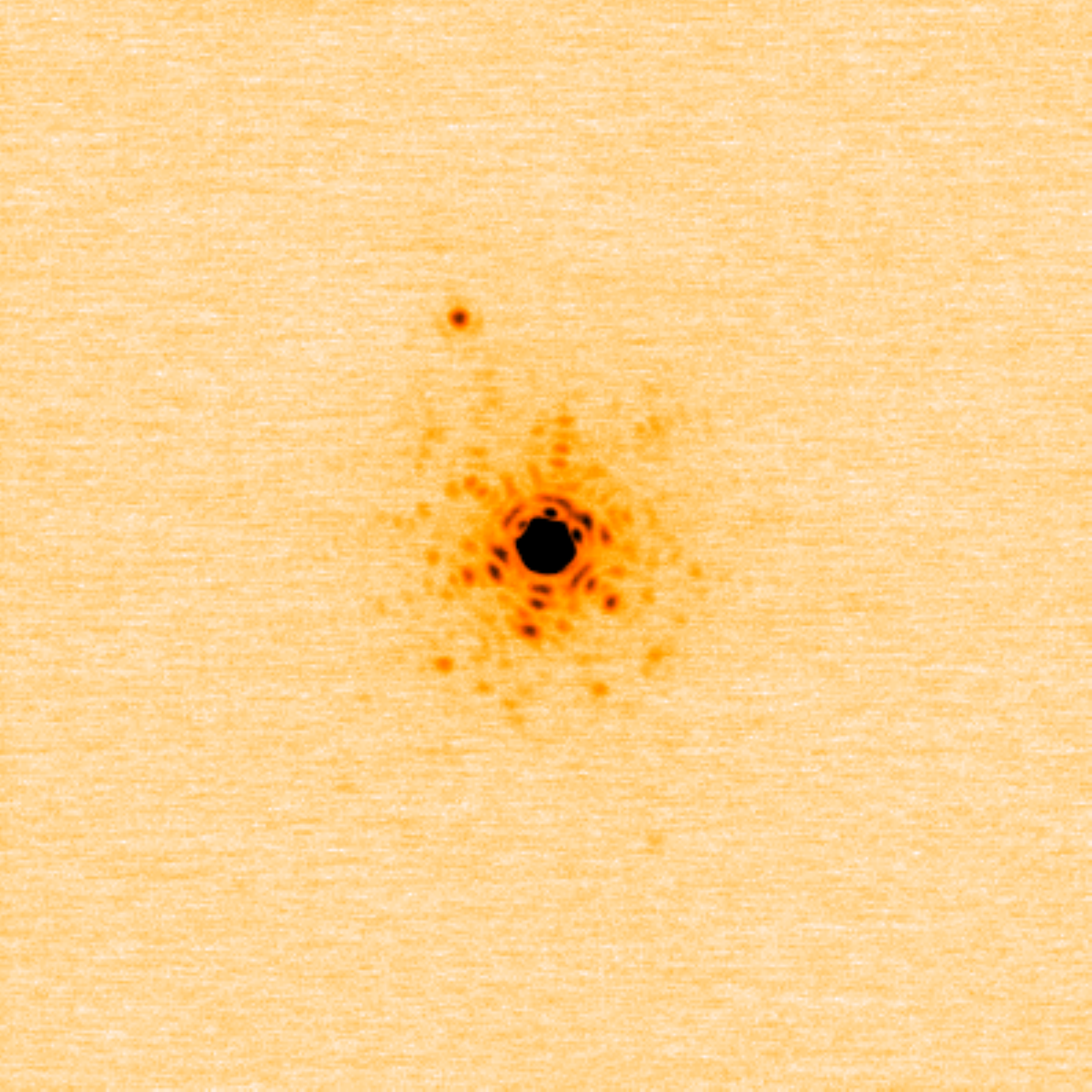}
\caption{ NIRC2 AO images in $J$-band ($1.260~\micron$) and
  $\rm{Br}\gamma$ ($2.165~\micron$), $4\arcsec\!\times\!4\arcsec$ in
  size, with a logarithmic flux scale. A faint companion to the
  northeast is clearly detected in both images, with separation
  $0\farcs8730 \pm 0\farcs0014$ and PA $20.86\dg \pm 0.07\dg$. North
  is up and east is left.
\label{fig:ao}}
\end{figure}

Adaptive optics imaging was obtained using the Near InfraRed Camera 2
(NIRC2) mounted on the Keck II $10$~m telescope on Mauna Kea, HI on UT
2014 September 4. Images were obtained in both $J$~($1.260~\micron$)
and Br$\gamma$~($2.165~\micron$; a good proxy for both $K$~and
$K_s$). NIRC2 has a field of view $10\arcsec\!\times\!10\arcsec$, a
pixel scale of about $0\farcs01\,{\rm pix}^{-1}$, and a rotator
accuracy of $0\fdg02$. The overlap region of the dither pattern of the
observations (i.e., the size of the final combined images) is $\mysim
4\arcsec\!\times\!4\arcsec$. In both filters, the FWHM of the stellar
PSF was better than $0\farcs05$, and the achieved contrasts were
$\Delta K \mysim 9$~($\Delta J \mysim 8$) beyond $0\farcs5$ (see
\reffigl{sens}).

A visual companion was detected in both images (\reffigl{ao}) with
separation $0\farcs8730 \pm 0\farcs0014$ and PA $20\fdg86 \pm 0\fdg07$
(east of north). Relative to \this, we calculate the companion to have
magnitudes ${\Delta}J = 5.59 \pm 0.06$~and ${\Delta}K = 5.16 \pm
0.02$, implying $J-K= 0.99 \pm 0.07$. Using the $J-K$ colors, we
estimate the magnitude in the {\em Kepler} bandpass to be $K_{\rm p}
\mysim 18.8$. The object was not detected in the speckle images
because they were taken with less aperture and the expected contrast
ratios are larger in $R$~and $I$---using the properties of the
companion as derived in the following section, we estimate $\Delta R
\mysim 6.7$~and $\Delta I \mysim 6.6$. These magnitudes are consistent
with non-detections in the speckle images, as plotted in
\reffigl{sens}.

\subsection{Properties of the Visual Companion}

The faint visual companion to \this~could be a background star or a
physically bound main-sequence companion. We argue that it is unlikely
to be a background star, and present two pieces of evidence to support
this conclusion. We first estimate the background stellar density in
the direction of \this~using the TRILEGAL stellar population synthesis
tool \citep{girardi:2005}: we expect $49,\!000$~sources per
deg$^2$~that are brighter than $K_s\mysim19$~(the detection limit of
our observation).  This translates to $0.06$ sources (of {\em any}
brightness and color) expected in our $16~{\rm arcsec}^2$~image, and
thus the a priori probability of a chance alignment is low.

\begin{figure}[!tb]
\includegraphics[width=0.48\textwidth]{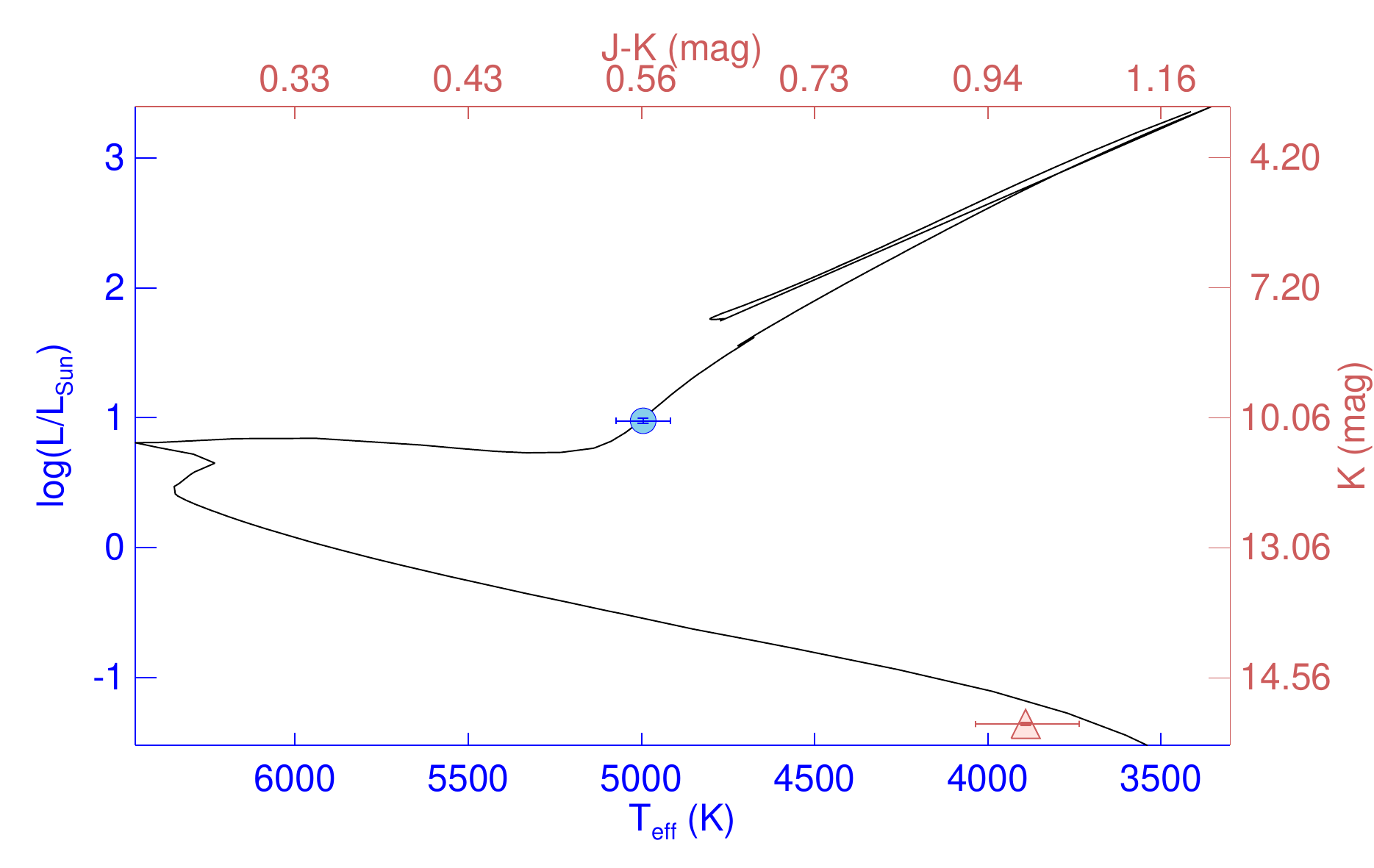}\hfill
\caption{ \this~(blue circle) placed on a $3.5$~Gyr, ${\rm
    [m/H]=-0.07}$ PARSEC isochrone \citep{bressan:2012}, plotted as a
    solid black line. The upper right (red) axes represent the
    $J-K$~and apparent $K$~mag corresponding to the $T_{\rm eff}$ and
    $\log{L/L\odot}$~on the lower left (blue) axes. The visual
    companion (red triangle) is placed on the plot according to its
    measured $J-K$~and $K$~mag. It lies very near the isochrone for
    the system, suggesting that it is indeed coeval with and
    physically bound to \this, rather than a background star.}
\label{fig:isochrone}
\end{figure}

Furthermore, since we know the properties of the primary star, we can
determine whether there exists a coeval main-sequence star that could
adequately produce the observed colors and magnitude
differences. Using the asteroseismically derived mass, radius, and age
of the primary (see \refsecl{seis}), we place the primary star on an
appropriate Padova PARSEC isochrone \citep{bressan:2012,chen:2014}. We
then use the observed magnitude differences between the stars to
search for an appropriate match to the companion in the isochrone. If
the visual companion is actually a background giant, it is unlikely
that it would happen to be at the right distance to match both the
colors and brightness of a physically bound companion. Therefore, we
do not expect a background giant star to lie on the
isochrone. However, we {\em do} find a close match to the observed
colors and magnitudes of the companion (see \reffigl{isochrone}),
further suggesting that it is not a background object, but truly a
physical companion, and this allows us to estimate its properties from
the isochrone.

We conclude that the companion is most likely a physically bound,
coeval M dwarf with a mass of $\mysim 0.52~\msun$ and an effective
temperature of $\mysim 3660$~K. The distance to the system ($\mysim
870$~pc; \refsecl{dist}), implies that the projected separation of the
companion is $\mysim 750$~AU. Using this as an estimate of the
semi-major axis, the binary orbital period is on the order of
$15,\!000$~yr. In reality, the semi-major axis may be smaller (if
we observed it near apastron of an eccentric orbit with the major axis
in the plane of the sky), or significantly larger (due to projections
into the plane of the sky and the unknown orbital phase).

We discuss the possibility of false positives due to this previously
undetected companion in \refsecl{fp}.

\section{Spectroscopic Follow-Up}
\label{sec:spec}

\subsection{Spectroscopic Observations}

We used the Tillinghast Reflector Echelle Spectrograph
\citep[TRES;][]{furesz:2008} mounted on the $1.5$-m Tillinghast
Reflector at the Fred L. Whipple Observatory (FLWO) on Mt. Hopkins, AZ
to obtain $84$ high resolution spectra of \this~between UT 2011 March
23 and 2014 June 18. TRES is a temperature-controlled, fiber-fed
instrument with a resolving power of $R \mysim 44,\!000$ and a
wavelength coverage of $\mysim 3850$--$9100$~\AA, spanning $51$ echelle
orders. Typical exposure times were $15$-$30$~minutes, and resulted in
extracted signal-to-noise ratios (S/Ns) between about $20$ and $45$ per
resolution element. The goal of the intial observations was to rule
out false positives involving stellar binaries as part of the {\em
Kepler} Follow-up Observing Program (KFOP, which has evolved into
CFOP\footnote{The {\em Kepler} Community Follow-up Observing Program,
CFOP, http://cfop.ipac.caltech.edu, publicly hosts spectra, images,
data analysis products, and observing notes for \this~and many other
Kepler Objects of Interest (KOIs).}), but upon analysis of the first
few spectra, it became clear that the planet was massive enough to
confirm with an instrument like TRES that has a modest aperture and
$\mysim 10~\ms$~precision \citep[see, e.g.,][]{quinn:2014}. By the
second observing season, an additional velocity trend was observed,
which led to an extended campaign of observations.

Precise wavelength calibration of the spectra was established by
obtaining ThAr emission-line spectra before and after each spectrum,
through the same fiber as the science exposures. Nightly observations
of the IAU RV standard star HD 182488 helped us track the achieved
instrumental precision and correct for any RV zero point drift. We
also shift the absolute velocities from each run so that the median RV
of HD 182488 is $-21.508~\kms$ \citep{nidever:2002}. This allows us to
report the absolute systemic velocity, $\gamma_{\rm abs}$. We are
aware of specific TRES hardware malfunctions (and upgrades) that
occurred during the timespan of our data that, in addition to small
zero point shifts (typically $<10\ms$), caused degradation (or
improvement) of RV precision for particular observing runs. For
example, the installation of a new dewar lens caused a zero point
shift after BJD 2455750, and a second shift (accompanied by
significant improvement in precision) occurred when the fiber
positioner was fixed in place on BJD 2456013. It will be important to
treat these with care so that the radial velocities are accurate and
each receives its appropriate weight in our analysis.

We also obtained five spectra with the FIber-fed Echelle Spectrograph
\citep[FIES;][]{frandsen:1999} on the $2.5$-m Nordic Optical Telescope
\citep[NOT;][]{djupvik:2010} at La Palma, Spain during the first
observing season (UT 2011 August 4 through 2011 October 7) to confirm
the initial RV detection before continuing to monitor the star with
TRES. Like TRES, FIES is a temperature-controlled, fiber-fed
instrument, and has a resolving power through the medium fiber of
${\rm R}\mysim 46,\!000$, a wavelength coverage of $\mysim
3600$--$7400$~\AA, and wavelength calibration determined from ThAr
emission-line spectra.

\subsection{Spectroscopic Reduction and Radial Velocity Determination}

We will discuss the reduction of spectra from both instruments
collectively but only briefly \citep[more details can be found
in][]{buchhave:2010} while detailing the challenges presented by our
particular data set. Spectra were optimally extracted, rectified to
intensity versus wavelength, and cross-correlated, order by order,
using the strongest exposure as a template. We used $21$~orders
(spanning $4290$--$6280$~\AA), rejecting those plagued by telluric
lines, fringing in the red, and low S/N in the blue. For each epoch,
the cross-correlation functions (CCFs) from all orders were added and
fit with a Gaussian to determine the relative RV for that epoch. Using
the summed CCF rather than the mean of RVs from each order naturally
weights the orders with high correlation coefficients more
strongly. Internal error estimates for each observation were
calculated as $\sigma_{\rm int}={\rm rms}(\vec{v})/\sqrt{N}$, where
$\vec{v}$ is the RV of each order, $N$ is the number of orders, and
rms denotes the root mean squared velocity difference from the
mean. These internal errors account for photon noise and the precision
with which we can measure the line centers, which in turn depends on
the characteristics of the \this~spectrum (line shapes, number of
lines, etc), but do {\em not} account for errors introduced by the
instrument itself.

The nightly observations of RV standards were used to correct for
systematic velocity shifts between runs and to estimate the
instrumental precision. The median RV of HD 182488 was calculated for
each run, which we applied as shifts to the \this~velocities, keeping
in mind that each shift introduces additional uncertainty. By also
applying the run-to-run offsets to the standard star RVs themselves,
we were able to evaluate the residual RV noise introduced by the
limited instrumental precision (separate from systematic zero point
shifts). After correction, the rms of the standard star RVs in each
run was consistent with the internal errors. This indicates that the
additional uncertainty introduced by run-to-run correction already
adequately accounts for the instrumental uncertainty, and we do not
need to explicitly include an additional error term to account for
it. The final error budget of \this~RVs was assumed to be the sum by
quadrature of all RV error sources---internal errors, run-to-run
offset uncertainties, and TRES instrumental precision: $\sigma_{\rm
RV}^2=\sigma_{\rm int}^2 + \sigma_{\rm run}^2 + \sigma_{\rm TRES}^2$,
where the final term $\sigma_{\rm TRES}=0$~because it is implicitly
incorporated into $\sigma_{\rm run}$. The final radial velocities are
listed in \reftabl{rv}. We recognize that stellar jitter or additional
undetected planets may also act as noise sources, and we address this
during the orbital fitting analysis in \refsecl{rv}.

\capstartfalse
\begin{deluxetable}{rrr|rrr}
  \tablewidth{0pt}
  \tablecaption{\this~Relative Radial Velocities
    \label{tab:rv}
  }
  \tablehead{
    \colhead{BJD} &
    \colhead{$v$} &
    \colhead{$\sigma_v$\tablenotemark{a}} &
    \colhead{BJD} &
    \colhead{$v$} &
    \colhead{$\sigma_v$\tablenotemark{a}} \\
    \colhead{(-2455000)} &
    \colhead{(\ms)} &
    \colhead{(\ms)} &
    \colhead{(-2455000)} &
    \colhead{(\ms)} &
    \colhead{(\ms)}
  }
  \startdata
    $ 644.00217$ & $235.4$ & $ 65.1$ & $1083.87676$ & $338.7$ & $36.7$ \\
    $ 722.95848$ & $453.1$ & $ 23.0$ & $1091.93504$ & $489.2$ & $44.3$ \\
    $ 727.87487$ & $452.2$ & $ 72.1$ & $1117.90004$ & $211.7$ & $41.6$ \\
    $ 734.86675$ & $544.4$ & $ 68.2$ & $1132.79275$ & $410.9$ & $42.4$ \\
    $ 755.93317$ & $ 47.2$ & $ 18.9$ & $1137.83647$ & $333.7$ & $21.8$ \\
    $ 757.82852$ & $197.1$ & $ 23.7$ & $1175.82536$ & $210.1$ & $62.5$ \\
    $ 758.88922$ & $121.7$ & $ 18.9$ & $1197.74504$ & $394.7$ & $22.3$ \\
    $ 760.95496$ & $193.8$ & $ 25.9$ & $1202.70980$ & $495.3$ & $20.5$ \\
    $ 764.92787$ & $270.3$ & $ 36.7$ & $1227.64276$ & $ 75.4$ & $23.1$ \\
    $ 768.78698$ & $262.1$ & $ 33.9$ & $1233.59012$ & $213.1$ & $32.0$ \\
    $ 768.80782$ & $277.4$ & $ 30.8$ & $1258.59157$ & $564.4$ & $24.3$ \\
    $ 822.70873$ & $197.0$ & $ 54.9$ & $1279.58992$ & $ 96.6$ & $28.9$ \\
    $ 825.63058$ & $265.0$ & $ 53.1$ & $1376.99835$ & $ 29.1$ & $25.1$ \\
    $ 826.64436$ & $310.1$ & $ 53.0$ & $1382.93180$ & $148.7$ & $23.9$ \\
    $ 827.71851$ & $362.0$ & $ 55.3$ & $1389.91148$ & $179.3$ & $26.7$ \\
    $ 828.61567$ & $358.8$ & $ 60.8$ & $1400.93674$ & $384.5$ & $31.0$ \\
    $ 829.72558$ & $358.7$ & $ 53.9$ & $1405.90249$ & $465.6$ & $30.5$ \\
    $ 830.69401$ & $446.6$ & $ 56.2$ & $1409.87680$ & $507.5$ & $28.4$ \\
    $ 837.59392$ & $588.0$ & $ 51.1$ & $1413.96428$ & $646.1$ & $37.2$ \\
    $ 840.65649$ & $603.7$ & $ 51.0$ & $1429.95861$ & $111.4$ & $23.7$ \\
    $ 841.70196$ & $706.7$ & $ 71.9$ & $1436.83822$ & $153.9$ & $22.7$ \\
    $ 842.57863$ & $574.4$ & $ 57.1$ & $1442.84894$ & $191.6$ & $26.9$ \\
    $ 843.59367$ & $437.4$ & $ 43.8$ & $1446.78723$ & $285.9$ & $21.2$ \\
    $ 844.64075$ & $346.9$ & $ 46.7$ & $1503.72052$ & $351.9$ & $35.0$ \\
    $ 845.68516$ & $157.7$ & $ 45.8$ & $1547.69956$ & $189.1$ & $19.5$ \\
    $ 846.62110$ & $ 91.9$ & $ 45.3$ & $1551.63468$ & $217.2$ & $20.0$ \\
    $ 852.61538$ & $ 64.2$ & $ 47.3$ & $1556.67843$ & $292.6$ & $20.9$ \\
    $ 854.62438$ & $ 57.6$ & $ 48.9$ & $1561.63231$ & $370.3$ & $18.2$ \\
    $ 856.71790$ & $132.6$ & $ 56.1$ & $1575.69266$ & $593.5$ & $20.0$ \\
    $ 858.64825$ & $105.2$ & $ 48.9$ & $1581.66560$ & $141.4$ & $18.9$ \\
    $1027.94566$ & $284.4$ & $ 26.4$ & $1586.63377$ & $ 40.5$ & $18.4$ \\
    $1046.89934$ & $594.1$ & $ 20.6$ & $1591.70580$ & $100.0$ & $21.4$ \\
    $1047.87727$ & $702.8$ & $ 38.6$ & $1729.01428$ & $553.2$ & $37.9$ \\
    $1048.86080$ & $693.7$ & $ 18.9$ & $1740.94781$ & $ 98.6$ & $33.0$ \\
    $1049.97884$ & $661.7$ & $ 28.7$ & $1743.99496$ & $ 34.8$ & $29.3$ \\
    $1051.99004$ & $631.1$ & $ 24.6$ & $1799.87657$ & $ 37.0$ & $26.3$ \\
    $1052.93268$ & $583.1$ & $ 30.0$ & $1816.87894$ & $248.9$ & $32.0$ \\
    $1053.90973$ & $461.0$ & $ 23.4$ & $1822.88725$ & $364.1$ & $24.3$ \\
    $1054.88376$ & $383.9$ & $ 29.3$ & $1826.79500$ & $419.8$ & $20.0$ \\
    $1055.88175$ & $282.3$ & $ 37.3$ & \tablenotemark{b}$777.59554$ & $-100.9$ & $10.7$ \\
    $1056.88078$ & $239.3$ & $ 26.4$ & \tablenotemark{b}$778.59121$ & $ -81.1$ & $10.5$ \\
    $1057.95303$ & $158.7$ & $ 24.6$ & \tablenotemark{b}$779.59677$ & $ -51.2$ & $14.1$ \\
    $1058.84781$ & $160.9$ & $ 33.7$ & \tablenotemark{b}$782.57231$ & $   0.0$ & $10.2$ \\
    $1074.90585$ & $267.8$ & $ 31.1$ & \tablenotemark{b}$842.41733$ & $  47.7$ & $10.2$ \\
    $1080.95245$ & $354.5$ & $ 30.6$ & & & \\
    [-2.4ex]
    \enddata
    \tablenotetext{a}{ Values reported here account for internal and
      instrumental error, as well as uncertainties in systematic zero
      point shifts applied to the velocities, but do {\em not} include
      the $20~\ms$~added during the orbital fit that is meant to
      encompass additional astrophysical noise sources (e.g., stellar
      activity or additional planets).}
    \tablenotetext{b}{The FIES observations have a different zero
      point, which is included as a free parameter in the orbital
      fit.}
\end{deluxetable}
\capstarttrue

\subsection{Spectroscopic Classification}
\label{sec:spec_class}

\capstartfalse
\begin{deluxetable}{lcc}
  \tablewidth{0pt}
  \tablecaption{Stellar Properties of \this
    \label{tab:stellar}
  }
  \tablehead{
    \multicolumn{1}{l}{Parameter} &
    \multicolumn{2}{c}{Value}
  }
  \startdata
  Asteroseismic & Grid-based Modeling & Frequency Modeling \\
  \hline \\ [-2ex]
  \Dnu\ (\uhz)           & $18.59\pm0.04$         & \ldots          \\
  \numax\ (\uhz)         & $266\pm3$              & \ldots          \\
  \logg\ (dex)           & $3.340\pm0.006$        & $3.345 \pm 0.006$ \\
  $\rho_\star$~(${\rm g~cm}^{-3}$)     
                         & $0.02723^{+0.00054}_{-0.00057}$ 
                                                  & $0.02794^{+0.00055}_{-0.00058}$ \\
  $R$ (\rsun)            & $4.12^{+0.12}_{-0.08}$ & $4.06^{+0.12}_{-0.08}$  \\
  $M$ (\msun)            & $1.35^{+0.10}_{-0.07}$ & $1.32^{+0.10}_{-0.07}$ \\
  Age (Gyr)              & $3.5^{+0.7}_{-0.8}$    & $4.2^{+0.8}_{-1.0}$   \\
  \hline \\ [-2ex]
  Spectroscopic & & \\
  \hline \\ [-2ex]
  $T_{\rm eff}$~(K)      & \multicolumn{2}{c}{$4995 \pm 78$}        \\
  $\log{g}$~(cgs)        & \multicolumn{2}{c}{$3.345 \pm 0.006$}    \\
  $[m/H]$                & \multicolumn{2}{c}{$-0.07 \pm 0.10$}     \\
  $v\sin{i_\star}$~(\kms)& \multicolumn{2}{c}{$2.7 \pm 0.5$}        \\
  \hline \\ [-2ex]
  Photometric & & \\
  \hline \\ [-2ex]
  $V$~(mag)\tablenotemark{a}              & \multicolumn{2}{c}{$12.465 \pm 0.060$}   \\
  $K_{\rm p}$~(mag)\tablenotemark{b}      & \multicolumn{2}{c}{$12.183 \pm 0.020$}   \\
  $J$~(mag)\tablenotemark{c}              & \multicolumn{2}{c}{$10.684 \pm 0.021$}   \\
  $H$~(mag)\tablenotemark{c}              & \multicolumn{2}{c}{$10.221 \pm 0.019$}   \\
  $K_{\rm s}$~(mag)\tablenotemark{c}      & \multicolumn{2}{c}{$10.121 \pm 0.017$}   \\
  \hline \\ [-2ex]
  Derived & & \\
  \hline \\ [-2ex]
  $L_\star$~(\lsun)\tablenotemark{d}      & \multicolumn{2}{c}{$9.206 \pm 0.010$}    \\
  $d$ (pc)               & \multicolumn{2}{c}{$870 \pm 20$}         \\
  $i_\star$~(\dg)        & \multicolumn{2}{c}{$90^{+0}_{-8}$}       \\
  $P_{\rm rot}$~(days)   & \multicolumn{2}{c}{$77 \pm 14$}          \\
   [-2.4ex]
  \enddata
  \tablenotetext{a}{From APASS, via UCAC4 \citep{zacharias:2013}.}
  \tablenotetext{b}{From the {\em Kepler} Input Catalog \citep{brown:2011}.}
  \tablenotetext{c}{From 2MASS \citep{skrutskie:2006}.}
  \tablenotetext{d}{Calculated using the relation $\left(\frac{L_\star}{\lsun}\right) = \left(\frac{R_\star}{R_\odot}\right)^2 \left(\frac{T_{\rm{eff},\star}}{T_{\rm{eff},\odot}}\right)^4$.}
  \tablecomments{While we have separated the stellar properties by
    observational technique, many of them are interdependent, for
    example, as described in \refsecs{spec}{seis}. We adopt the
    asteroseismic results from detailed frequency modeling (rather
    than the grid-based approach) in subsequent analyses.}
\end{deluxetable}
\capstarttrue

We initially determined the spectroscopic stellar properties
(effective temperature, $T_{\rm eff}$; surface gravity, $\log{g}$;
projected rotational velocity, $v\sin{i}$; and metallicity, [m/H])
using Stellar Parameter Classification \citep[SPC;][]{buchhave:2012},
with the goal of providing an accurate temperature for the
asteroseismic modeling (see \refsecl{seis}). SPC cross-correlates an
observed spectrum against a grid of synthetic spectra, and uses the
correlation peak heights to fit a three-dimensional surface in order
to find the best combination of atmospheric parameters ($v\sin{i}$ is
fit iteratively since it only weakly correlates with the other
parameters). We used the CfA library of synthetic spectra, which are
based on Kurucz model atmospheres \citep{kurucz:1992}. SPC, like other
spectroscopic classifications, can be limited by degeneracy between
$T_{\rm eff}$, $\log{g}$, and [m/H] \citep[see a discussion
in][]{torres:2012}, but asteroseismology provides a nearly independent
measure of the surface gravity (depending only weakly on the effective
temperature and metallicity). This allows one to iterate the two
analyses until agreement is reached, generally requiring only $1$
iteration \citep[see, e.g.,][]{huber:2013a}. In our initial analysis,
we found $T_{\rm eff}=5072 \pm 55$~K, $\log{g}=3.49 \pm 0.11$, ${\rm
[m/H]}=-0.02 \pm 0.08$, and $v\sin{i}=2.5 \pm 0.5~\kms$. After
iterating with the asteroseismic analysis and fixing the final
asteroseismic gravity, we find similar values: $T_{\rm eff}=4995 \pm
78$~K, $\log{g}=3.345 \pm 0.006$, ${\rm [m/H]}=-0.07 \pm 0.10$, and
$v\sin{i}=2.7 \pm 0.5~\kms$. We adopt the values from the combined
analysis, and these final spectroscopic parameters are listed in
\reftabl{stellar}.

\section{Asteroseismology of \this}
\label{sec:seis}

\subsection{Background}

Cool stars exhibit brightness variations due to oscillations driven by
near-surface convection \citep{houdek:1999,aerts:2010}, which are a
powerful tool to study their density profiles and evolutionary
states. A simple asteroseismic analysis is based on the average
separation of modes of equal spherical degree (\Dnu) and the frequency
of maximum oscillation power (\numax), using scaling relations to
estimate the mean stellar density, surface gravity, radius, and mass
\citep{kjeldsen:1995,stello:2008,kallinger:2010,belkacem:2011}. \citet{huber:2013a}
presented an asteroseismic analysis of \this\ by measuring \Dnu\ and
\numax\ using three quarters of SC data, combined with an SPC analysis
\citep{buchhave:2012} of high-resolution spectra obtained with the
FIES and TRES spectrographs. The results showed that \this\ is an
evolved star just beginning to ascend the red-giant branch (RGB), with
a radius of $R=4.16\pm0.12$\,\rsun\ and a mass of
$1.35\pm0.10$\,\msun\ (\reftabl{stellar}, \reffigl{isochrone}).

Compared to average oscillation properties, individual frequencies
offer a greatly increased amount of information by probing the
interior sound speed profile. In particular, evolved stars oscillate
in mixed modes, which occur when pressure modes excited on the surface
couple with gravity modes confined to the core
\citep{aizenman:1977}. Mixed modes place tight constraints on
fundamental properties such as stellar age, and provide the
possibility to probe the core structure and rotation
\citep{bedding:2011,beck:2012,mosser:2012a}. Importantly, relative
amplitudes of individual oscillation modes that are split by rotation
can be used to infer the stellar line-of-sight inclination
\citep{gizon:2003}, providing valuable information on the orbital
architectures of transiting exoplanet systems
\citep[e.g.,][]{chaplin:2013,huber:2013b,benomar:2014,lund:2014,vaneylen:2014}. In
the following section we expand on the initial asteroseismic analysis
by \citet{huber:2013a} by performing a detailed individual frequency
analysis based on all eight quarters (Q9--17) of {\em Kepler}
short-cadence data.

\subsection{Frequency Analysis}

\begin{figure*}[!htb]
\centering
\includegraphics[width=\textwidth]{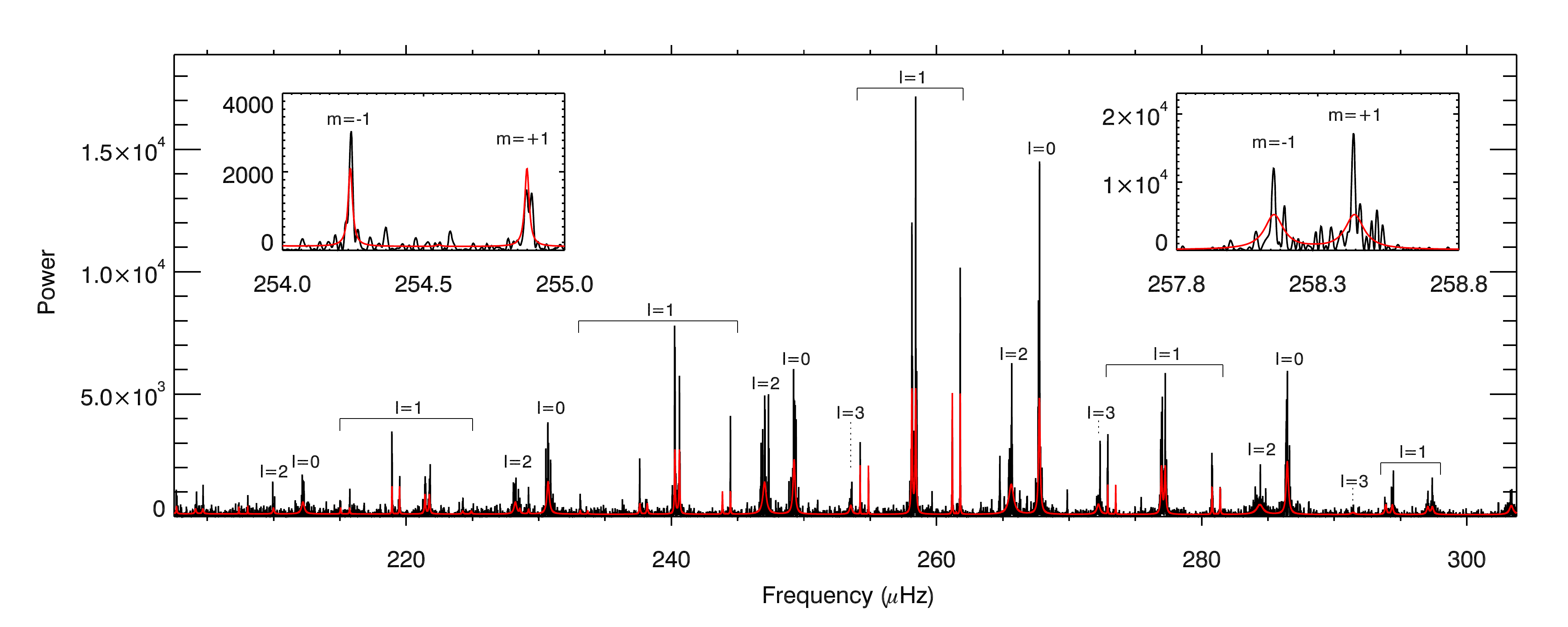}
\caption{Power spectrum of \this. The red curve shows the fit to
the power spectrum. Insets show a close-up of two rotationally split
$\ell=1$ mixed modes. The left inset shows a mixed mode dominated by
$g$-mode characteristics, with lower amplitudes, narrower linewidths
and a larger rotational splitting relative to the mixed mode on the
right, which has strong $p$-mode characteristics.}
\label{fig:powspec}
\end{figure*}

The time series was prepared for asteroseismic analysis from the raw
{\em Kepler} target pixel data using the {\em Kepler Asteroseismic
Science Operations Center} (KASOC) filter \citep{handberg:2014}. The
KASOC filter removes instrumental and transit signals from the light
curve, which may produce spurious peaks in the frequency domain. The
power spectrum, shown in \reffigl{powspec}, was computed using a
Lomb--Scargle periodogram \citep{lomb:1976,scargle:1982} calibrated to
satisfy Parseval's theorem.

The pattern of oscillation modes in the power spectrum is typical of
red giants, with $\ell=0$ modes of consecutive order being
approximately equally spaced by $\Delta\nu$, adjacent to $\ell=2$
modes. In addition to the $\ell=0, 2$ pairs, several $\ell=1$ mixed
modes are observed in each radial order which, on inspection, are the
outer components of rotationally split triplets corresponding to the
$m=\pm1$ modes. This indicates that the star is seen equator-on (see
\refsecl{inc}).

The relative $p$- and $g$-mode behavior of each mixed mode depends on
the strength of the coupling between the oscillation cavities in the
stellar core and envelope. Detecting the $\ell=1$ modes with the
greatest $g$-mode character may be challenging because they have low
amplitudes, and overlap in frequency with the $\ell=0$ and $\ell=2$
modes. Adding to the possible confusion, mixed $\ell=2$ and $\ell=3$
modes may also be present, although the weaker coupling between the
$p$- and $g$-modes results in only the most $p$-like modes having an
observable amplitude.

\capstartfalse
\begin{deluxetable*}{lccccc}
  \centering
  \tablewidth{0pt}
  \tablecaption{Oscillation Frequencies of \this
    \label{tab:oscillation}
  }
  \tablehead{
    \colhead{Order} &
    \colhead{$\ell=0$} &
    \multicolumn{2}{c}{$\ell=1$} &
    \colhead{$\ell=2$} &
    \colhead{$\ell=3$} \\
    \colhead{} &
    \colhead{} &
    \colhead{$m=0$} &
    \colhead{$\nu_s$} &
    \colhead{} &
    \colhead{}
  }
  \startdata
  9  & $194.111^{+0.018}_{-0.014}$  & $202.500^{+0.006}_{-0.008}$ & $0.184^{+0.014}_{-0.013}$ & $209.990^{+0.010}_{-0.023}$ & \\
     &                              & $204.392^{+0.015}_{-0.010}$ & $0.274^{+0.018}_{-0.022}$ & & \\
     &                              & $207.713^{+0.006}_{-0.007}$ & $0.335^{+0.009}_{-0.008}$ & & \\
 10  & $212.214^{+0.026}_{-0.020}$  & $215.394^{+0.008}_{-0.007}$ & $0.342^{+0.014}_{-0.015}$ & $228.236^{+0.019}_{-0.019}$ & \\
     &                              & $219.215^{+0.006}_{-0.008}$ & $0.284^{+0.007}_{-0.008}$ & & \\
     &                              & $221.586^{+0.011}_{-0.019}$ & $0.165^{+0.014}_{-0.011}$ & & \\
     &                              & $224.590^{+0.016}_{-0.008}$ & $0.325^{+0.017}_{-0.017}$ & & \\
     &                              & $228.905^{+0.020}_{-0.013}$ & $0.328^{+0.022}_{-0.017}$ & & \\
 11  &  $230.704^{+0.009}_{-0.008}$ & $233.410^{+0.013}_{-0.017}$ & $0.280^{+0.015}_{-0.018}$ & $247.031^{+0.013}_{-0.016}$ & $253.534^{+0.018}_{-0.034}$ \\
     &                              & $237.904^{+0.014}_{-0.011}$ & $0.279^{+0.010}_{-0.010}$ & & \\
     &                              & $240.439^{+0.011}_{-0.009}$ & $0.169^{+0.008}_{-0.009}$ & & \\
     &                              & $244.157^{+0.015}_{-0.009}$ & $0.302^{+0.009}_{-0.019}$ & & \\
 12  &  $249.259^{+0.016}_{-0.009}$ & $254.553^{+0.005}_{-0.006}$ & $0.313^{+0.005}_{-0.005}$ & $265.636^{+0.011}_{-0.043}$ & $272.212^{+0.015}_{-0.006}$ \\
     &                              & $258.288^{+0.006}_{-0.008}$ & $0.143^{+0.010}_{-0.011}$ & & \\
     &                              & $261.490^{+0.003}_{-0.004}$ & $0.298^{+0.005}_{-0.006}$ & & \\
 13  &  $267.757^{+0.017}_{-0.016}$ & $273.216^{+0.007}_{-0.007}$ & $0.301^{+0.007}_{-0.007}$ & $284.389^{+0.013}_{-0.011}$ & $291.410^{+0.008}_{-0.008}$ \\
     &                              & $277.110^{+0.006}_{-0.008}$ & $0.140^{+0.012}_{-0.014}$ & & \\
     &                              & $281.091^{+0.006}_{-0.006}$ & $0.304^{+0.007}_{-0.008}$ & & \\
 14  &  $286.457^{+0.006}_{-0.009}$ & $294.131^{+0.015}_{-0.014}$ & $0.246^{+0.012}_{-0.011}$ & $303.352^{+0.021}_{-0.015}$ & \\
     &                              & $297.229^{+0.012}_{-0.017}$ & $0.173^{+0.016}_{-0.014}$ & & \\
 15  &  $305.284^{+0.015}_{-0.014}$ & $315.134^{+0.032}_{-0.011}$ & $0.207^{+0.026}_{-0.026}$ & & \\
   [-2.4ex]
  \enddata
  \tablecomments{ All frequencies are in units of \uhz. The
  $\ell=1$~modes are presented as the central frequency of the
  rotationally split mode profile ($m=0$), along with the value of the
  rotational splitting between the $m=0$~and $m= \pm 1$~components,
  $\nu_s$.}
\end{deluxetable*}
\capstarttrue

The first step to fitting the oscillation modes and extracting their
frequencies is to correctly identify the modes present. Fortunately,
the $\ell=1$ mixed modes follow a frequency pattern that arises from
coupling of the $p$-modes in the envelope, which have approximately
equal spacing in frequency ($\Delta\nu$), to $g$-modes in the core,
which are approximately equally spaced in period ($\Delta\Pi$). This
pattern is well described by the asymptotic relation for mixed modes
\citep{mosser:2012b}. We calculated the asymptotic mixed mode frequencies
by fitting this relation to several of the highest-amplitude $\ell=1$
modes. From these calculations, nearby peaks could be associated with
$\ell=1$ mixed modes. In this way we have been able to identify both
of the $m=\pm1$ components for 21 out of 27 $\ell=1$ mixed modes
between $200$~and $320$~\uhz.

Following a strategy that has been implemented in the mode fitting of
other {\em Kepler} stars \citep[e.g.,][]{appourchaux:2012}, three teams
performed fits to the identified modes. The mode frequencies from each
fit were compared to the mean values, and the fitter that differed
least overall was selected to provide the frequency solution. This
fitter performed a final fit to the power spectrum to include modes
that other fitters had detected, but were absent from this fitter's
initial solution.

The final fit was made using a Markov chain Monte Carlo (MCMC) method
that performs a global fit to the oscillation spectrum, with the modes
modeled as Lorentzian profiles \citep{handberg:2011}. Each $\ell=1$
triplet was modeled with the frequency splitting and inclination angle
as additional parameters to the usual frequency, height, and width
that define a single Lorentzian profile. Owing to the differing
sensitivity of the $\ell=1$ mixed modes to rotation at different depths
within the star, each $\ell=1$ triplet was fitted with an independent
frequency splitting, although a common inclination angle was used. We
discuss the rotation of the star further in \refsecl{inc}.

The measured mode frequencies are given in \reftabl{oscillation}. The
values of the $\ell=1$ modes are presented as the central frequency of
the rotationally split mode profile, which corresponds to the value of
the $m=0$ component, along with the value of the rotational splitting
between the $m=0$ and $m=\pm1$ components, $\nu_s$. Revised values of
\Dnu\ and \numax\ can be obtained from the measured mode frequencies
and amplitudes. We find \Dnu\ $=~18.59\pm0.04$~\uhz\ and
\numax\ $=~266\pm3$~\uhz, both of which are in agreement with the
values provided by \citet{huber:2013a}. Additionally, we measure the
underlying $\ell=1$\ $g$-mode period spacing, $\Delta\Pi_1$, to be
$89.9\pm0.3$\,s, which is consistent with a red giant branch star with
a mass below $\sim1.6\,$\msun\ \citep[e.g.,][]{stello:2013}.

\subsection{Host Star Inclination}\label{sec:inc}

The line-of-sight inclination of a rotating star can be determined by
measuring the relative heights of rotationally split modes
\citep{gizon:2003}. A star viewed pole-on produces no visible
splitting, while stars viewed with an inclination near $i=45^{\circ}$
would produce a frequency triplet. \reffigl{powspec} shows that
all dipole modes observed for \this\ are split into doublets, which we
interpret as triplets with the central peak missing, indicating a
rotation axis nearly perpendicular to the line of sight (inclination
$i=90^{\circ}$).

\begin{figure}[!tb]
\includegraphics[width=0.48\textwidth]{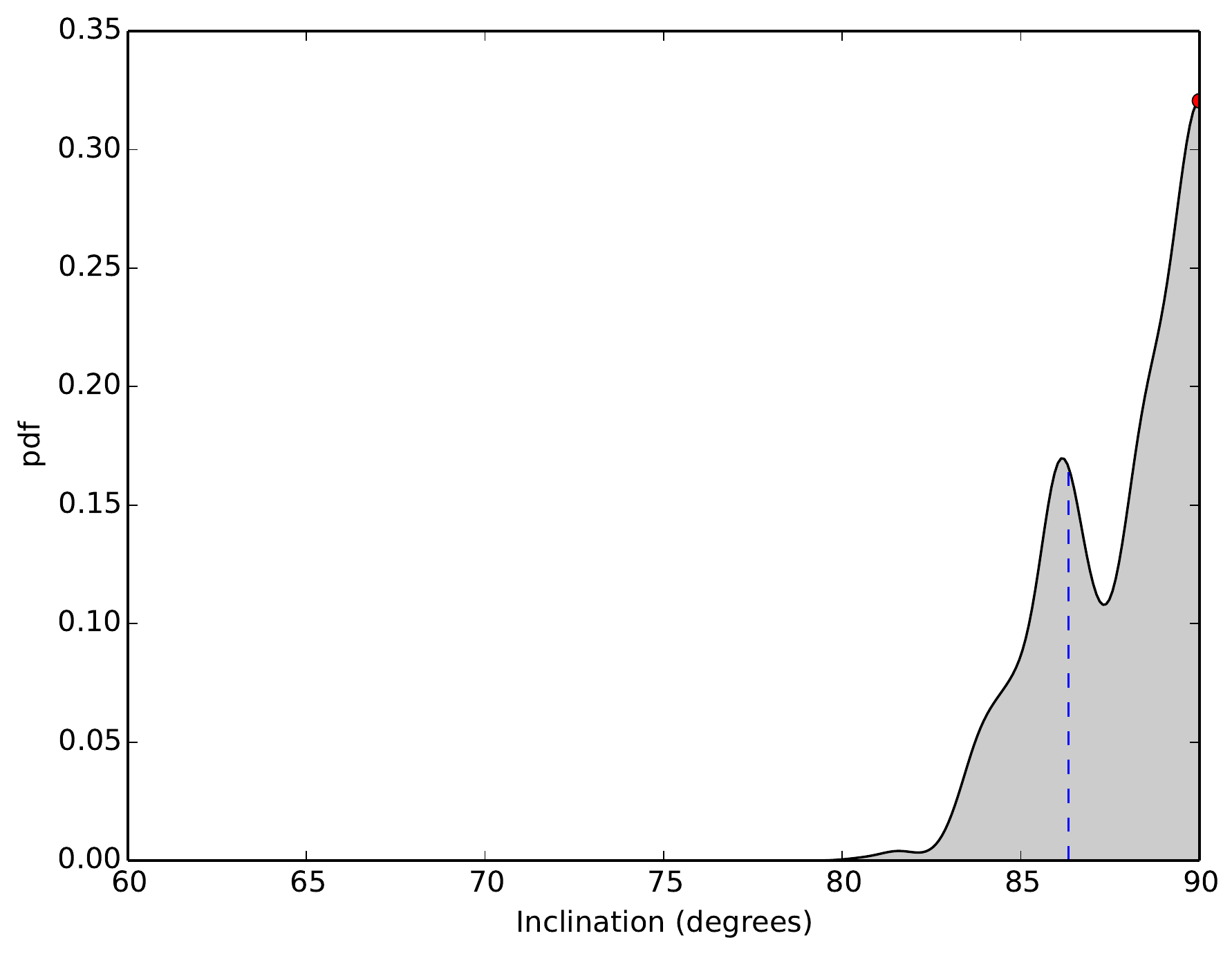}\hfill
\caption{Posterior of the host star line-of-sight inclination derived
from the MCMC analysis of the oscillation power spectrum of
\this. The red point indicates the mode of the distribution, and
the blue dashed line indicates the limit of the 68\% highest
probability density region.}
\label{fig:incl}
\end{figure}

To measure the inclination of \this, we included rotationally split
Lorentzian profiles for each of the 21 dipole modes in the global MCMC
fit of the power spectrum. \reffigl{incl} shows the posterior
distribution of the stellar inclination. The mode of the posterior
distribution and $68.3\%$~highest probability density region is
$90.0^{+0.0}_{-3.7}$\ deg.

%% The inclination estimate is based on three important assumptions: the
%% inclination is the same for p-dominated and g-dominated mixed modes,
%% that there is equipartition of energy between modes with the same $n$
%% and $\ell$, and that the modes are well-resolved.

%% While the former assumption has been shown to be the case for
%% Kepler-56 \citep{huber:2013b}, the second is only valid if the modes
%% of the same $n$ and $\ell$ and different $m$ probe the same part of
%% the star. 

The inclination estimate is based on three important assumptions: the
inclination is the same for $p$-dominated and $g$-dominated mixed modes,
that there is equipartition of energy between modes with the same $n$
and $\ell$, and that the modes are well-resolved.

To test the first assumption, we performed additional fits to
individual $\ell=1$~modes using the Python implementaiton of the
nested sampling algorithm MultiNest, pyMultiNest
\citep{feroz:2013,buchner:2014}. No significant difference was found
between the inclination angle for $p$-dominated and $g$-dominated mixed
modes. We therefore use the results of our global MCMC
fit. \citet{huber:2013b} similarly found no difference between the
inclination angle of $p$-dominated and $g$-dominated mixed modes in
{\em Kepler}-56. \citet{beck:2014} have shown that these modes actually have
slightly different pulsation cavities. They identified asymmetric
rotational splittings between $m=0$ and $m=\pm1$ modes in the red
giant KIC\,5006817, which results from the modes having varying $p$- and
$g$-mode characteristics. Besides the effect on the rotational
splitting, there is also a small impact on mode heights and
lifetimes. \citet{beck:2014} further note that the asymmetries are
mirrored about the frequency of the uncoupled $p$-modes. This means that
the heights of the $m=\pm1$ components relative to the $m=0$ component
will change in opposite directions, so the effect can be mitigated by
forcing the $m=\pm1$ components to have the same height in the fit, as
well as by performing a global fit to all modes, as we have done.

Unless mode lifetimes are much shorter than the observing baseline,
the Lorentzian profiles of the modes will not be well-resolved, and
the mode heights will vary. We determined the effect on the measured
inclination angle in the manner of \citet{huber:2013b}, by
investigating the impact on simulated data with similar properties to
the frequency spectrum of \this. Taking this effect into account in
the determination of our measurement uncertainty, we find a final
value of $i=90^{+0}_{-8}$\ deg. The asteroseismic analysis
therefore shows directly that the spin axis of the host star is in the
plane of the sky.

\subsection{Modeling}

Two approaches may be used when performing asteroseismic modeling. The
first is the so-called {\it grid-based method}, which uses
evolutionary tracks that cover a wide range of metallicities and
masses, and searches for the best fitting model using \Dnu, \numax,
\teff, and \feh\ as constraints
\citep[e.g.][]{stello:2009,basu:2010,gai:2011,chaplin:2014}. The
second is {\it detailed frequency modeling}, which uses individual
mode frequencies instead of the global asteroseismic parameters to
more precisely determine the best-fitting model
\citep[e.g.,][]{metcalfe:2010,jiang:2011}. For comparison, we have
modeled \this\ using both approaches.

For the grid-based method we used the Garching Stellar Evolution Code
\citep{weiss:2008}. The detailed parameters of this grid
are described by \citet{silvaaguirre:2012} and its coverage has been
now extended to stars evolved in the RGB phase. The spectroscopic
values of \teff\ and \feh\ found in \refsecl{spec_class}, and our new
asteroseismic measurements of \Dnu\ and \numax\ were used as inputs in
a Bayesian scheme as described in \citet{silvaaguirre:2014}. Note that
while [Fe/H] is the model input, our spectroscopic analysis yields an
estimate of [m/H]. We have assumed that the two are equivalent for
\this\ (i.e., that the star has a scaled solar composition). If this
assumption is invalid, it may introduce a small bias in our results,
which are given in \reftabl{stellar}. The values of mass and radius
agree well with those from \citet{huber:2013a}, who also used the
grid-based method. We note that a comparison of results provided by
several grid-pipelines by \citet{chaplin:2014} found typical
systematic uncertainties of $3.7\%$~in mass, $1.3\%$~in radius, and
$12\%$~in age across an ensemble of main-sequence and subgiant stars
with spectroscopic constraints on \teff\ and \feh.

We performed two detailed modeling analyses using separate stellar
evolution codes in order to better account for systematic
uncertainties. The first analysis modeled the star using the
integrated astero extension within MESA \citep{paxton:2013}.  After an
initial grid search to determine the approximate location of the the
global minimum, we found the best-fitting model using the build-in
simplex minimization routine, which automatically adjusted the mass,
metallicity, and the mixing length parameter. The theoretical
frequencies were calculated using GYRE \citep{townsend:2013} and were
corrected for near-surface effects using the power-law correction of
\citet{kjeldsen:2008} for radial modes. The non-radial modes in red
giants are mixed with $g$-mode characteristics in the core, so they are
less affected by near-surface effects. To account for this,
MESA-astero follows \citet{brandao:2011} in scaling the correction
term for non-radial modes by $Q_{n,\ell}^{-1}$, where $Q_{n,\ell}$ is
the ratio of the inertia of the mode to the inertia of a radial mode
at the same frequency.

The second analysis was performed with the Aarhus Stellar Evolution
Code \citep[ASTEC;][]{cd08a}, with theoretical frequencies calculated
using the Aarhus adiabatic oscillation package
\citep{cd08b}. The best-fitting model was found in a similar
manner as the first analysis, although the mixing length parameter was
kept fixed at a value of $\alpha=1.8$.

\begin{figure}
\centering
\includegraphics[width=0.48\textwidth]{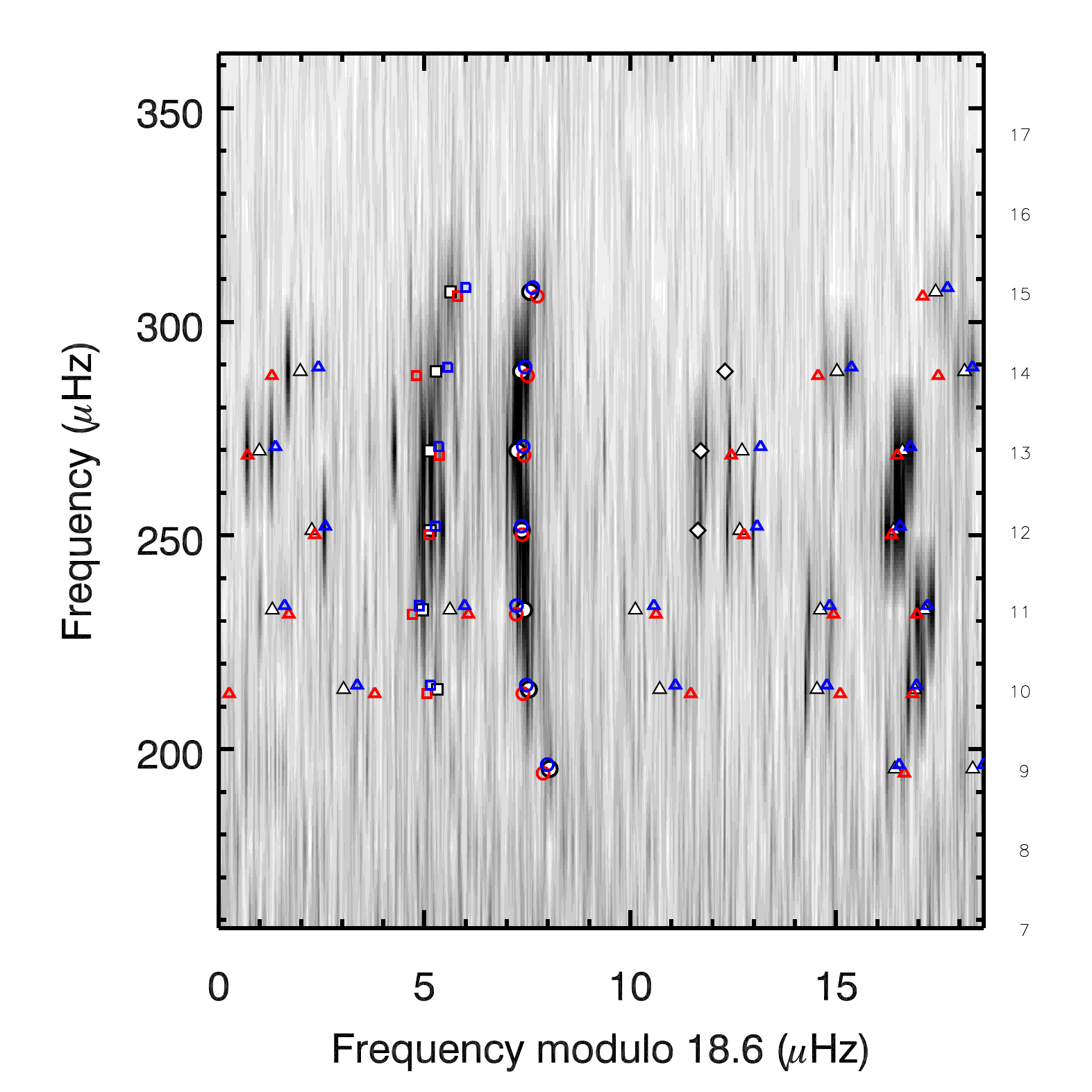}
\caption{\'Echelle diagram of \this~showing observed frequencies in
  white. Modes are identified as $\ell=0$~(circles),
  $\ell=1$~(triangles), $\ell=2$~(squares) and
  $\ell=3$~(diamonds). Frequencies of the best-fitting MESA and ASTEC
  models are indicated by the red and blue open symbols,
  respectively. For reference, a gray-scale map of the power spectrum
  is shown in the background. Numbers to the right of the plot
  indicate the radial order of the $\ell=0$~modes.}
\label{fig:echelle}
\end{figure}

\reffigl{echelle} shows the best-fitting models compared to the
observed frequencies in an \'echelle diagram. Both analyses found an
asteroseismic mass and radius of $M=1.32^{+0.10}_{-0.07}$\,\msun and
$R=4.06^{+0.12}_{-0.08}$\,\rsun, but differ in the value of the age,
with the best-fitting MESA and ASTEC models having ages of
$4.2^{+0.8}_{-1.0}$ and $2.9^{+0.6}_{-0.7}$\,Gyr,
respectively. Uncertainties were estimated by adopting the fractional
uncertainties of the grid-based method, thereby accounting for
systematic uncertainties in model input physics and treatment of
near-surface effects. The consistency between the detailed model
fitting results and grid-based results demonstrates the precise
stellar characterization that can be provided by
asteroseismology. Throughout the remainder of the paper we adopt the
results obtained with the MESA code, though we use an age of
$3.5$\,Gyr, which is consistent with both detailed frequency analyses
as well as the age from the grid-based modeling.

\subsection{Distance and Reddening}
\label{sec:dist}

The stellar model best-fit to the derived stellar properties provides
color indices that may be compared against measured values as a
consistency check, and as a means to determine a photometric distance
to the system. Given the physical stellar parameters derived from the
asteroseismic and spectroscopic measurements, the PARSEC isochrones
\citep{bressan:2012} predict $J-K_{\rm s}=0.564$, in good agreement
with the measured 2MASS colors ($J-K_{\rm s}=0.563 \pm 0.028$). While
this indicates the dust extinction along the line of sight is probably
low, we attempt to correct for it nonetheless using galactic dust
maps. The mean of the reddening values reported by
\citet{schlafly:2011} and \citet{schlegel:1998}---$E(B-V)=0.079 \pm
0.008$---is indeed low, implying extinction in the infrared of ${\rm
A_J} = 0.070$, ${\rm A_H} = 0.045$, and ${\rm A_{K_s}} =
0.028$. Applying these corrections, the observed 2MASS index becomes
$J-K_{\rm s}=0.521$, now slightly inconsistent at the
$1$-$\sigma$~level with the PARSEC model colors. We note that if only
a small fraction of the dust column lies between us and \this, it
would lead to a slight over-correction of the magnitudes and
distance. The distances derived in the two cases (using $J, H, {\rm
and}~K_s$ magnitudes) are $878 \pm 9$~pc (no extinction) and $859 \pm
13$~pc (entire column of extinction). Because the colors agree more
closely {\em without} an extinction correction, it is tempting to
conclude that only a small fraction of the extinction in the direction
of \this~actually lies between us and the star. However, in the
direction to the star (i.e., out of the galactic plane), it seems
unlikely that a significant column of absorbers would lie {\em beyond}
$\mysim 1$~kpc. In reality, the model magnitudes are probably not a
perfect match for the star and the appropriate reddening for this star
is probably between $0$~and that implied by the full column
($0.079$). The derived distance does not depend strongly on the value
we adopt for reddening, and we choose to use the mean of the two
distance estimates and slightly inflate the errors: $d = 870 \pm
20$~pc.

\begin{figure*}[!tb]
\centering
\includegraphics[width=0.2\textwidth]{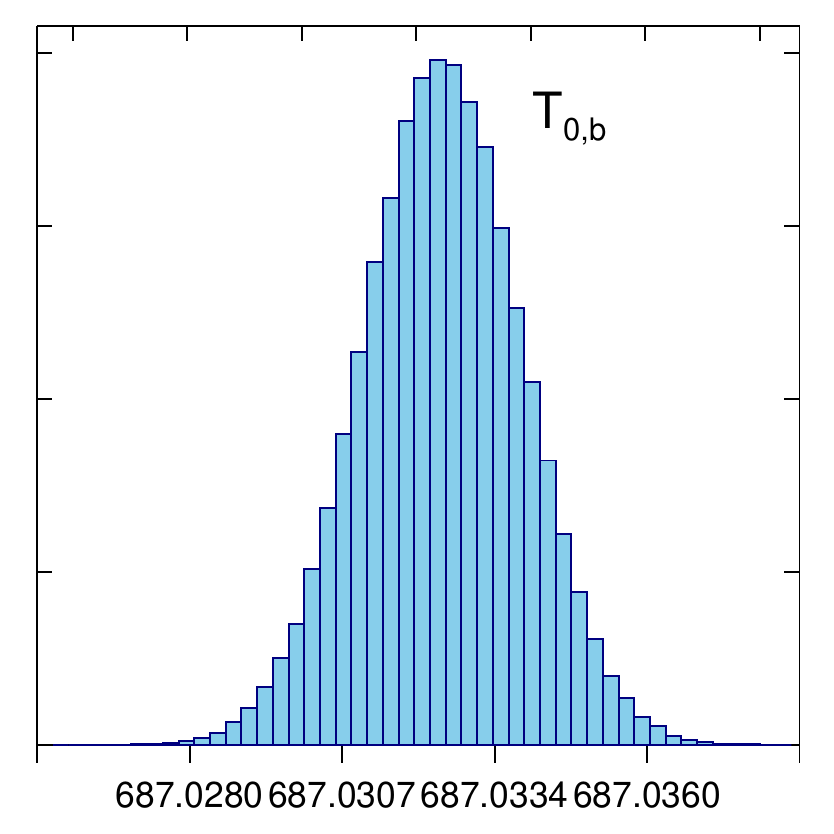}\hfill
\includegraphics[width=0.2\textwidth]{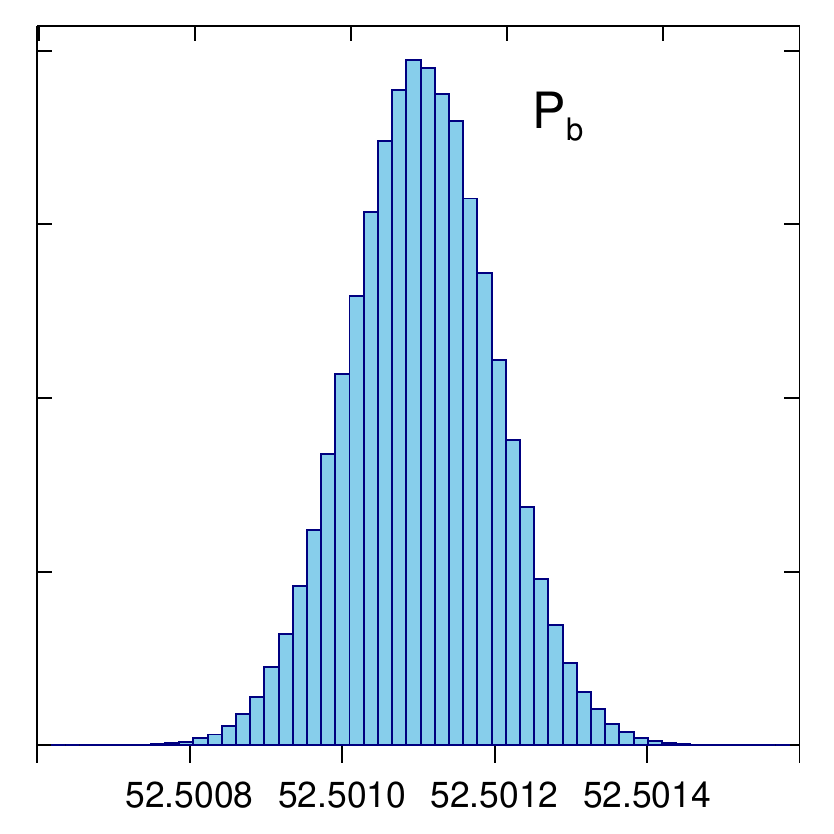}\hfill
\includegraphics[width=0.2\textwidth]{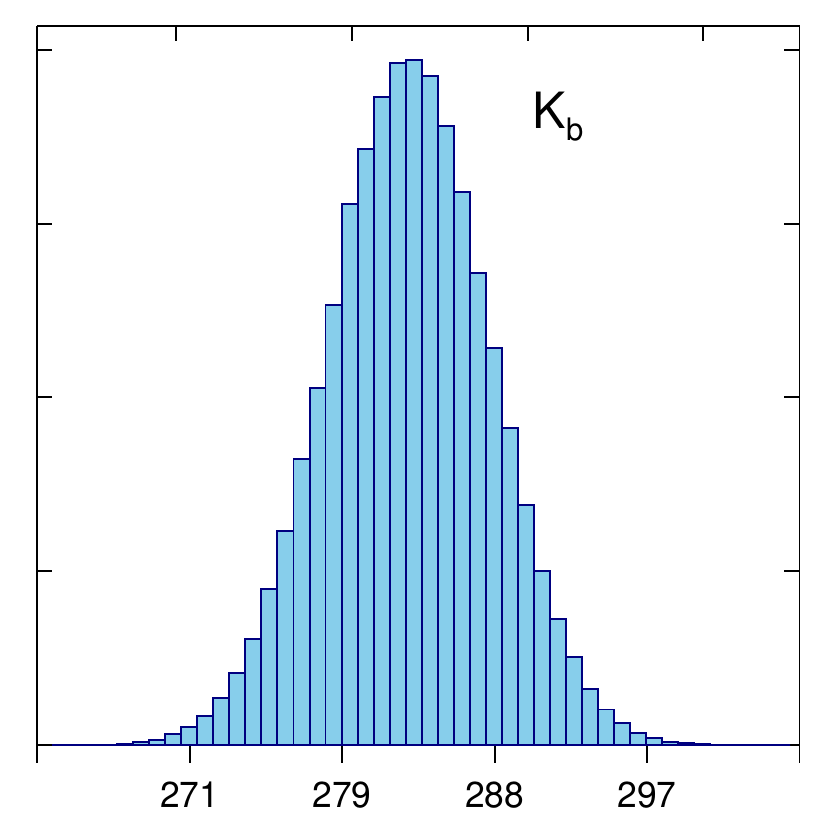}\hfill
\includegraphics[width=0.2\textwidth]{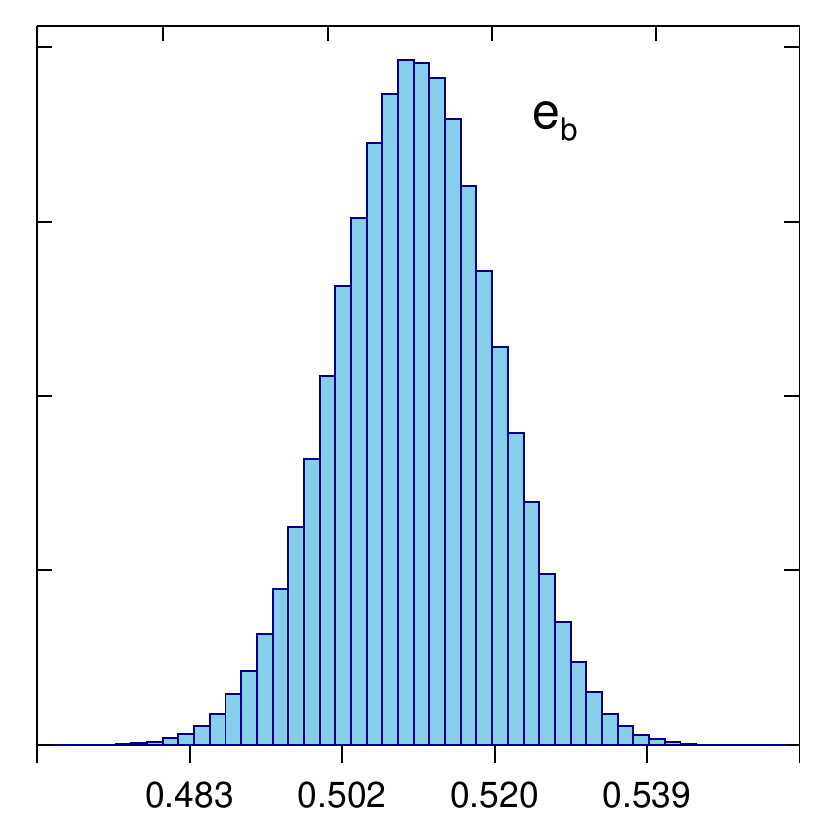}\hfill
\includegraphics[width=0.2\textwidth]{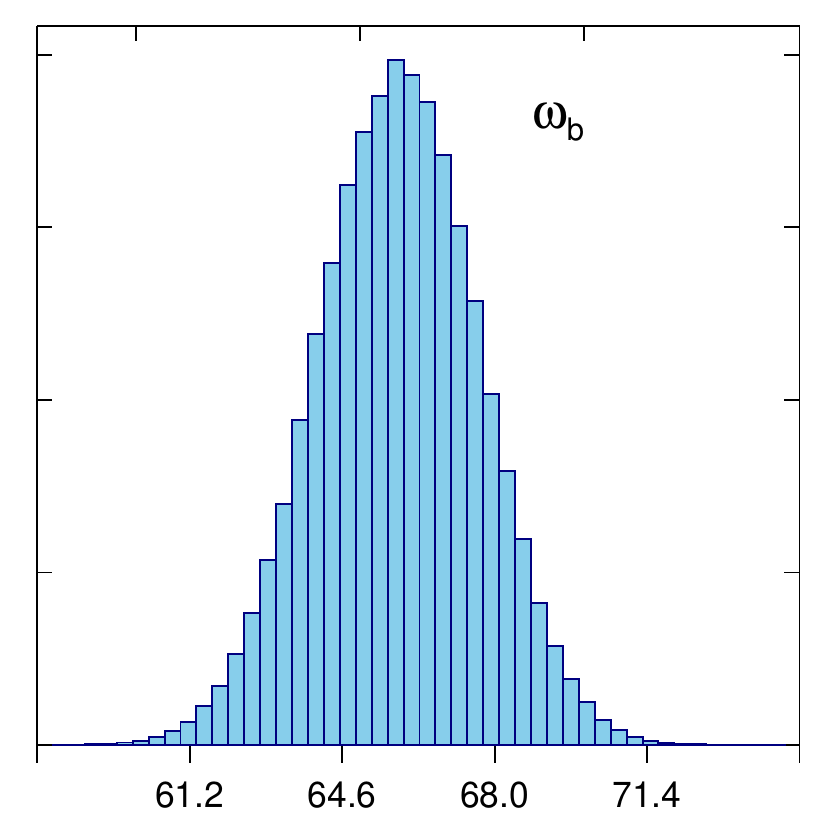}\\
\includegraphics[width=0.2\textwidth]{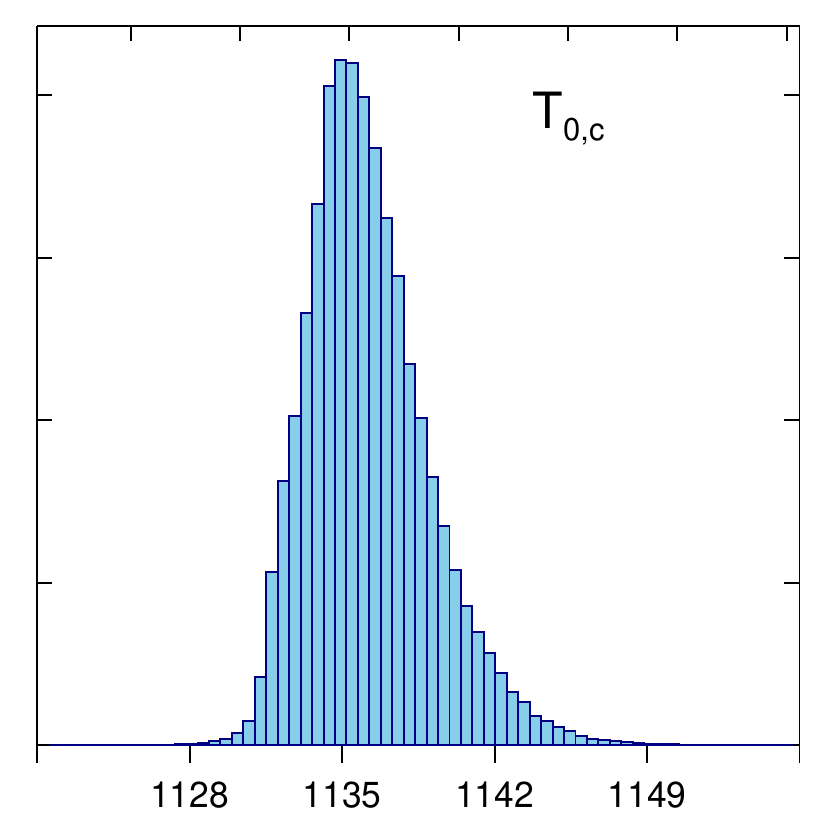}\hfill
\includegraphics[width=0.2\textwidth]{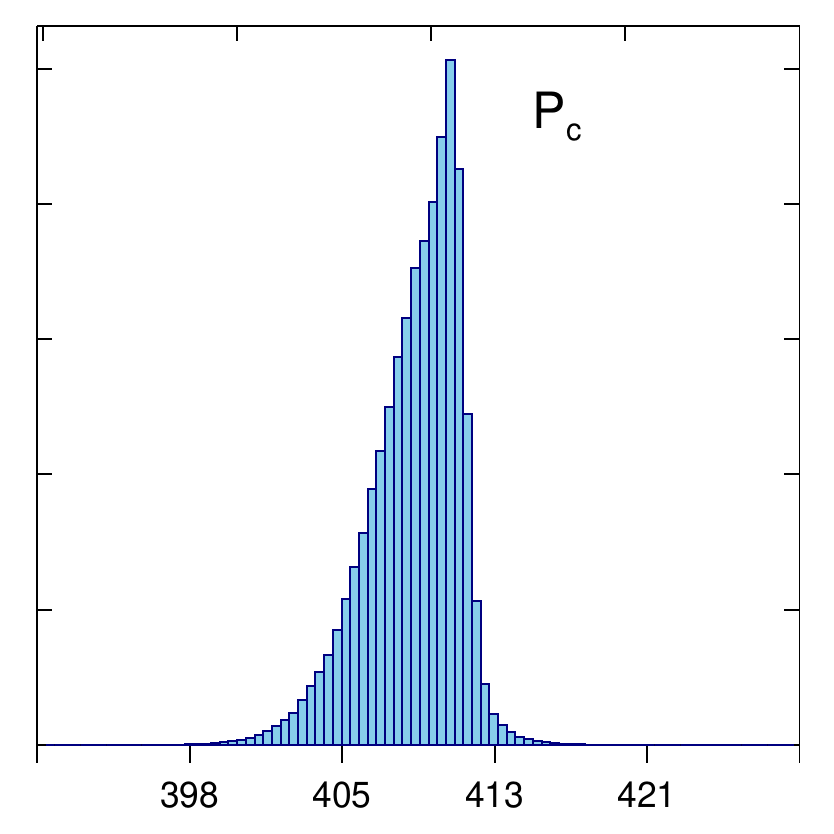}\hfill
\includegraphics[width=0.2\textwidth]{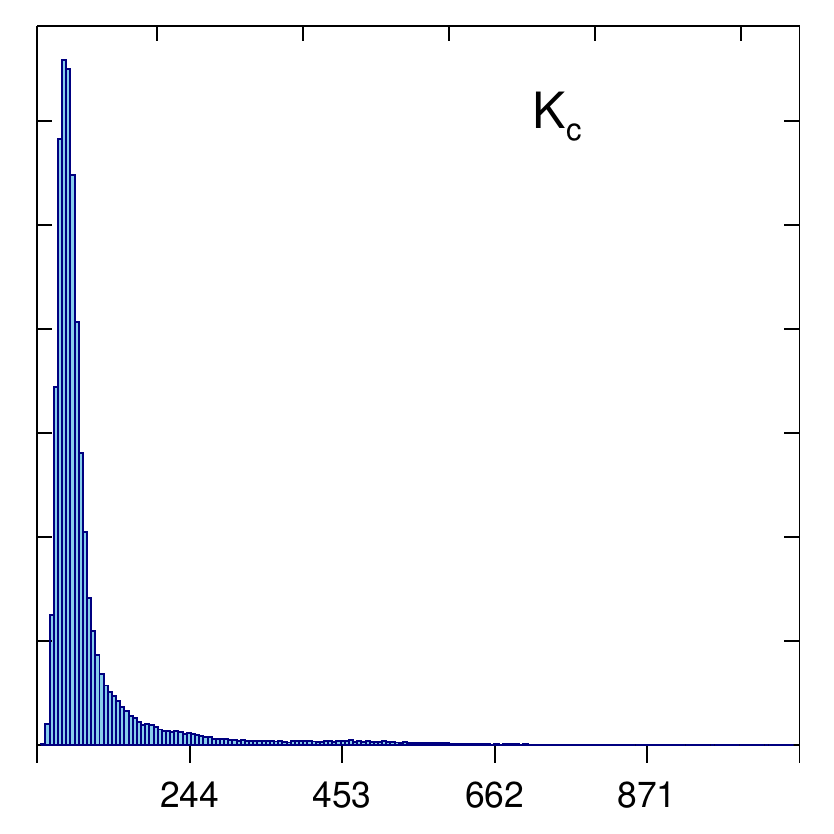}\hfill
\includegraphics[width=0.2\textwidth]{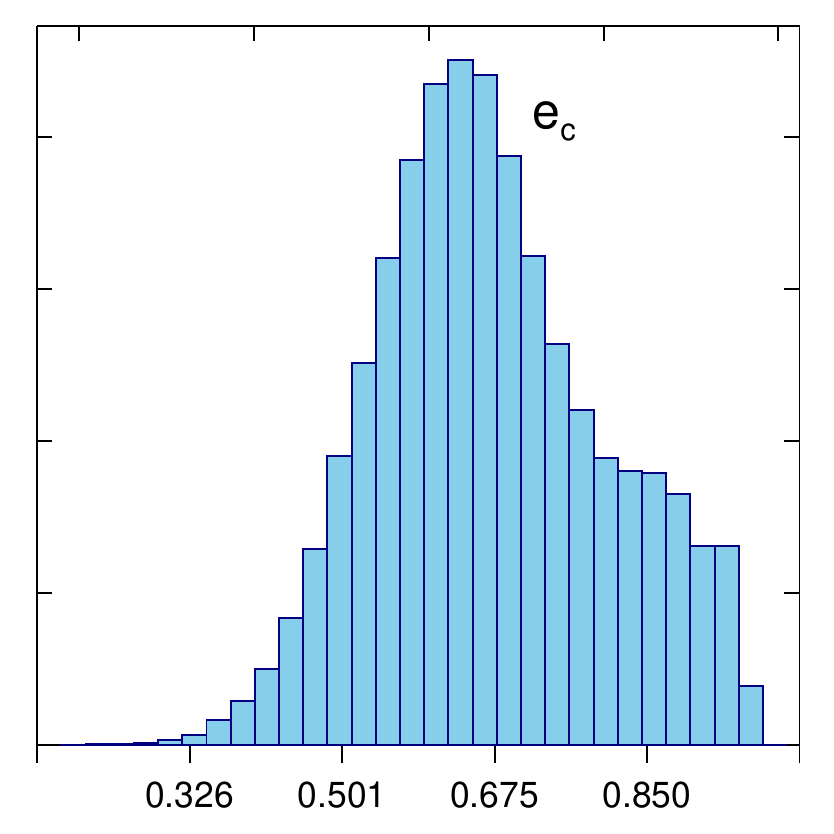}\hfill
\includegraphics[width=0.2\textwidth]{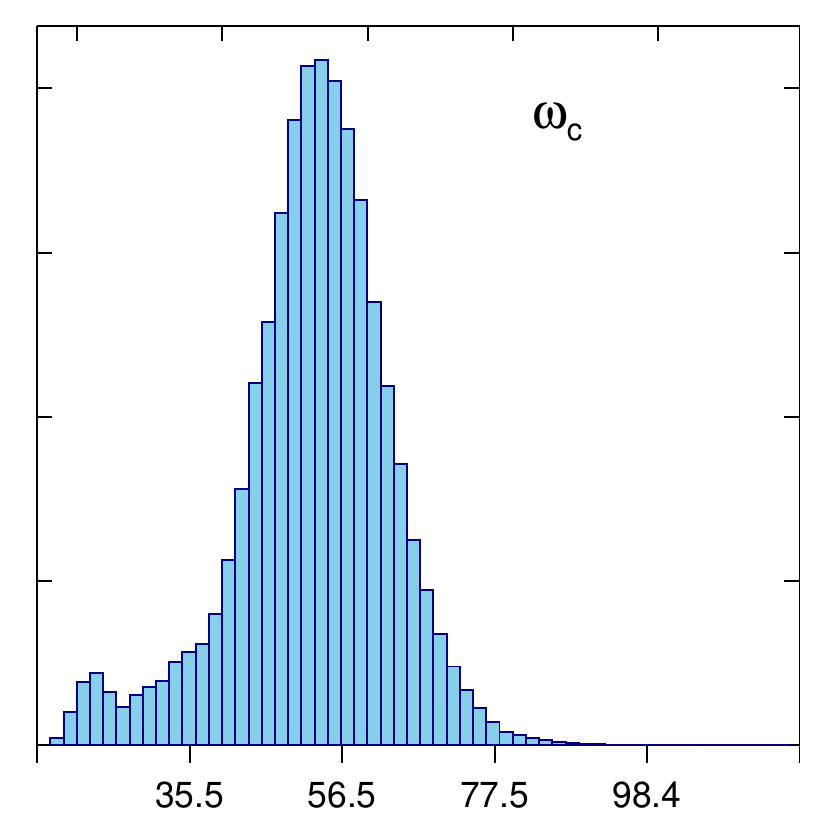}
\caption{ MCMC marginalized posterior distributions of the
  radial-velocity orbital parameters for the \this~system. While the
  inner planet (top row) has a well-constrained solution, it is clear
  from the marginalized distributions that there is a tail of high
  eccentricity outer orbits (bottom row) that cannot be ruled out by
  the current RV data. This tail can also be seen in the joint
  mass-eccentricity posterior shown in the stability analysis of
  \reffigl{stability}.
\label{fig:post}}
\end{figure*}

\section{Orbital Solution}
\label{sec:rv}

After recognition of the signature of the non-transiting planet in the
\this\ RVs (the outer planet was identified both visually and via
periodogram analysis), they were fit with two Keplerian orbits using a
MCMC algorithm with the Metropolis-Hastings rule
\citep[][]{metropolis:1953,hastings:1970} and a Gibbs sampler \citep[a
review of which can be found in][]{casella:1992}. Twelve parameters
were included in the fit: for each planet, the times of inferior
conjunction $T_{\rm 0}$, orbital periods $P$, radial-velocity
semi-amplitudes $K$, and the orthogonal quantities
$\sqrt{e}\sin{\omega}$ and $\sqrt{e}\cos{\omega}$, where $e$ is
orbital eccentricity and $\omega$ is the longitude of periastron; the
systemic velocity, $\gamma_{\rm rel}$, in the arbitrary zero point of
the TRES relative RV data set; and the FIES RV offset,
${\Delta}RV_{\rm FIES}$. (The absolute systemic velocity, $\gamma_{\rm
abs}$, was calculated based on $\gamma_{\rm rel}$~and the offset
between relative and absolute RVs, which are discussed in
\refsecl{spec}.) We applied Gaussian priors on $T_{\rm 0,b}$ and
$P_{\rm b}$ based on the results of the light curve fitting.

\capstartfalse
\begin{deluxetable}{lrr}[!h]
  \tablewidth{0.48\textwidth}
  \tablecaption{\this~Planetary Properties
    \label{tab:planetary}
  }
  \tablehead{
    \colhead{Parameter} &
    \colhead{Without N-Body} &
    \colhead{With N-Body}
  }
  \startdata
  Inner Planet & & \\
  ~$T_{\rm 0,b}$~(BJD)      & $2455949.5374^{+0.0018}_{-0.0016}$  & $2455949.5374^{+0.0011}_{-0.0012}$  \\
  ~$P_{\rm b}$~(days)       & $52.501134^{+0.000070}_{-0.000107}$ & $52.501129^{+0.000067}_{-0.000053}$ \\
  ~$K_{\rm b}$~(\ms)        & $285.9^{+4.1}_{-4.7}$               & $286.8^{+4.7}_{-4.0}$               \\
  ~$\sqrt{e_{\rm b}}\cos{\omega_{\rm b}}$
                            & $0.294^{+0.022}_{-0.015}$           & $0.311^{+0.018}_{-0.016}$           \\
  ~$\sqrt{e_{\rm b}}\sin{\omega_{\rm b}}$
                            & $0.6482^{+0.0134}_{-0.0094}$        & $0.645^{+0.012}_{-0.011}$           \\
  ~$e_{\rm b}$              & $0.5121^{+0.0084}_{-0.0107}$        & $0.5134^{+0.0098}_{-0.0089}$        \\
  ~$\omega_{\rm b}$~(deg)   & $65.6^{+1.5}_{-1.8}$                & $64.1^{+1.6}_{-1.5}$                \\   
  ~$i_b$~(deg)              & $88.17^{+0.61}_{-0.33}$             & $88.17^{+0.61}_{-0.33}$             \\
  ~$M_{\rm b}$~(\mjup)      & $5.41^{+0.30}_{-0.19}$              & $5.41^{+0.32}_{-0.18}$              \\
  ~$R_{\rm b}$~(\rjup)      & $1.145^{+0.036}_{-0.039}$           & $1.145^{+0.036}_{-0.039}$           \\
  ~$\rho_{\rm b}~({\rm g\,cm}^{-3})$ 
                            & $4.46^{+0.36}_{-0.29}$              & $4.46^{+0.37}_{-0.29}$              \\
  ~$a_{\rm b}$~(AU)         & $0.301^{+0.016}_{-0.011}$           & $0.301^{+0.016}_{-0.011}$           \\
  ~$\langle F_{\rm b} \rangle$~($\langle F_\oplus \rangle$)\tablenotemark{a}
                            & $118 \pm 10$                        & $118 \pm 10$ \\
  Outer Planet & & \\
  ~$T_{\rm 0,c}$~(BJD)      & $2456134.9^{+3.0}_{-2.1}$           & $2456139.3^{+3.6}_{-2.9}$           \\
  ~$P_{\rm c}$~(days)       & $411.0^{+0.9}_{-3.2}$               & $406.2^{+3.9}_{-2.5}$               \\
  ~$K_{\rm c}$~(\ms)        & $73^{+25}_{-15}$                    & $62.1^{+6.1}_{-5.8}$                \\
  ~$\sqrt{e_{\rm c}}\cos{\omega_{\rm c}}$
                            & $0.47^{+0.12}_{-0.14}$              & $0.336^{+0.115}_{-0.076}$           \\
  ~$\sqrt{e_{\rm c}}\sin{\omega_{\rm c}}$
                            & $0.648^{+0.072}_{-0.073}$           & $0.602^{+0.048}_{-0.071}$           \\
  ~$e_{\rm c}$              & $0.64^{+0.14}_{-0.13}$              & $0.498^{+0.029}_{-0.059}$           \\
  ~$\omega_{\rm c}$~(deg)   & $53.2^{+8.7}_{-9.5}$                & $60.8^{+7.0}_{-11.2}$               \\   
  ~$M_{\rm c} \sin{i_{\rm c}}$~(\mjup)     
                            & $2.63^{+0.43}_{-0.35}$              & $2.43^{+0.22}_{-0.24}$              \\
  ~$a_{\rm c} \sin{i_{\rm c}}$~(AU)      
                            & $1.188^{+0.062}_{-0.042}$           & $1.178^{+0.063}_{-0.042}$           \\
  ~$\langle F_{\rm c} \rangle$~($\langle F_\oplus \rangle$)\tablenotemark{a}
                            & $8.5^{+2.1}_{-1.2}$                 & $7.7^{+0.7}_{-0.8}$                 \\
  Other Parameters & & \\
  ~$\gamma_{\rm rel}$~(\ms) & $313.4^{+5.1}_{-4.5}$               & $306.7^{+2.6}_{-2.7}$               \\
  ~$\gamma_{\rm abs}$~(\kms)
                            & $-35.22 \pm 0.19$                   & $-35.22 \pm 0.19$                   \\
  ~$\Delta RV_{\rm FIES}$~(\ms) 
                            & $-493.5^{+8.4}_{-10.4}$             & $-490.5^{+8.6}_{-7.7}$              \\
  ~RV Jitter (\ms)          & $20$                                & $20$                                \\
    [-2.4ex]
  \enddata
  \tablenotetext{a}{The time-averaged incident flux, $\langle F
  \rangle = \sigma T_{\rm eff}^4 \left(\frac{R_\star}{a}\right)^2
  \left(\frac{1}{1-e^2}\right)^{1/2}$.}
  \tablecomments{The first set of parameters comes from the the
    photometric, radial-velocity, and asteroseismic analyses, and the
    second set incorporates additional constraints from the stability
    analysis of our N-body simulations. We adopt the properties
    derived with constraints from the N-body simulations.}
\end{deluxetable}
\capstarttrue

\begin{figure*}[!htb]
\centering
\includegraphics[height=0.38\textwidth,clip,trim={0cm 0.3cm 0cm 0.5cm}]{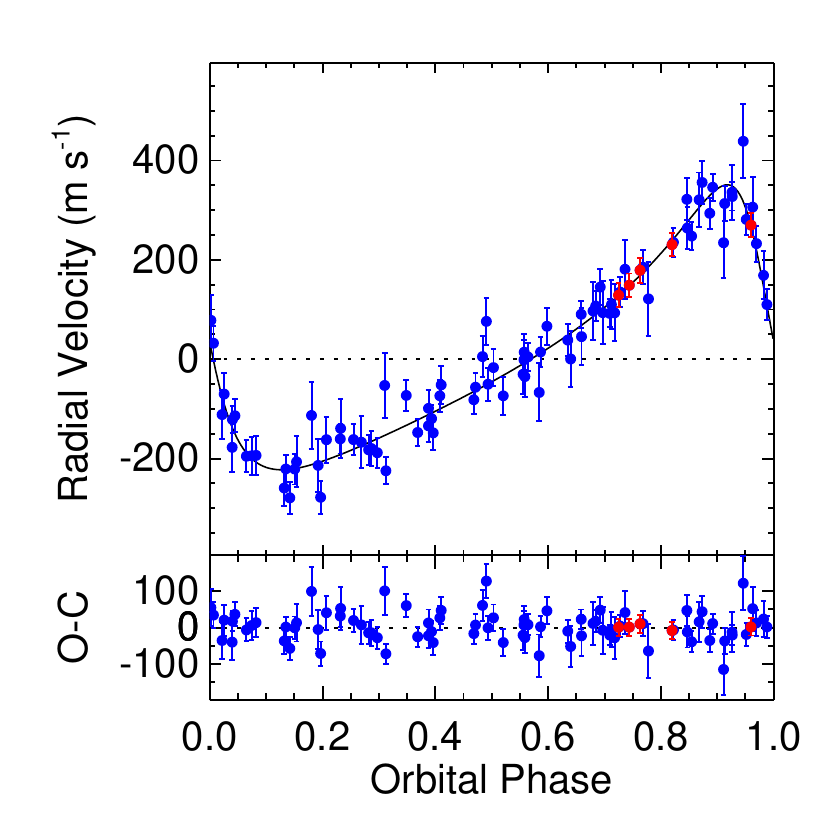}
\includegraphics[height=0.38\textwidth,clip,trim={1.12cm 0.3cm 0cm 0.5cm}]{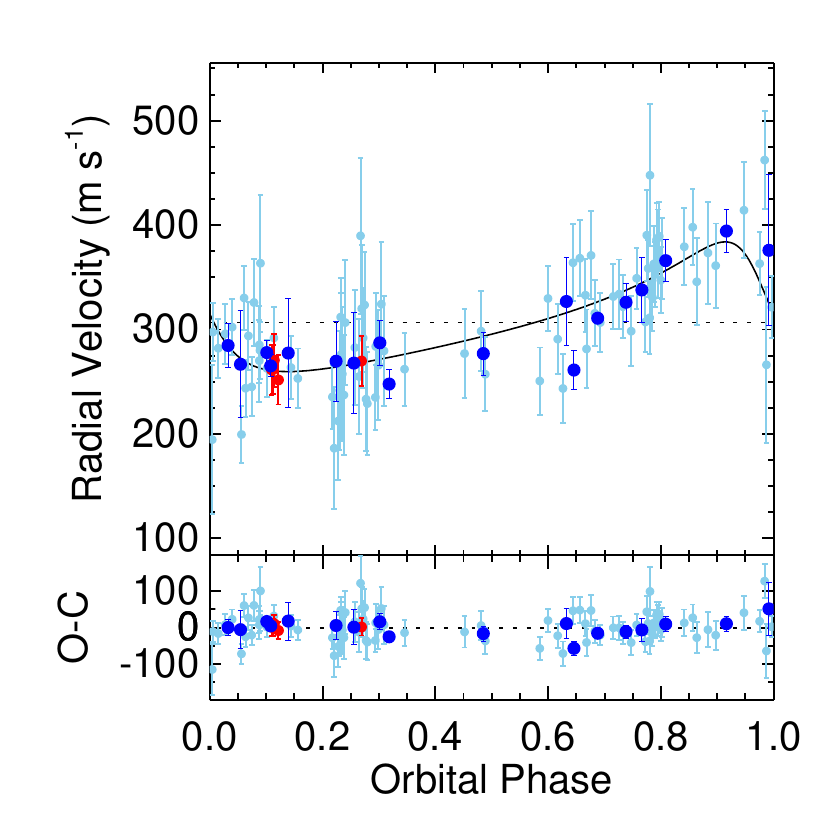}
\includegraphics[height=0.38\textwidth,clip,trim={0.5cm 0.3cm 0.5cm 0.5cm}]{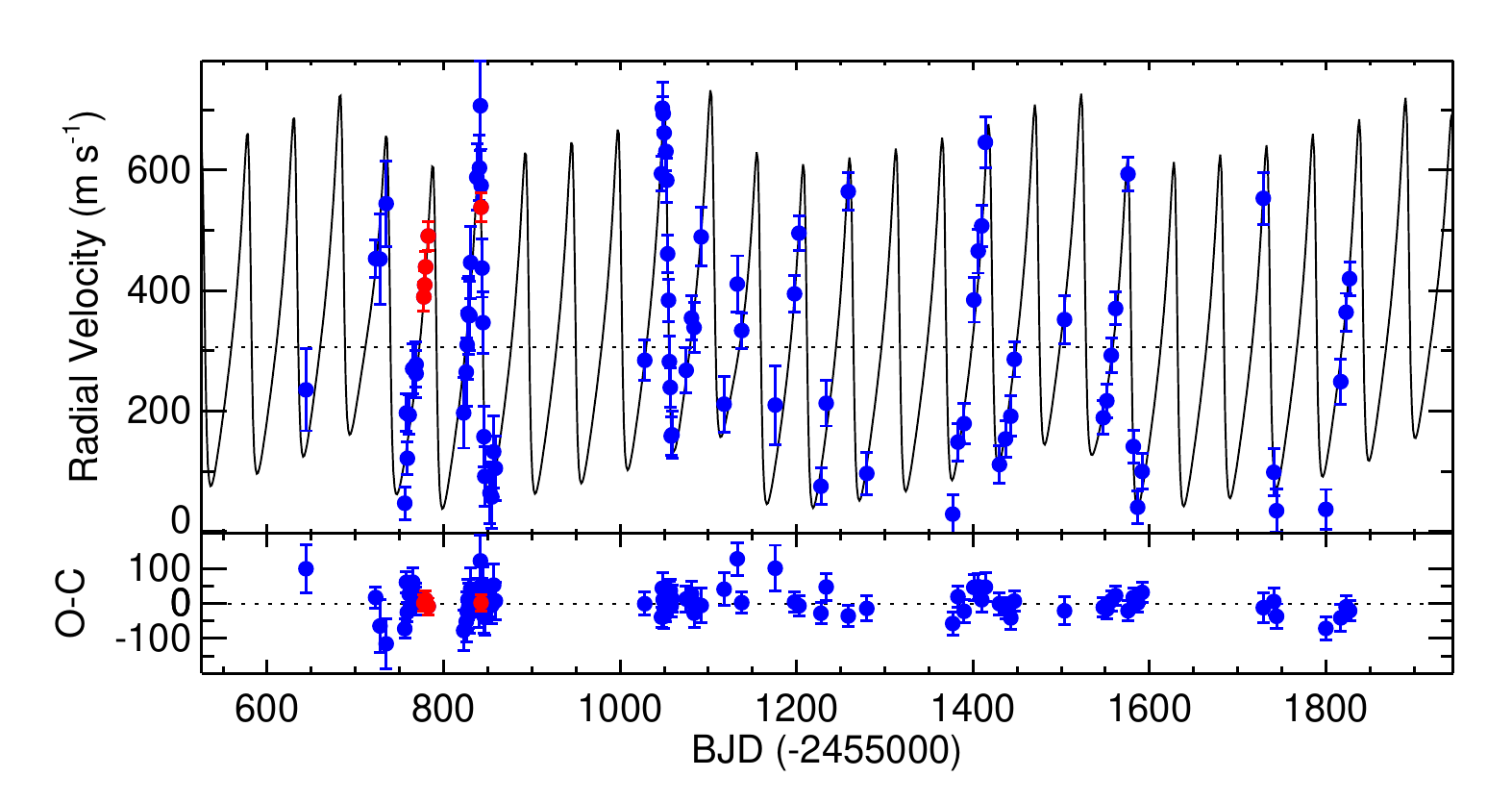}
\caption{The individual components of the orbital fit corresponding to
  planets b (top left) and c (top right), and the combined RV
  timeseries (bottom). TRES RVs are plotted in blue and FIES RVs are
  plotted in red. For planet c, we plot the individual RVs in light
  blue and the binned data in dark blue. Velocity residuals are
  plotted in the corresponding lower panels. The plotted error bars
  include both the internal errors and the $20$\,\ms\ jitter that was
  added to the uncertainties during the fitting process.
\label{fig:orbit}}
\end{figure*}

We ran a chain with $1.01 \times 10^7$~steps, treating the first
$10^5$ realizations as burn-in and thinning the chain by saving every
tenth entry, for a final chain length of $10^6$. The marginalized
posterior distributions are shown in \reffigl{post}. It is apparent
that while the parameters of the inner planet are very well
constrained, there is a high-eccentricity tail of solutions for the
outer planet that cannot be ruled out by RVs alone (we investigate
this further using N-body simulations in \refsecl{nbody}). Because
several of the posteriors are non-Gaussian, we cannot simply adopt the
median and central $68.3\%$~confidence interval as our best fit
parameters and $1$-$\sigma$~errors as we normally might. Instead, we
adopt best fit parameters from the mode of each distribution, which we
identify from the peak of the probability density function (PDF). We
generate the PDFs using a Gaussian kernel density estimator with
bandwidths for each parameter chosen according to Silverman's rule. We
assign errors from the region that encloses $68.3\%$~ of the PDF, and
for which the bounding values have identical probability
densities. That is, we require the $\pm1$-$\sigma$~values to have
equal likelihoods. The resulting orbital solution using these
parameters has velocity residuals larger than expected from the
nominal RV uncertainties. We attribute this to some combination of
astrophysical jitter (e.g., stellar activity or additional undetected
planets) and imperfect treatment of the various noise sources
described in \refsecl{spec}. An analysis of the residuals does not
reveal any significant periodicity, but given our measurement
precision, we would not expect to detect any additional planets unless
they were also massive gas giants, or orbiting at very small
separations. To account for the observed velocity residuals, we re-run
our MCMC with the inclusion of an additional RV jitter term. Tuning
this until $\chi^2$ is equal to the number of degrees of freedom, we
find an additional $20~\ms$~jitter is required. While part of this
jitter may be due to instrumental effects, we do not include separate
jitter terms for the TRES and FIES RVs because the FIES data set is
not rich enough to reliably determine the observed scatter. We report
the best fit orbital and physical planetary parameters in
\reftabl{planetary}, which also includes the set of parameters
additionally constrained by dynamical stability simulations, as
described in the N-body analysis of \refsecl{nbody}. The corresponding
orbital solution is shown in \reffigl{orbit}.

\section{False Positive Scenarios}
\label{sec:fp}

An apparent planetary signal (transit or RV) can sometimes be caused
by astrophysical false positives. We consider several scenarios in
which one of the \this~planetary signals is caused by something
other than a planet, and we run a number of tests to rule these out.

For an object with a deep transit, such as \this{b}, one may worry
that the orbiting object is actually a small star, or that the signal
is caused by a blend with an eclipsing binary
system. \citet{sliski:2014} noted that there are very few planets
orbiting evolved stars with periods shorter than $100$~days (\this{b}
would be somewhat of an outlier), and also found via AP that either
the transit signal must be caused by a blend, or it must have
significant eccentricity ($e>0.488$). Without additional evidence,
this would be cause for concern, but our radial velocity curve
demonstrates that the transiting object is indeed orbiting the target
star, that its mass is planetary, and, consistent with the prediction
of \citet{sliski:2014}, its eccentricity is $0.5134$.

If a planet does not transit, as is the case for \this{c}, determining
the authenticity of the planetary signal is less straightforward. An
apparent radial-velocity orbit can be induced by a genuine planet,
spots rotating on the stellar surface \citep{queloz:2001}, or a
blended stellar binary \citep{mandushev:2005}. Both of these false
positive scenarios should manifest themselves in the shapes of the
stellar spectral lines. That is, spots with enough contrast with the
photosphere to induce apparent RV variations will also deform the line
profiles, as should blended binaries bright enough to influence the
derived RVs. A standard prescription for characterizing the shape of a
line is to measure the relative velocity at its top and bottom; this
difference is referred to as a line bisector span \citep[see,
e.g.,][]{torres:2005}. To test against the scenarios described, we
computed the line bisector spans for the TRES spectra. We do find a
possible correlation with the RVs of the outer planet, having a
Spearman's rank correlation value of $-0.21$ and a significance of
$94\%$. While this is a potential concern and we cannot conclusively
demonstrate the planetary nature of the $407$-day signal, we find it
to be the most likely interpretation. In the following paragraphs, we
explain why other interpretations are unlikely.

With the discovery of a close stellar companion (see \refsecl{ao}), it
is reasonable to ask how that might affect interpretation of the outer
planetary signal. That is, if the stellar companion is itself a binary
with a period of $\mysim 400$~days, could it cause the RV signal we
observe? The answer is unequivocally no, as the companion is far too
faint in the optical compared to the primary star ($\Delta V\mysim 7$)
to contribute any significant light to the spectrum, let alone induce
a variation of $\mysim 100~\ms$~or affect the bisector spans. If there
is also a brighter visual companion inside the resolution limit of our
high resolution images ($\mysim 0\farcs05 \simeq 45$\ AU projected
separation), it could be a binary with a $400$-day period responsible
for the RV and bisector span variations on that timescale. However,
such a close physically bound binary may pose problems for formation
of the $52.5$-day planet, and the a priori likelihood of a background
star bright enough to cause the observed variations within
$0.05\arcsec$~ is extremely low; a TRILEGAL simulation suggests
$\mysim 5 \times 10^{-6}$~background sources should be expected, and
only a small fraction of those would be expected to host a binary with
the correct systemic velocity.

To rule out spot-induced velocity variation, we examine the {\em
Kepler} light curve for evidence of spot activity. From the measured
$v\sin{i_\star}$, $i_\star$, and $R_\star$, the stellar rotation
period is $77 \pm 14$~days. Not only is this inconsistent with the
observed outer orbital period ($407$~days), but we detect no
significant photometric signal near either of these periods. For
\this~($v\sin{i_\star}=2.7~\kms$), a spot must cover $\mysim4$\%--$5$\%
of the stellar surface to induce the observed RV amplitude
\citep{saar:1997}, and such a spot would have been apparent in the
high precision {\em Kepler} light curve.

As we have shown in \refsecl{seis}, \this~exhibits strong
oscillations, so one may wonder whether these could induce the
observed RV signal for the outer planet. Oscillations on a $400$-day
timescale are intrinsically unlikely, as there is no known driving
mechanism that could cause them in giant stars; the well-known
stochastically driven oscillations are confined to much higher
frequencies. Furthermore, if there were such a mechanism, a mode with
$\mysim 50~\ms$ velocity semi-amplitude should cause a photometric
variation of $\mysim 1.3$\,mmag \citep[see][]{kjeldsen:1995}, which is
clearly ruled out by the {\em Kepler} data.

Upon examining all of the evidence available, we conclude that both
detected orbits are caused by bona fide planets orbiting the primary
star.

\section{Orbital Stability Analysis}
\label{sec:nbody}
\subsection{Methodology}
\label{sec:nbody_method}

Following the Keplerian MCMC fitting procedure described in
\refsecl{rv}, we wish to understand whether these posterior solutions
are dynamically stable---i.e., whether they describe realistic
systems which could survive to the $3.5$~Gyr age of the system. When
presenting our results in the sections that follow, we use the
planet--planet separation as an easily visualized proxy for stability:
given the semi-major axes of the two planets---$\mysim 0.3$~and
$\mysim 1.2$\,AU---separations greater than $\mysim 10$\,AU are clear
indications that the system has suffered an instability and the
planets subsequently scattered. If some solutions do prove to be
unstable, it will lead to further constraints on the orbital elements,
and thus the planetary masses. We perform integrations of both
coplanar and inclined systems. We first explore the less
computationally expensive coplanar case to understand the behavior of
the system, and then extend the simulations to include inclination in
the outer orbit, which has the additional potential to constrain the
mutual inclination of the planets.

To understand this stability, we utilize the integration algorithm
described in \citet{payne:2013}.  This algorithm uses a symplectic
method in Jacobi coordinates, making it both accurate and rapid for
systems with arbitrary planet-to-star mass ratios.  It uses
calculations of the tangent equations to evaluate the Lyapunov
exponents for the system, providing a detailed insight into whether
the system is stable (up to the length of the simulation examined) or
exhibits chaos (and hence instability).

For the coplanar systems, we take the ensemble of $10^6$~solutions
generated in \refsecl{rv} (which form the basis of the reported
elements in column 2 of \reftabl{planetary}) and convert these to
Cartesian coordinates, assuming that the system is coplanar and
edge-on ($90\dg$), and hence both planets have their minimum masses.
We then evolve the systems forward for a fixed period of time (more
detail supplied in \refsecl{nbody_coplanar} below) and examine some
critical diagnostics for the system (e.g., the Lyapunov time, and the
planet--planet separation) to understand whether the system remains
stable, or whether some significant instability has become apparent.

For the inclined systems, we assume that the inner (transiting) planet
is edge-on, and hence retains its measured mass.  However, the outer
planet is assigned an inclination that is drawn randomly from a
uniform distribution $0\dg < i_{\rm c} < 90\dg$, and its mass is
scaled by a factor $1 / \sin{i_{\rm c}}$.  As such, the outer
planet can have a mass that is significantly above the minimum values
used in the coplanar case.  The longitude of ascending node for the
outer planet is drawn randomly from a uniform distribution between $0$
and $2\pi$.  We then proceed as in the coplanar case, integrating the
systems forward in time to understand whether the initial conditions
chosen can give rise to long-term stable systems.

\citet{rauch:1999} demonstrated that $\sim 20$~timesteps per
inner-most orbit is sufficient to ensure numerical stability in
symplectic integrations.  As the inner planet has a period $\mysim
50$~days, we use a timestep of $1$~day in all of our simulations,
ensuring that our integrations will comfortably maintain the desired
energy conservation and hence numerical accuracy.

\subsection{Coplanar Stability}
\label{sec:nbody_coplanar}

\begin{figure}[!tb]
\includegraphics[height=3.5in,angle=270,trim={0cm 2cm 0cm 2cm},clip]{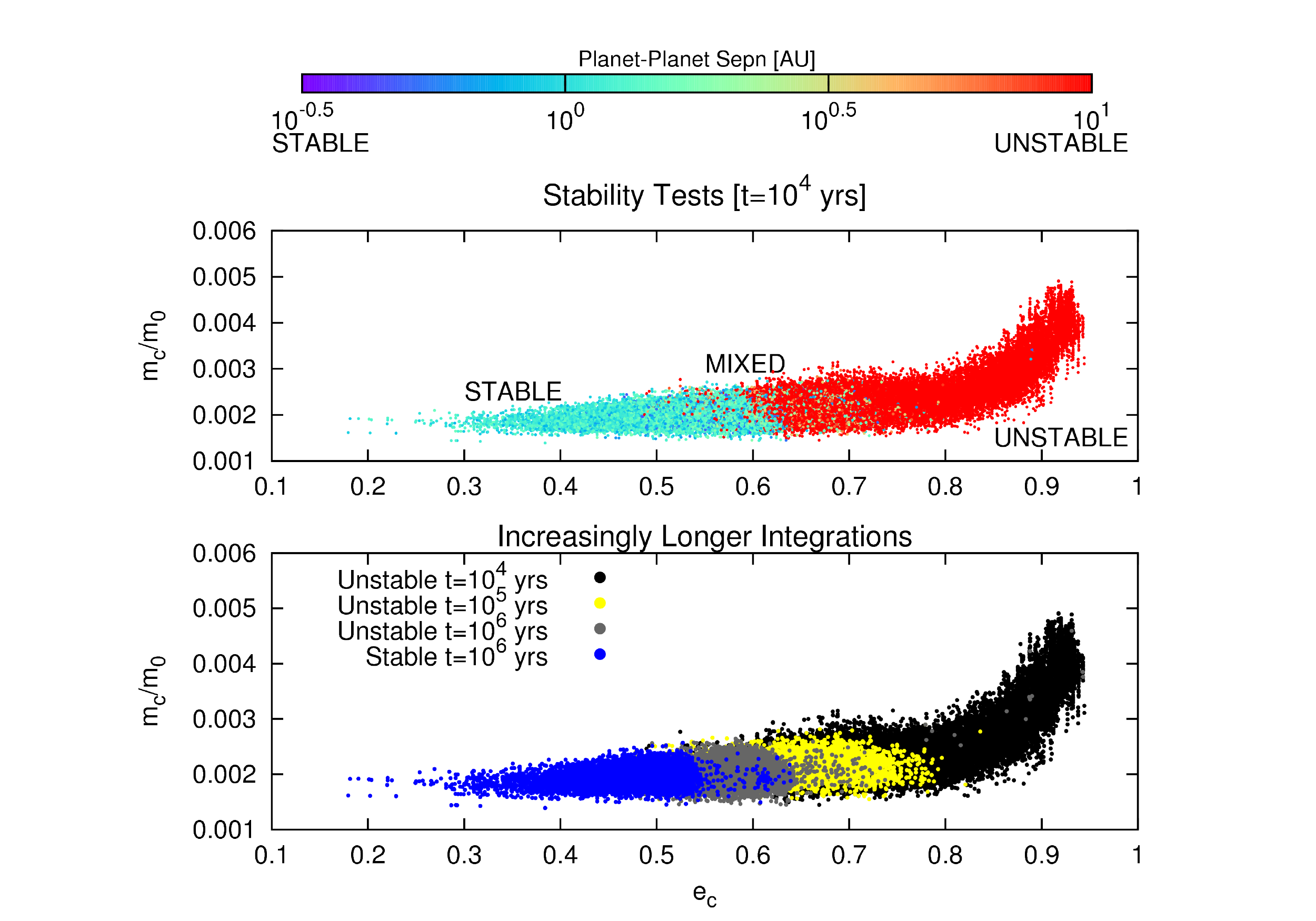}
\caption{ Top: in the mass-eccentricity plane for the outer planet
($m_c,e_c$), we plot the separation between the two planets in the
system at $t=10^4$\,yr using the color scheme at the top of the plot.
A separation greater than $\sim 10$\,AU is an indication that the
system has suffered an instability and the planets subsequently
scattered.  Systems with eccentricity $e_c \gsim 0.75$ are clearly
unstable, while those with $e_c \lsim 0.55$ are clearly stable for
$t=10^4$\,yr.
Bottom: in the $(m_c,e_c)$ plane, we plot the systems which are
unstable at $10^4$\,yr in black, $10^5$\,yr in yellow and $10^6$\,yr
in gray, and those that remain \emph{stable} at $10^6$\,yr in blue.
The systems which are long-term stable in these coplanar simulations
occupy a region of parameter space confined to eccentricities
$e_c\lsim 0.55$.
\label{fig:stability}}
\end{figure}

We begin by taking a random selection of $10^5$ of the $10^6$
solutions from \refsecl{rv} and integrating them for a period of
$10^4$\ yr ($3 \times 10^6$ timesteps).  While this is {\em not} a
particularly long integration period compared with the period of the
planets ($\mysim 50$ and $\mysim 400$~days), we demonstrate that even
during this relatively short integration, approximately half of the
systems become unstable.  Tellingly, the unstable systems all tend to
be the systems in which the outer planet has particularly high
eccentricity.  We illustrate this in \reffigl{stability}, where we
plot the mass-eccentricity plane for the outer planet ($m_{\rm
c},e_{\rm c}$) and plot the separation between the two planets in the
system at $t=10^4$\ yr.  As described above, separations greater than
$\mysim 10$\,AU indicate that the system has suffered an
instability. This initial simulation clearly demonstrates that at
$t=10^4$\ yr essentially all systems with $e_{\rm c} > 0.8$ are
unstable, all those with $e_{\rm c} < 0.45$ are stable, and those with
$0.45 < e_{\rm c} < 0.8$ are ``mixed,'' with some being stable and
some being unstable.

Given this promising demonstration that the dynamical integrations can
restrict the set of solutions, we go on to integrate the systems for
increasingly longer periods of time.  To save on integration
time/cost, we only select the stable solutions from the previous step,
and then extend the integration time by an order of magnitude, perform
a stability analysis, and repeat.  By this method, the $10^5$ systems
at $t=0$ are reduced to $\mysim 4.4 \times 10^4$ stable systems at
$10^4$\,yr, $\mysim 2.5 \times 10^4$ stable systems at $10^5$\,yr, and
$\mysim 9.4 \times 10^3$ stable systems at $10^6$\,yr. We illustrate
in \reffigl{stability} the successive restriction of the parameter
space in the mass-eccentricity plane for the outer planet ($m_{\rm
c},e_{\rm c}$) as the integration timescales increase.  We find that
the long-term stable systems occupy a significantly smaller region of
parameter space in the $m_{\rm c},e_{\rm c}$ plane.  In particular, we
see the eccentricity of the outer planet is restricted to $e_{\rm c}
\lsim 0.55$.

Using the $\mysim 9.4 \times 10^3$ systems which remained stable at
$10^6$\,yr, we use the same method as in \refsecl{rv} to determine a
new set of best-fit orbital paramters. That is, we use a Gaussian
kernel density estimator to smooth the posterior distribution and
select the mode (i.e., the value with the largest probability
density). The errors correspond to values with equal probability
density that enclose the mode and $68.3\%$~of the PDF. The most
significant changes in the best-fit parameters occurred for the outer
planet, most notably for eccentricity and velocity semi-amplitude,
resulting in smaller and more symmetric error bars and a revision in
the best-fit minimum mass for planet c.

\subsection{Inclined-system Stability}
\label{sec:nbody_inclined}

\begin{figure}[t]
\includegraphics[height=3in,angle=0,trim={2cm 1cm 0cm 3cm},clip,angle=270]{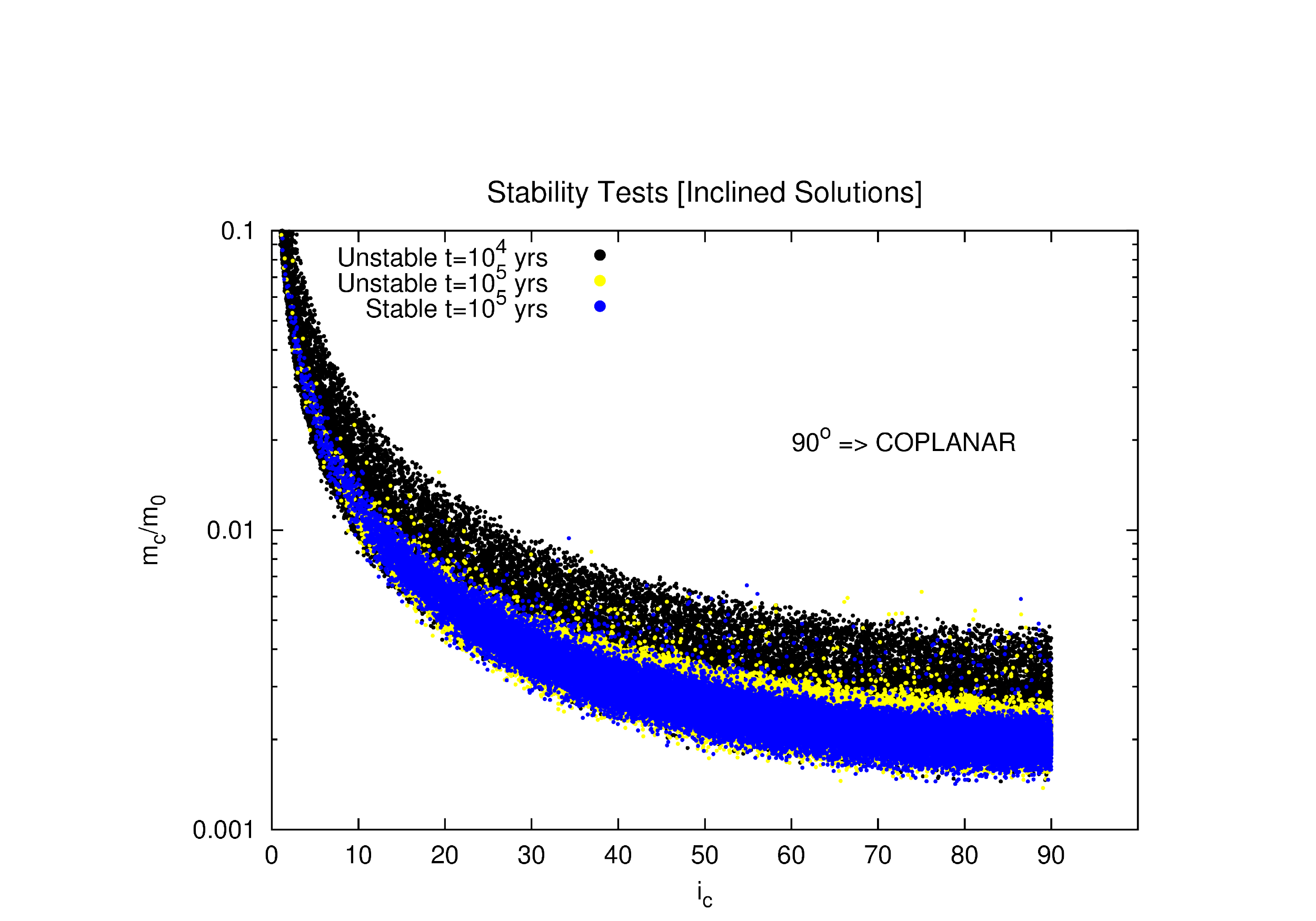}
\caption{ Stability of systems in the mass-orbital inclination plane
  of the outer planet ($m_{\rm c}, i_{\rm c}$), with black points
  going unstable within $10^4$\,yr and yellow points within
  $10^5$\,yr, and blue points representing stable systems after
  integrating for $10^5$\,yr. The results mirror those of the coplanar
  case---at all inclinations, the low-mass (low-eccentricity) systems
  remain stable, while the high-mass (high-eccentricity) systems do
  not survive long integrations.  As some systems remain stable even
  for highly inclined orbits ($i_{\rm c} \mysim 0\dg$), we are unable
  to put strong constraints on the mutual inclination of the planets.
\label{fig:inclined}}
\end{figure}

In a manner similar to the coplanar analysis of
\refsecl{nbody_coplanar}, we begin by taking all $10^6$~solutions from
\refsecl{rv} and integrating them for a period of $10^4$\,yr ($3
\times 10^6$~timesteps).  However, in these inclined system
integrations, the outer planet has an inclination that is not edge-on,
and hence a mass for the outer planet that is inflated compared to its
minimum edge-on value, as described in \refsecl{nbody_method}.

A much larger fraction of the inclined systems become unstable in the
first $10^4$\,yr, and we find that the $N=10^6$~systems at $t=0$ are
reduced to $N \mysim 1.5 \times 10^5$~stable systems at $10^4$\,yr.
Integrating these stable systems while keeping the pre-assigned
inclinations for the outer planet (i.e., we do {\em not} re-randomize
the inclinations) we integrate this subset on to $10^5$\,yr, and find
that the number of stable systems subsequently reduces to $N \mysim
10^4$. We plot these stability results as a function of mass and
inclination of the outer planet in \reffigl{inclined}.

We find that the vast majority of the stable solutions are (somewhat
unsuprisingly) restricted to the range of parameter space with
relative inclinations $\lsim 70\dg$~(i.e., $i_{\rm c} \gsim 20\dg$~in
\reffigl{inclined}). There is a small population with significantly
higher eccentricities and relative inclinations that remains stable at
$t=10^5$\,yr, and it is possible that these systems exhibit strong
Kozai oscillations, but their long-term ($>10^5$\,yr) behavior has not
been investigated. We emphasize that the inclinations and longitudes
of the ascending node were assigned {\em randomly}, and hence it is
possible that specifically chosen orbital alignments could allow for
enhanced stability in certain cases. Unfortunately, because some
systems remain stable even for highly inclined outer orbits
($i_c\mysim 0\dg$), we are unable to strongly constrain the mutual
inclination of the planetary orbits.

We also note that the presence of a distant stellar companion (as
detected in our AO images) or additional, as yet undetected, planets
(as may be suggested by the excess RV scatter) could influence the
long-term stability of the systems simulated herein. The evolution of
the star on the red giant branch is also likely to affect the orbital
evolution (especially for the inner planet; see \refsecl{disc}), but
it is not important over the $10^6$-year timescales of these
simulations. Simulating the effect of poorly characterized or
hypothetical orbits, or the interactions between expanding stars and
their planets, is beyond the scope of the current paper, and instead
we simply remind the reader of these complications. We present
planetary properties derived with and without constraints from our
simulations (\reftabl{planetary}) so that the cautious reader may
choose to adopt the more conservative (and poorly determined)
parameters of the outer orbit in any subsequent analysis.

\section{Discussion}
\label{sec:disc}

The properties of the star and planets of \this~are unusual in several
ways among the known exoplanets, which makes it a valuable system to
study in detail. For example, it is a planetary system around an
intermediate mass star (and an evolved star), it hosts a planet of
intermediate period, and it hosts at least one very massive transiting
planet. Close examination of the system may provide insight into the
processes of planet formation and orbital evolution in such
regimes. \this~is also the first planet orbiting a giant star to have
its eccentricity independently determined by RVs and photometry, which
helps address a concern that granulation noise in giants can inhibit
such photometric measurements.

\subsection{Comparing the Eccentricity from AP versus RVs}

AP provides an independent technique for measuring orbital
eccentricities with photometry alone, via the so-called photoeccentric
effect \citep{dawson:2012}, and can be used as a tool to evaluate the
quality of planet candidates \citep{tingley:2011}. It was originally
envisioned as a technique for measuring eccentricities
\citep{kipping:2012}, but other effects, such as a background blend,
can also produce AP effects \citep{kipping:2014a}. Given that we here
have an independent radial velocity orbital solution, there is an
opportunity to compare the two independent solutions, allowing us to
comment on the utility of AP.

\begin{figure}
\centering
\includegraphics[width=0.48\textwidth]{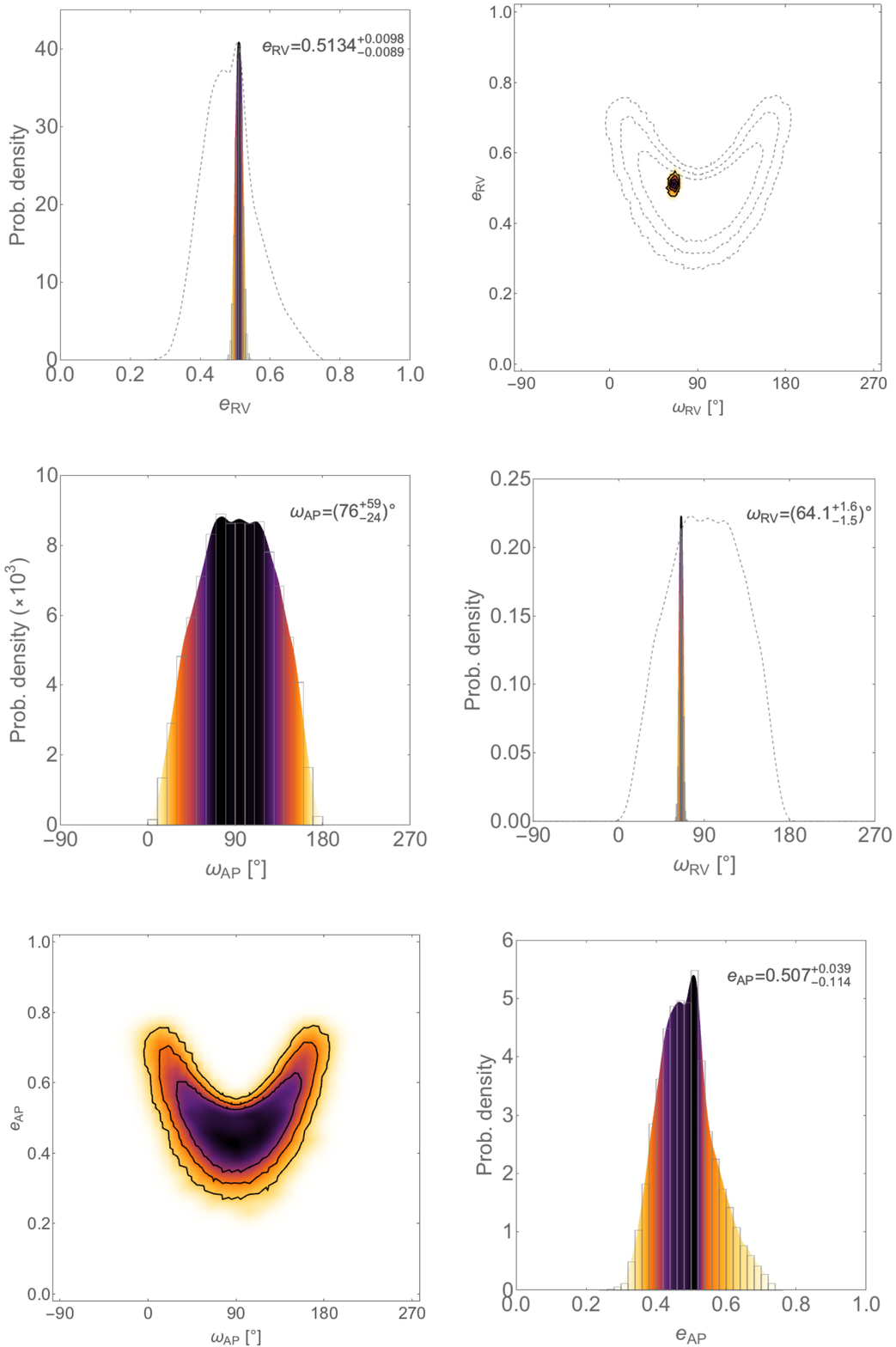}
\caption{ The marginalized and joint posterior distributions of
$e$~and $\omega$, as determined by AP and RVs. The three panels in the
lower left are from AP; the three in the upper right (the narrow
histograms and well-constrained joint distribution) are from RVs,
which also show the AP distributions overplotted as thin gray dotted
lines for comparison.
\label{fig:ew}}
\end{figure}

\this{b} was previously analyzed as part of an ensemble AP analysis by
\citet{sliski:2014}. In that work, the authors concluded that \this{b}
displayed a strong AP effect, either because the candidate was a
false-positive or because it exhibited a strong photoeccentric effect
with $e\geq0.488_{-0.051}^{+0.025}$. Our analysis, using slightly more
data and a full $\omega$ marginalization is excellent agreement with
that result, finding $e=0.507^{+0.039}_{-0.114}$ and
$\omega=76^{+59}_{-24}$\,deg. The AP results may be compared to that
from our radial velocity solution, where again we find excellent
agreement, since RVs yield $e=0.5134^{+0.0098}_{-0.0089}$ and
$\omega=64.1^{+1.6}_{-1.5}$\,deg. The results may be visually compared
in \reffigl{ew}. We note that this is not the first time AP and RVs
have been shown to yield self-consistent results, with
\citet{dawson:2012} demonstrating the same for the Sun-like star HD
17156b. However, this is the first time that this agreement has been
established for a giant host star. This is particularly salient in
light of the work of \citet{sliski:2014}, who find that the AP
deviations of giant stars are consistently excessively large. The
authors proposed that many of the KOIs around giant stars were
actually orbiting a different star, with the remaining being eccentric
planets around giant stars. \this{b} falls into the latter category,
consistent with its original ambiguous categorization as being either
a false positive or photoeccentric.

The analysis presented here demonstrates that AP can produce accurate
results for giant stars. This indicates that the unusually high AP
deviations of giant stars observed by \citet{sliski:2014} cannot be
solely due to time-correlated noise caused by stellar granulation,
which has recently been proposed by \citet{barclay:2014} to explain
the discrepancy between AP and radial velocity data for
{\em Kepler}-91b. However, time-correlated noise may still be an important
factor in giant host stars which are more evolved than \this~(such as
{\em Kepler}-91), for which granulation becomes more pronounced
\citep{mathur:2011}. To further investigate these hypotheses, we
advocate for further observations of giant planet-candidate host stars
to resolve the source of the AP anomalies.

\subsection{A Benchmark for Compositions of Super-Jupiters}

\begin{figure}[!t]
\includegraphics[width=0.48\textwidth]{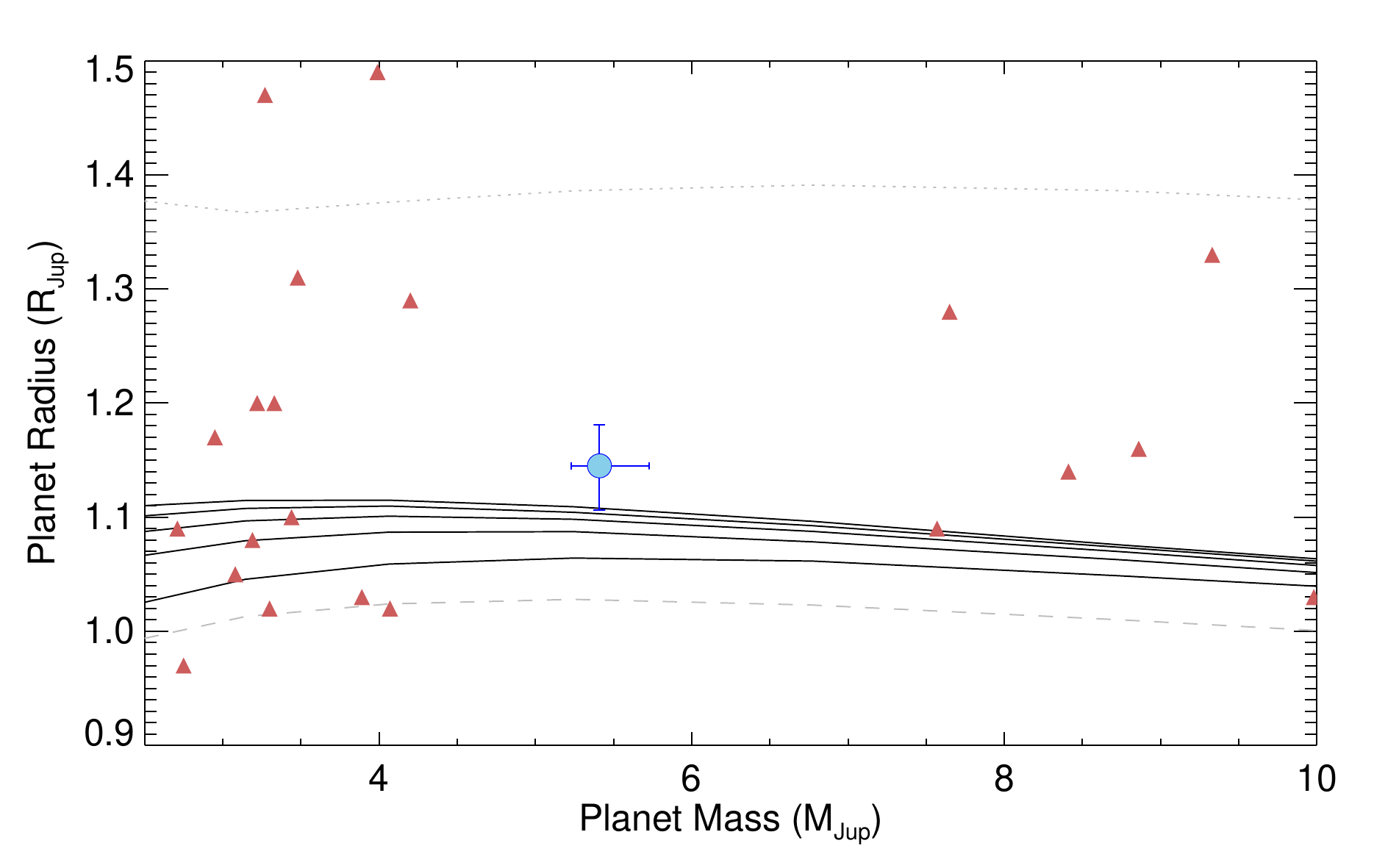}\hfill
\caption{ The masses and radii of known giant planets (red triangles)
  and \this{b} (blue circle). Overplotted as solid black lines are a
  series of planetary models \citep{fortney:2007} with age $3.5$~Gyr,
  an appropriate semi-major axis, and core masses $0$, $10$, $25$,
  $50$, and $100~\mearth$~(from top to bottom). Also plotted are
  models for young ($100$\,Myr), hot ($0.02$\,AU), coreless planets
  (dotted gray line); and old ($10$\,Gyr), cool ($1$\,AU) planets with
  massive ($100~\mearth$) cores (dashed gray line). These roughly
  illustrate the range of sizes possible for a given mass.
\label{fig:massradius}}
\end{figure}

\this{b} has a measured mass, radius, and age, which allows us to
investigate its bulk composition. Among transiting planets---i.e.,
those for which interior modeling is possible---there are only $11$
more massive than $4.5~\mjup$. We also point out the gap in the center
of the plot in \reffigl{massradius}; no planets with masses between
$4$~and $7.25~\mjup$ also have measured radii, so \this{b} immediately
becomes a valuable data point to modelers. Moreover, most of the
super-Jupiters are highly irradiated, which further complicates
modeling and interpretation of planetary structure. \this{b} receives
only about $20\%$~of the insolation of a $3$-day hot Jupiter orbiting
a Sun-like star. Because of this, it may prove to be an important
benchmark for planetary interior models---for example, as a means of
checking the accuracy of models of super-Jupiters without the
complication of high levels of incident flux. In \reffigl{massradius},
we compare the mass and radius of \this{b} to age- and
insolation-appropriate planetary models \citep{fortney:2007} of
varying core mass. We interpolate the models to a planet with the age
of \this~($3.5$~Gyr) in a circular orbit of $0.092$~AU around a
Sun-like star (the insolation of which is identical to the
time-averaged insolation of \this{b} on its more distant eccentric
orbit around a more luminous star). The radius is apparently slightly
inflated, but is somewhat consistent ($1$-$\sigma$) with that of a
planet lacking a core of heavy elements. This is not an iron-clad
result, though; the radius even agrees with the prediction for a
planet with a $50~\mearth$~core to within
$1.5$-$\sigma$. Nevertheless, we interpret this as evidence that
\this{b} most likely has only a small core of heavy elements.

\subsection{Jupiters Do (Briefly) Orbit Giants Within $0.5$\,AU}

\begin{figure}[!t]
\includegraphics[width=0.48\textwidth]{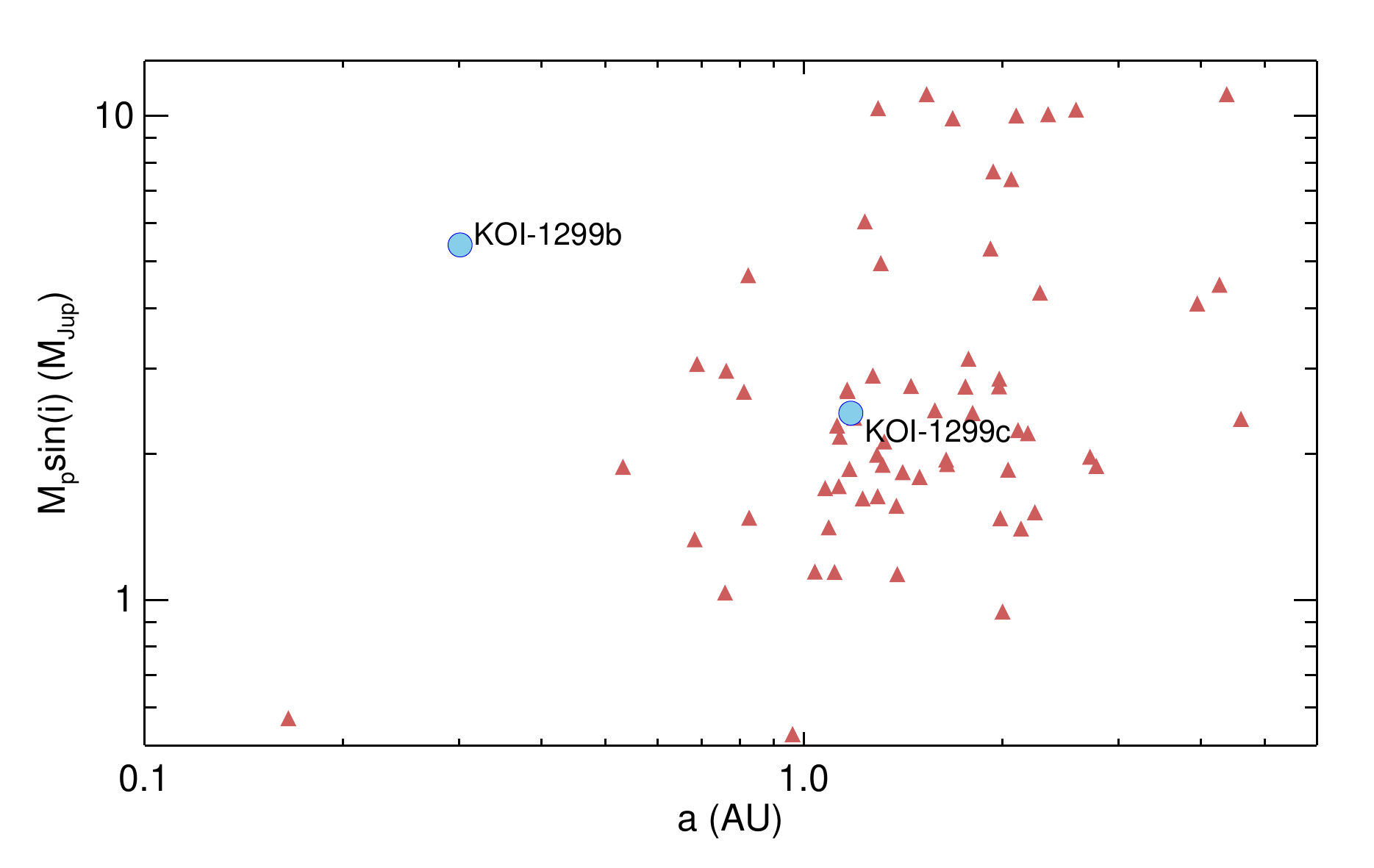}\hfill
\caption{ All planets with minimum masses $M_{\rm p}\sin{i} >
  0.5~\mjup$ orbiting giant stars ($\log{g_\star} < 3.9$), plotted as
  red triangles. \this{b} and c are labeled and plotted as large blue
  circles.
\label{fig:mass_semi}}
\end{figure}

As discussed previously, we do not know of many transiting giant
planets orbiting giant stars. This is not because giant planets
orbiting the progenitors to giant stars are rare; due in part to
survey strategies, giant stars are, on average, more massive than the
known main-sequence planet hosts, and massive stars seem to be {\em
more} likely to harbor massive planets. However, because very few
giant stars host planets inside $1$\,AU, the a priori probability of a
transit is low for these systems. In fact, there are only two other
giant planets with a measured mass ($M_p > 0.5~\mjup$) transiting a
giant star ($\log{g_\star} < 3.9$)---{\em Kepler}-56c \citep{huber:2013b},
and the soon-to-be-swallowed hot Jupiter {\em Kepler}-91b
\citep{barclay:2014,lillo:2014a,lillo:2014b}. This makes \this{b}
important in at least two respects: it is a rare fully characterized
giant planet transiting a giant star, and it is a short-period outlier
among giant planets orbiting giant stars (see
\reffigl{mass_semi}). \this{b} thus gives us a new lens through which
to examine the dearth of short and intermediate period Jupiters
($<0.5$\,AU) around giant stars. \this{c}, on the other hand, appears
to be very typical of Jupiters around evolved stars, in both mass and
orbital separation.

After noticing that \this{b} sits all alone in the $M_{\rm p}$~versus
$a$~parameter space, it is natural to wonder why, and we suggest two
potential explanations, each of which may contribute to this apparent
planetary desert. In the first, we consider that \this{b} may simply
be a member of the tail of the period distribution of planets around
massive and intermediate-mass main-sequence stars. That is, perhaps
massive main-sequence stars simply harbor very few planets with
separations less than $1$\,AU. If this occurrence rate is a smooth
function of stellar mass, then because \this~would most accurately be
called intermediate mass, it might not be so surprising that it
harbors a planet with a separation of $0.3$\,AU while its more massive
counterparts do not. If this is the case, then as we detect more giant
planets orbiting giant stars (and main-sequence stars above the Kraft
break), we can expect to find a sparsely populated tail of planets
interior to $1$\,AU. This may ultimately prove to be responsible for
the observed distribution, and would have important implications for
giant planet formation and migration around intermediate mass stars in
comparison to Sun-like stars, but it cannot be confirmed now. It will
require additional planet searches around evolved stars and
improvements in detecting long period planets orbiting rapidly
rotating main-sequence stars.

In the second scenario, we consider that \this{b} may be a member of a
more numerous group of planets interior to $\mysim 0.5$\,AU that exist
around main-sequence F- and A-type stars, but do not survive through
the red giant phase when the star expands to a significant fraction of
an AU. If this is the case, then \this{b} is fated to be swallowed by
its star through some combination of expansion of the stellar
atmosphere and orbital decay, and we only observe it now because
\this~has only recently begun its ascent up the red giant
branch. Suggestively, the resulting system that would include only
\this{c} would look much more typical of giant stars. In this
scenario, planets must begin their orbital decay very soon after the
star evolves off of the main sequence, or we would expect to observe
many more of them. Assuming, then, that \this{b} is close enough to
its host to have started its orbital decay, we may be able to detect
evidence of tidal or magnetic SPI \citep[see,
e.g.,][]{shkolnik:2009}. As the star expands those interactions would
strengthen, which should lead to more rapid orbital decay. While
typically tidal interaction leading to circularization and orbital
decay has been modeled to depend only on tides in the planet,
\citet{jackson:2008} show that tides in the star (which are strongly
dependent on the stellar radius) can also influence orbital
evolution. This provides a tidal mechanism for more distant planets to
experience enhanced orbital evolution as the star expands. Recent
simulations by \citet{strugarek:2014} further demonstrate that in some
scenarios (especially with strong stellar magnetic fields and slow
rotation), magnetic SPI can be as important to orbital migration as
tidal SPI, and can lead to rapid orbital decay. \this~does rotate
slowly, and the results of a two-decade survey by
\citet{konstantinova-antova:2013} reveal that giants can exhibit field
strengths up to $\mysim 100$~G. Encouraged by these findings, we
search for evidence of SPI in the \this~system in \refsecl{tidal_ev}.

The hypothetical scenario in which planets inside $\mysim 0.5$\,AU
around massive and intermediate-mass stars ultimately get destroyed
may also help explain the properties of the observed distribution of
planets orbiting evolved stars. If planets form on circular orbits,
then in order to excite large eccentricities, a planet must experience
dynamical interaction with another planet
\citep[e.g.,][]{rasio:1996,juric:2008} or another star
\citep[e.g.,][]{fabrycky:2007,naoz:2012}. More massive stars (like
many of those that have become giants) are more likely than lower
mass, Sun-like stars to form giant planets, and they are also more
likely to have binary companions
\citep[e.g.,][]{derosa:2014,raghavan:2010}, so one might expect the
typical planet orbiting a giant star to be {\em more} eccentric, not
less. However, the most eccentric planets around evolved stars would
have pericenter distances near, or inside the critical separation at
which planets get destroyed. This could lead either to destruction by
the same mechanisms as the close-in planets, or to partial orbital
decay and circularization. The consequence of either process would be
more circular orbits on average and an overabundance of planets at
separations near the critical separation ($0.5\,{\rm AU} \lsim a \lsim
1\,{\rm AU}$), which is indeed observed. For similar arguments and
additional discussion, see \citet{jones:2014}. Counterarguments
articulated by those authors include that massive subgiants do not
seem to host many planets inside $1$\,AU either, but subgiants are not
large enough to have swallowed them yet. This might point toward a
primordial difference in the planetary period distribution between
Sun-like and more massive stars, rather than a difference that evolves
over time.

\subsection{Implications for Giant Planet Formation and Migration}

Assuming that giant planets form beyond the ice line \citep[located at
$\mysim 4.9$\,AU for a star with $M\mysim 1.4~\msun$, using the
approximation of][]{ida:2005}, both \this{b} and c have experienced
significant inward migration. At first glance, their large orbital
eccentricities would suggest that they have experienced gravitational
interactions, perhaps during one or more planet--planet scattering
events that brought them to their current orbital distances, or
through the influence of an outer companion (like that detected in our
AO images). However, the apparent alignment between the stellar spin
axis and the orbit of the inner planet presents a puzzle; if the
(initially circular, coplanar) planets migrated via multi-body
interactions, we would expect both the eccentricities {\em and} the
inclinations to grow. It is possible, of course, that by chance the
inner planet remained relatively well-aligned after scattering, or
that the planets experienced coplanar, high-eccentricity migration,
which has recently been suggested as a mechanism for producing hot
Jupiters \citep{petrovich:2014}. Measuring the mutual inclination
between the planets could lend credence to one of these
scenarios. Unfortunately, because we are unable to constrain the
mutual inclination of the planets, we cannot say whether they are
mutually well-aligned (which would argue against planet-planet
scattering) or not.

\begin{figure}[!t]
\includegraphics[width=0.48\textwidth]{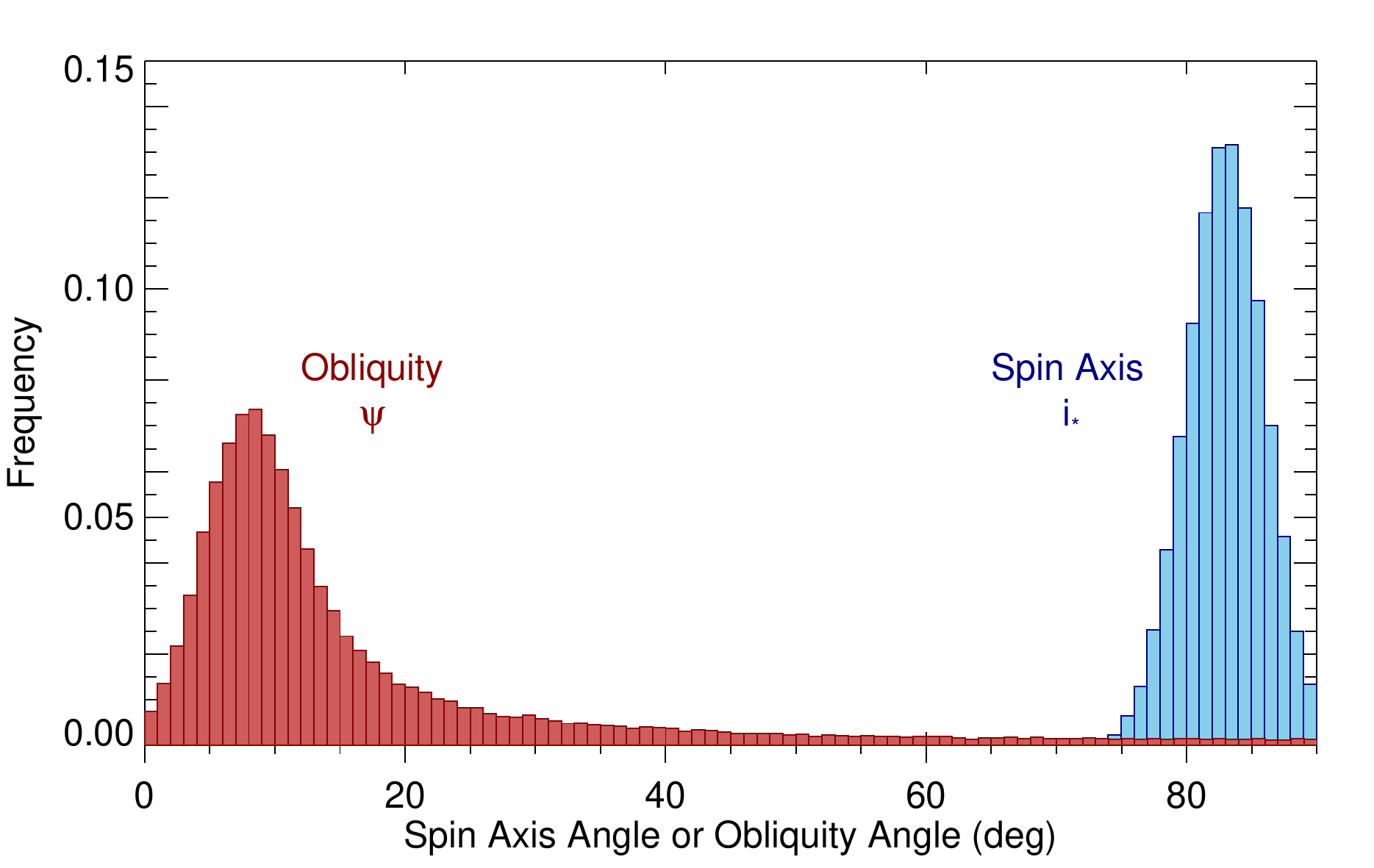}\hfill
\caption{ An underlying distribution of true obliquities, $\psi$,
  corresponding to a hypothetical measured stellar inclination,
  $i_\star$. A system with edge-on rotation has $i_\star = 90\dg$, and
  $\psi = 0\dg$~indicates perfect alignment between stellar spin and
  planetary orbital angular momentum. The long tail of high
  obliquities illustrates that even with a nearly edge-on stellar
  spin, a transiting planet may still be misaligned.
\label{fig:obliquity}}
\end{figure}

At this point, it is prudent to reiterate that the stellar spin and
orbital angular momenta are measured to be nearly aligned only as
projected along the line of sight. The true obliquity is unknown, and
may be significantly non-zero. In \reffigl{obliquity}, we illustrate
this by simulating the underlying obliquity distribution that
corresponds to an example $i_\star$~distribution with a mean of
$83\dg$, similar to that measured from asteroseismology. To generate
this obliquity distribution, we draw from $i_\star$ and a random
uniform azimuthal angle, $\theta$, and calculate the true obliquity,
$\psi$.  As is apparent from the $\psi$~distribution, while it is
likely that the true obliquity is low, there is a tail of highly
misaligned systems (representing those with spin axes lying nearly in
the plane of the sky but not coincident with the orbital angular
momentum) that could produce the stellar inclination we
observe. Ideally, we would also measure the sky-projected spin--orbit
angle, $\lambda$, in order to detect any such conspiring misalignment
and calculate the true obliquity. However, there is not an obvious way
to do this: the slow rotation, long period, and lack of star spots
render current techniques to measure the sky-projected obliquity
ineffective. We are left with an inner planet that is {\em likely},
but not certainly, aligned with the stellar spin axis, and an unknown
inclination of the outer orbital plane.

The likely alignment of the inner planet seems at odds with the large
orbital eccentricities and the assumption that multi-body interactions
are responsible for their migration, but we have thus far neglected
interaction between the planets and the host star. In their
investigation of hot Jupiter obliquities, \citet{winn:2010} suggested
that well-aligned hot Jupiters may have been misaligned previously,
and that subsequent tidal interaction may have realigned the stellar
rotation with the planetary orbit, perhaps influencing the convective
envelope independently of the interior. This idea was furthered by
\citet{albrecht:2012}, who found that systems with short
tidal-dissipation timescales are likely to be aligned, while those
with long timescales are found with a wide range of obliquities. In
the following section, we explore the possibility that \this{b} has
undergone a similar evolution, obtaining an inclined orbit through its
inward migration, but subsequently realigning the stellar spin.

\subsection{Evidence for Spin--Orbit Evolution}
\label{sec:tidal_ev}

During the main-sequence lifetime of \this, the current position of
the planets would be too far from the star to raise any significant
tides, but the star is now several times its original size, such that
at periastron, the inner planet passes within $7.7$~stellar radii of
the star. The unusually large planetary mass also helps strengthen
tidal interactions. Following \citet{albrecht:2012} \citep[who in turn
used the formulae of][]{zahn:1977}, for \this~we calculate the tidal
timescale for a star with a convective envelope:

\begin{equation}
  \tau_{\rm CE} = (10^{10}~{\rm yr}) q^{-2} \left(\frac{a/R_\star}{40}\right)^{6}.
\end{equation}

For a radiative atmosphere, as \this~would have had on the main
sequence, the appropriate timescale would be

\begin{equation}
  \tau_{\rm RA} = (1.25 \times 10^9~{\rm yr}) q^{-2} (1 + q)^{-5/6} \left(\frac{a/R_\star}{6}\right)^{17/2}.
\end{equation}

\begin{figure}[!t]
\includegraphics[width=0.48\textwidth]{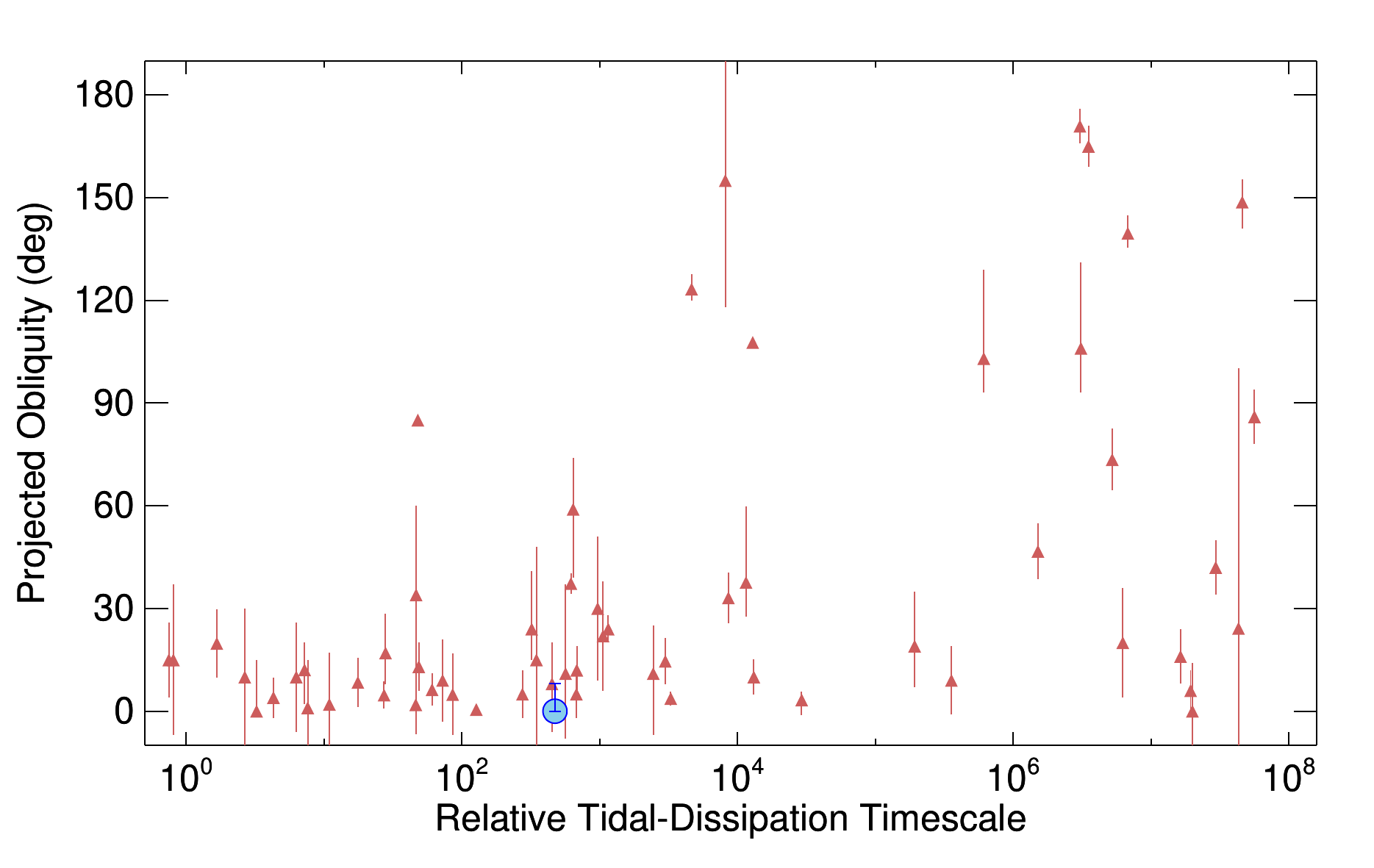}\hfill
\caption{ An adapted version of Figure 24 from
  \citet{albrecht:2012}. We plot the projected obliquity of giant
  planets versus the relative tidal timescale. We include planets from
  \citet{albrecht:2012} as well as $8$~more recent spin--orbit
  measurements from the literature (red triangles) and \this{b} (large
  blue circle). Because it sits in a region with well-aligned hot
  Jupiters, we posit that \this{b} has also realigned the spin of its
  host star.
\label{fig:tidal}}
\end{figure}

We repeat the words of caution from \citet{albrecht:2012}: these
equations are calibrated to star--star interactions rather than
star--planet interactions, and are timescales for spin--orbit
synchronization, not realignment. As such, the timescales are assumed
to be valid only as a relative metric. The authors arbitrarily divided
the resulting values by $5 \times 10^9$~to express this point, and to
set the values on a convenient scale. In \reffigl{tidal}, we have
reproduced a version of Figure 24 of \citet{albrecht:2012}, showing
the relative tidal dissipation timescales for all hot Jupiters with
measured spin--orbit angles. We have excluded the same systems they
did, and added $9$~recent projected obliquity measurements, including
that of \this{b}. Their result still holds true; systems with short
tidal dissipation timescales are well aligned, while those with long
tidal timescales display a wide range of obliquities. We also note
that recent work by \citet{valsecchi:2014}, which includes a more
detailed treatment of the convection and stellar evolution of each
star, similarly concludes that the observed obliquity distribution can
be explained by tidal evolution.  While the angle measured for
\this{b} is the line of sight projection rather than the sky-plane
projection, and the efficiency of realignment may be different for
evolved stars given their different internal structure, \this{b} does
have a short tidal timescale, and it sits in a region with hot Jupiter
systems that are mostly well-aligned. We interpret this result as
evidence that the spin axis of \this~has been realigned to the orbit
of the inner planet by the same mechanism responsible for hot Jupiter
realignment.

Alternative mechanisms have been proposed to explain the alignment for
hot Jupiters orbiting cool stars. One such alternative, suggested by
\citet{dawson:2014}, is that the orbital alignment trend with stellar
temperature is due to strong versus weak magnetic breaking (for stars
below and above the Kraft break), rather than tidal dissipation
efficiency as in \citet{albrecht:2012}. This has the attractive
quality of being part of a theory that is also able to predict the
observed mass cut-off for retrograde planets and the trend between
planet mass and host star rotation periods. In the \citet{dawson:2014}
framework, one may still conclude that \this~would realign during its
red giant phase, but this depends in part on the unknown magnetic
properties of the star: the \citet{konstantinova-antova:2013} finding
of strong magnetic fields in some giants may indeed indicate that
\this~possesses a short enough magnetic braking timescale to allow for
realignment in this framework, too.

In the event that the envelope and the core have different spin axis
inclinations, this might manifest itself in the asteroseismic
modes. Because $g$-mode-dominated $\ell=1$~modes are most sensitive to
core rotation while $p$-mode-dominated mixed modes are most sensitive to
the surface, one might obtain different inclination measurements from
the different modes. However, our analysis of individual
$\ell=1$~modes in \refsecl{inc} showed they were consistent with
$\mysim 90\dg$~inclination for both core and envelope. This would
argue {\em against} a realigned envelope for \this, but a modest
misalignment might be undetectable given the degraded precision when
fitting modes independently. Furthermore, as shown for the obliquity
and spin axis in \reffigl{obliquity}, even if the two spin axes both
lie in the sky plane, there is a chance they could be significantly
misaligned. Regardless of its efficacy in the case of \this,
asteroseismology may be able to detect a significant misalignment
between core and envelope in other systems hosting giant planets with
short tidal timescales. If detected, this would be important evidence
in the interpretation of giant planet migration, and we suggest this
be explored for other systems.

\begin{figure}[!t]
\includegraphics[width=0.48\textwidth]{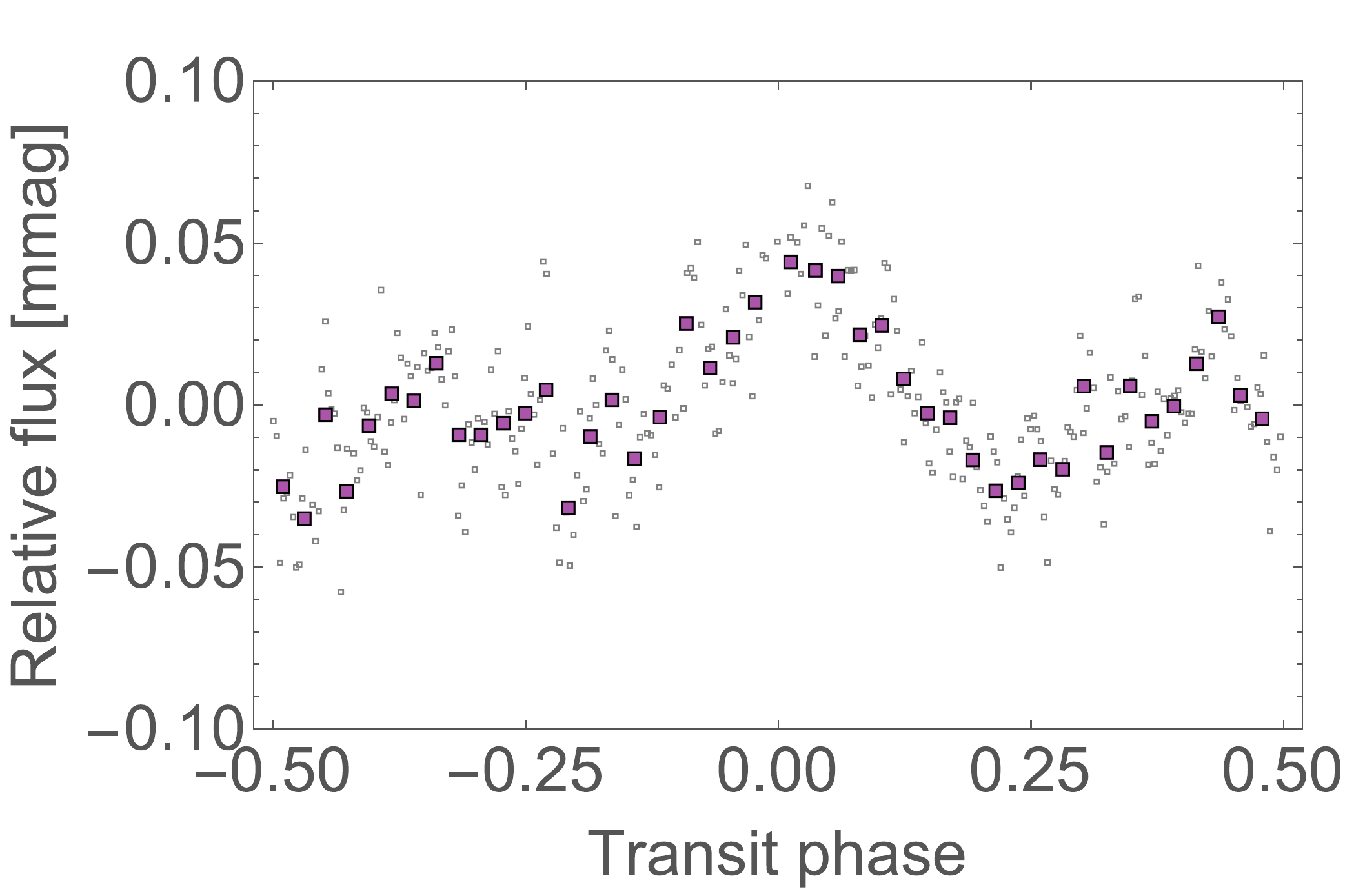}\hfill
\caption{ The {\em Kepler} light curve, binned and folded to the
  period of the transiting planet, with the transits removed. Given
  that periastron occurs just before transit and the stellar rotation
  period is longer than the orbital period, the brightening just after
  transit (orbital phase 0) is what would be expected for a bright
  spot that is tidally excited during the periastron passage of the
  planet.
\label{fig:fold_spot}}
\end{figure}

While the evidence presented thus far for realignment of the
\this~system is circumstantial, there may be additional evidence to
support the scenario. If tidal forces are strong enough to realign the
spin axis of the host star, they may also excite activity on the
stellar surface. SPI need not be limited to tidal interaction,
though. At closest approach, the planet is only $7.7$~stellar radii
from the star, and thus magnetic interaction may also be possible,
since the planet passes inside the Alfv\'en radius ($\mysim
10\,R_\star$). We return to the light curve to look for signs of these
interactions. When folded on the orbital period, an interesting
feature emerges (see \reffigl{fold_spot}). Just after transit, we
detect a brightening of $\mysim 0.05$\,mmag that maintains a coherent
shape and phase throughout the mission. The simplest explanation for a
light curve feature with the period of the planet is a phenomenon
associated with the planet. We suggest that SPI excites a bright spot
on the stellar photosphere each orbit during periastron
passage. Periastron occurs $1.24$~days before transit, and because the
rotation period is longer than the orbital period, the spot should
rotate onto the meridian and reach maximum apparent brightness
slightly after transit, as is observed.

One might argue that if the planet has realigned the stellar spin, it
may have also synchronized it to the orbital period---this would
cause {\em any} long-lived bright spot to create a peak in the light
curve when folded to the orbital period. However, to obtain the
observed phasing, the spot would have to conveniently occur at the
longitude that coincides with the close approach of the planet. There
is also no reason to expect a long-lived bright spot on the star,
irrespective of its phasing. Furthermore, there exists evidence that
the spin is {\em not} synchronized to the orbital period anyway. We
have measured $R_\star$, $i_\star$, and $v\sin{i_\star}$, so we can
estimate the rotation period to be $P_{\rm rot}\mysim 77 \pm 14$~days,
which is inconsistent with the orbital period ($P_{\rm orb} =
52.5$~days). Thus, a long-lived rotating star spot should not produce
a coherent photometric signal when folded to the orbital
period. (Similarly, we must conclude that if $P_{\rm rot} \neq P_{\rm
orb}$, any planet-induced spot must decay in brightness relatively
quickly, or the brightening feature would appear at different phases
in each orbit, washing out the coherent signal that we observe.) We do
caution that our $v\sin{i_\star}$ measurement for this slowly rotating
giant could be biased, for example, due to the unknown macroturbulent
velocity of \this. However, even if our measurement of the rotational
velocity is incorrect and the rotation period {\em is} synchronized to
the orbit, we believe a bright star spot phased with the periastron
passage of the planet would more easily be attributed to SPI than
coincidence.

\begin{figure}[!t]
\includegraphics[width=0.48\textwidth]{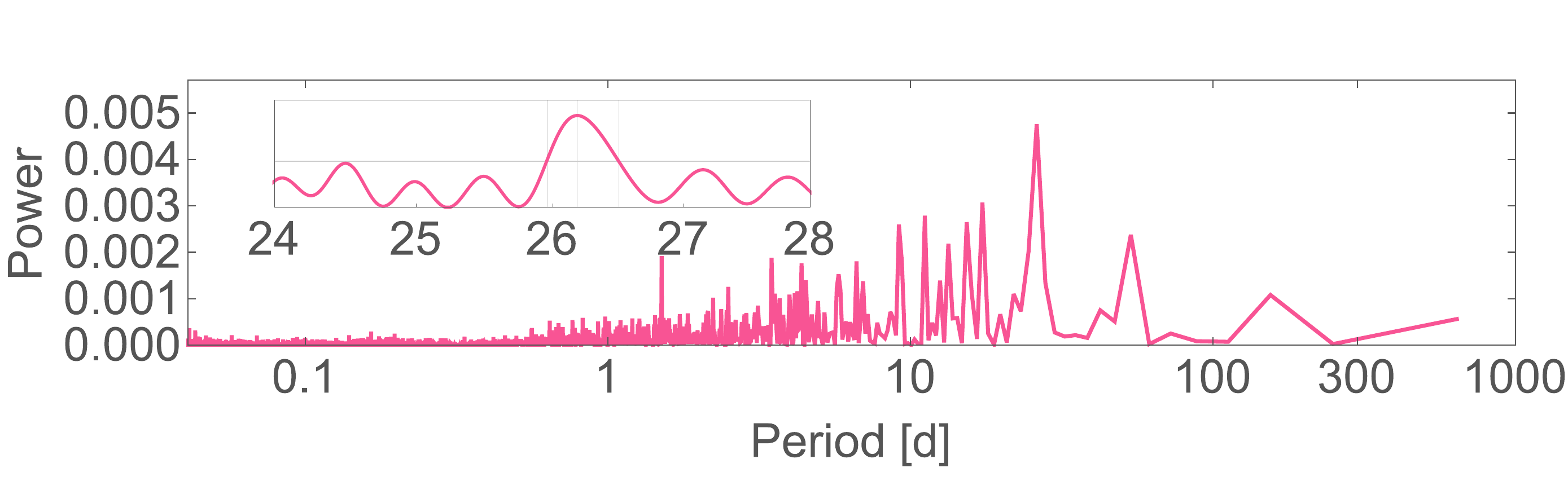}\hfill
\caption{ A periodogram of the {\em Kepler} light curve, after
  transits were removed. The maximum power occurs at a period of
  $\mysim 26.2$~days, or half of the orbital period of the inner
  planet.
\label{fig:periodogram}}
\end{figure}

One additional worry would be that granulation could produce
correlated noise that happens to fold coherently on the orbital
period. We find this to be an unlikely scenario given the periodogram
of the {\em Kepler} data (see \reffigl{periodogram}). If granulation
produced light curve amplitudes similar to that observed, we would
expect several periods to show similar power simply by
chance. Instead, we observe the peak power at the second harmonic of
the orbital period ($\mysim 26.2$~days), and very little power at
other periods. It is not particularly surprising that the second
harmonic, rather than the first, is the strongest peak, especially
given that there is some additional structure in the light curve at
phases other than around transit.

The observant reader might notice a possible brightening of lower
amplitude at phase $\mysim 0.4$~of the folded light curve. We are not
confident that this is real, but if it is, it holds a wealth of
information. As mentioned above, any excited spot must decay in
brightness quickly, or much of the star would soon be covered in such
spots, obfuscating the coherent signal that we observe. If, however, a
spot excited during periastron lasted only slightly more than one
rotation period, we would expect to see a second peak in the folded
light curve, with an amplitude dependent on the spot lifetime and
phase dependent on the rotation period. For \this, the second
peak $\mysim 1.4$~orbits after the first would imply a rotation period
of $\mysim 1.4 \times 52.5~{\rm days} \approx 73.5$~days, perfectly
consistent with our independently estimated rotation period. The
amplitude of the second peak is roughly half that of the first, so we
could conclude that the characteristic lifetime (half-life) of such a
spot is roughly the same as the rotation period, $\tau_{\rm spot}
\mysim 73.5$~days. We reiterate that we are not confident that the
data here is strong enough to make such conclusions; we merely note
the consistent rotation period it would imply.

\section{Summary}
\label{sec:summary}

We have presented herein the discovery of the \this~planetary
system, consisting of two giant planets orbiting a red giant star, and
a faint visual companion that is probably a physically bound M dwarf
with an orbital separation of at least $750$\,AU. The inner planet
($P=52.5$~days) transits the star, allowing a more detailed study of
its properties.

An asteroseismic analysis of the host star allows precise measurements
of the stellar mass, radius, age, and spin axis inclination. This in
turn leads to precise planetary properties, especially for the
transiting inner planet. N-body simulations have helped constrain the
properties of the outer planet by invoking stability arguments to rule
out a large fraction of orbital solutions.

\this{b} is among the most massive transiting planets orbiting any
type of star. Furthermore, it is not highly irradiated like many of
the more common short-period transiting planets, experiencing only
$\mysim 20\%$ of the insolation that a $3$-day hot Jupiter orbiting a
Sun-like star would receive. This makes it an interesting benchmark
for interior models, and a good test of planetary inflation at
moderate insolation. It appears to be slightly inflated compared to
the models, but is marginally consistent with having no core (or a
small core) of heavy elements.

The eccentricities of both planets are high, which is suggestive of
migration through multi-body interactions. Puzzlingly, the transiting
planet is likely well aligned with the stellar spin, which is not an
expected property of systems that have experienced such
interactions. However, subsequent tidal or magnetic interaction with
the host star may reconcile the two results. We find that because of
the large stellar radius and planetary mass, the tidal dissipation
timescale of the system is similar to that of hot Jupiters that are
well-aligned. Those planets are thought to be well-aligned due to
reorientation of the angular momenta after the initial inward
migration; we conclude that the same process that realigns hot Jupiter
systems may have also realigned this system, despite its long period.

Under the assumption that the star and planet are interacting strongly
enough to realign the stellar spin, we searched for evidence of this
interaction in the {\em Kepler} photometry. The star is
photometrically quiet on long timescales, which allowed us to detect a
low amplitude brightening soon after periastron. We conclude that the
most likely explanation is a bright spot excited (either tidally or
magnetically) by the planet during its close approach to the stellar
photosphere. This evidence for ongoing SPI further supports the
obliquity realignment scenario.

Finally, we note that \this{b} is an outlier among giant planets
orbiting giant stars. It is one of only three such planets orbiting
within $0.5$\,AU of its host star. Either it is intrinsically rare, or it
is one of a more common class of planets orbiting interior to $1$\,AU
that get destroyed relatively quickly once the star reaches the red
giant branch. Given that \this~has only recently started its ascent up
the RGB---most red giants are in a more advanced evolutionary
state---and there is already evidence for SPI, it is plausible that
many more red giants initially hosted planets similar to \this{b} that
have subsequently been engulfed. Further investigation of this group
of planets will provide more clarity and ultimately have strong
implications for giant planet formation around intermediate and high
mass stars, which are more difficult to study on the main sequence
because of their high temperatures and rapid rotation.

\acknowledgements 

We thank Russel White and an anonymous referee for valuable discussion
and feedback. S.N.Q. is supported by the NSF Graduate Research
Fellowship, Grant DGE-1051030. D.W.L. acknowledges partial support
from NASA's Kepler mission under Cooperative Agreement NNX11AB99A with
the Smithsonian Astrophysical Observatory. D.H. acknowledges support
by the Australian Research Council's Discovery Projects funding scheme
(project number DEI40101364) and support by NASA under Grant
NNX14AB92G issued through the Kepler Participating Scientist
Program. M.J.P. gratefully acknowledges the NASA Origins of Solar
Systems Program grant NNX13A124G. Funding for the Stellar Astrophysics
Centre is provided by The Danish National Research Foundation (Grant
DNRF106). The research is supported by the ASTERISK project
(ASTERoseismic Investigations with SONG and Kepler) funded by the
European Research Council (Grant agreement no.: 267864). The research
leading to the presented results has also received funding from the
European Research Council under the European Community's Seventh
Framework Programme (FP7/2007-2013) / ERC grant agreement no 338251
(StellarAges).

This publication makes use of data products from the Two Micron All
Sky Survey, which is a joint project of the University of
Massachusetts and the Infrared Processing and Analysis
Center/California Institute of Technology, funded by the National
Aeronautics and Space Administration and the National Science
Foundation. This research has also made use of the APASS database,
located at the AAVSO web site. Funding for APASS has been provided by
the Robert Martin Ayers Sciences Fund.

{\it Facilities:} \facility{Kepler}, \facility{FLWO:1.5m (TRES)},
\facility{NOT (FIES)}, \facility{Keck:II (NIRC2)}, \facility{WIYN
(DSSI)}

\end{document}